\documentclass[a4paper,11pt]{article}
\usepackage{iftex}
\ifPDFTeX
\pdfoutput=1
\fi
\usepackage{jheppub}
\usepackage[T1]{fontenc}
\usepackage{lmodern}
\usepackage{bm}
\usepackage{braket}
\usepackage{xcolor}
\usepackage{tikz}
\usetikzlibrary{shapes,arrows,positioning,automata,backgrounds,calc,er,patterns,snakes,decorations.pathmorphing,decorations.markings}
\usepackage[compat=1.1.0]{tikz-feynman}
\tikzstyle{wave}=[decoration={aspect=1,segment length=2.1mm,amplitude=0.3mm,snake},decorate]

\newcommand{\dd}{{\rm d}}

\newcommand{\K}{{\bm k}}

\allowdisplaybreaks[1]
\newcommand{\GenericSourceBlob}{%
 \begin{gathered}
 \begin{tikzpicture}[scale=.8,baseline=(current bounding box.center)]
  \draw[thick] (-.34,.36)--(-.72,.88) (.34,.36)--(.72,.88);
  \draw[thick] (-.34,-.36)--(-.72,-.88) (.34,-.36)--(.72,-.88);
  \fill[gray!25] (0,0) circle (.5);
  \foreach \x in {-.72,.72}{
   \fill (\x,.88) circle (1.5pt);
   \fill (\x,-.88) circle (1.5pt);
  }
  \node at (0,.88) {$\cdots$};
  \node at (0,-.88) {$\cdots$};
 \end{tikzpicture}
 \end{gathered}%
}

\begin{document}
\title{Five-loop static two-body potential in massless scalar-tensor gravity}
\author{Gabriel Luz Almeida,}
\author{Zhengwen Liu,}
\author{and Jia Zhang}
\affiliation{School of Physics \& Shing-Tung Yau Center, Southeast University,\\
Nanjing 210096, China}
\emailAdd{galmeida@seu.edu.cn}
\emailAdd{zhengwen.liu@seu.edu.cn}
\emailAdd{jia.zhang@seu.edu.cn}
\abstract{We compute the static contribution to the conservative two-body potential in massless scalar-tensor theories through fifth post-Newtonian (5PN) order, with the strong-field scalar responses of the compact objects encoded in scalar-dependent masses. Within post-Newtonian effective field theory (EFT), evaluating the 1PN--5PN static terms requires one- through five-loop massless two-point Feynman integrals. Although the additional scalar modifies both source couplings and integrand numerators, the required scalar master topologies remain those of the corresponding static problem in general relativity. Moreover, independent reflection symmetries of the two static scalar sectors extend the odd-PN factorization mechanism to the present theory, so that no genuinely new five-loop master integrals are required. Our results reproduce the known static interaction through 2PN, provide an independent EFT determination at 3PN, and extend the explicit potential to 4PN and 5PN. We verify the general-relativity limit and body-exchange symmetry; regulator poles and residual transcendental terms cancel in the complete result, leaving rational coefficients. These high-precision results provide ingredients for increasingly precise gravitational-wave tests of gravity beyond general relativity.}

\maketitle

\section{Introduction}
\label{sec:intro}

Gravitational-wave observations of compact binaries connect the relativistic two-body problem with measurements of gravity in strong-field systems~\cite{LIGOScientific:2016aoc,LIGOScientific:2021djp}.
Turning these observations into tests of gravity requires theoretical models whose accuracy keeps pace with the information carried by the data. The scientific programs of future observatories sharpen this need, motivating systematic improvements in binary dynamics both within and beyond general relativity (GR)~\cite{Maggiore:2019uih,LISA:2022kgy}. For a specified alternative theory, this program begins in part with the conservative interaction, which determines the orbital dynamics to be combined with the radiation generated by the binary.

The appropriate description of that dynamics depends on the stage of coalescence, conventionally divided into inspiral, merger, and ringdown. For gravitationally bound binaries in the weak-field, slow-motion inspiral, the post-Newtonian (PN) expansion provides a controlled description based on the virial counting $v^2\sim GM/r$. Here $v$ is the relative orbital velocity, $r$ the binary separation, and $M=m_1+m_2$ the total mass, with $m_1$ and $m_2$ the bodies' masses and $G$ the gravitational constant.
Merger generally requires strong-field numerical methods, whereas the subsequent ringdown can be described by perturbations and quasinormal modes of the final compact object.
The conservative PN interaction studied here thus contributes to the analytic description of inspiral.

Within the inspiral regime, the effective field theory (EFT) formulation of the PN expansion, originally introduced as nonrelativistic general relativity (NRGR)~\cite{Goldberger:2004jt}, separates the internal, orbital, and radiation scales and organizes the classical calculation in terms of worldline and bulk interactions. With finite-size information encoded in worldline operators, the potential can then be computed using Feynman diagrams. This organization connects the classical gravitational problem to algebraic and multiloop methods developed in quantum field theory (QFT), while retaining the physical distinction between classical spatial momentum integrations and quantum field loops.

That connection is especially useful in the static sector. More generally, at $n$PN, terms in the local dynamics can be classified schematically as $G^p(v^2)^{n+1-p}$, with the masses and powers of the separation understood. The static term is the highest-$G$ contribution, $G^{n+1}v^0$, and maps onto an $n$-loop massless two-point function at fixed spatial momentum transfer. It therefore carries the highest potential-loop order encountered at that PN order.
Isolating this sector reduces the kinematic complexity while retaining an important part of the multiloop challenge. In particular, the static 5PN problem lies at order $G^6$ and involves five spatial loops.

The GR static sector has accordingly provided a testing ground for these techniques. The four-loop contribution at 4PN was organized in terms of massless two-point master integrals in Ref.~\cite{Foffa:2016rgu}, and two complementary computations subsequently obtained the 5PN static potential \cite{Foffa:2019hrb,Blumlein:2019zku}. More recently, the program has reached the six-loop static interaction at 6PN and the 7PN static interaction obtained through factorization, with the Kaluza--Klein (KK) parametrization playing an important role in exposing the structure underlying factorization~\cite{Brunello:2025static,Brunello:2026static}.
These developments also revealed a structural difference between even and odd PN orders: odd-PN static diagrams do not require new prime bulk building blocks, namely diagrams that cannot be split into lower-order static pieces at composite worldline insertions.
In particular, as investigated in \cite{Brunello:2026static}, the reflection symmetry of the static KK scalar gives a correlation-function interpretation of this factorization.

These advances motivate extending the high-PN static program to massless scalar-tensor (ST) theories, a natural class in which to study additional gravitational degrees of freedom. Such theories supplement the metric with a scalar degree of freedom and include Brans--Dicke gravity \cite{Brans:1961sx} and the broader formulation of Damour and Esposito-Far\`ese~\cite{Damour:1992we}. Despite the simple two-derivative bulk action, compact objects can have nontrivial scalar responses, including spontaneous scalarization \cite{Damour:1993hw,Damour:1996ke}. At orbital scales this response is described by scalar-dependent masses $M_A(\varphi)$, where $\varphi$ is the fundamental scalar and $A=1,2$ labels the compact objects, following the skeletonized description of compact bodies \cite{1975ApJEardley,Damour:1992we}. Their derivatives about the asymptotic background supply the source couplings in the EFT, allowing the orbital calculation to be carried out without fixing a particular compact-object model. We use this description to determine the static point-particle potential through 5PN.

This effective description has already supported calculations of conservative PN dynamics well beyond the leading scalar exchange. The 1PN description was developed by Damour and Esposito-Far\`ese~\cite{Damour:1992we}, who also developed a diagrammatic 2PN treatment~\cite{Damour:1996field}, and Mirshekari and Will derived the equations of motion through 2.5PN, including the conservative 2PN terms~\cite{Mirshekari:2013vb}.
Bernard obtained the 3PN equations of motion and their associated Lagrangian and conserved quantities \cite{Bernard:2018hta,Bernard:2018ivi}; these results also underlie 3PN effective-one-body Hamiltonians \cite{Jain:2022nxs,Julie:2022qux}. Within EFT, the massless ST framework was developed in Ref.~\cite{Kuntz:2019zef}, and the conservative dynamics through 2PN was computed with the temporal KK parametrization in Ref.~\cite{Almeida:2024uph}. The latter supplies a direct lower-order comparison for the present static calculation.

The same effort to develop analytic binary dynamics also extends to scalars with curvature couplings, notably in Einstein--scalar--Gauss--Bonnet (ESGB) gravity. In that setting, analytic binary dynamics has been developed through PN descriptions of nonspinning bodies~\cite{BertiJulie2019,Almeida:2024uph} and EFT calculations including black-hole spin~\cite{Almeida:2024cqz}.
Alongside these PN results, post-Minkowskian (PM) calculations expand in $G$ without a slow-velocity expansion. A 3PM radial action with a perturbative scalar-charge truncation was obtained in \cite{WilsonGerow:2025conservative}. Recent PM-EFT work has obtained the conservative scattering angle through third PM order in massless ST theories~\cite{Bernard:2026PMEFT}, while the conservative impulse and scattering angle have also been computed through that order in ST and ESGB gravity~\cite{Almeida:2026abr}. These scattering results thus provide information complementary to the bound-inspiral PN expansion considered here.

Beyond these perturbative regimes, numerical-relativity simulations have explored dynamical scalarization in neutron-star binaries and its effect on late inspiral and merger~\cite{Barausse:2012da,Shibata:2013pra}, as well as black-hole--neutron-star coalescence in ST gravity~\cite{Ma:2023sok}. Massive-scalar studies address scalarization along quasi-equilibrium binary sequences~\cite{Kuan:2023hrh} and the subsequent dynamics of post-merger remnants~\cite{Lam:2024wpq}. Perturbative EFT descriptions of black-hole quasinormal modes, in turn, investigate ringdown in broader scalar-tensor theories admitting scalar hair~\cite{Franciolini:2018uyq}. These complementary approaches place the conservative interaction within a wider inspiral--merger--ringdown program, where it provides an essential ingredient for waveform modeling.

Our calculation covers every static point-particle sector from $G$ through $G^6$.
The $G$--$G^3$ terms reproduce the known results through 2PN, while $G^4$ supplies an independent EFT determination of the 3PN static sector, whose conservative dynamics is already available in other formulations. The $G^5$ and $G^6$ terms extend the explicit static potential to 4PN and 5PN. To our knowledge, the explicit general ST point-particle static potential at 4PN and 5PN has not previously been obtained. The novelty at these orders concerns this highest-$G$ sector and does not amount to a determination of the complete 4PN or 5PN dynamics.

To carry out this extension, our computation uses \texttt{QGRAF} for diagram generation, custom symbolic routines for classical two-body filtering, topology identification and momentum rerouting, \texttt{FORM} for index contractions and algebraic processing, and \texttt{FIRE 6.5} for integration-by-parts (IBP) reduction \cite{Nogueira:1991ex,Vermaseren:2000nd,Smirnov:2023FIRE65}.
The additional massless scalar modifies the source couplings and integrand numerators, but the required scalar master topologies are inherited from GR~\cite{Foffa:2016rgu,Foffa:2019hrb,Blumlein:2019zku}. The two static scalars have the same propagator momentum dependence, and the bulk vertices $\sigma^k\phi^2$ have counterparts $\sigma^k\Phi^2$, explaining the reuse of the GR denominator families.
We adopt the KK parametrization, which has proved particularly useful for high-PN static calculations in GR.
Thus, the static bulk theory considered here also retains independent reflection symmetries in its two scalar sectors: the gravitational scalar arising from the KK decomposition and the additional scalar-tensor degree of freedom. These symmetries preserve the selection rule responsible for odd-PN factorization, so the 5PN sector requires no genuinely new five-loop master integrals.

The resulting potential can be tested in several complementary ways. In addition to reproducing the known static ST results through 2PN, it reduces to the corresponding GR potential when the scalar couplings are removed. We also verify body-exchange symmetry and, at each order from 1PN through 5PN, finiteness and the absence of residual transcendental constants. The rational final coefficients result from combining the regulated contributions with their full dimensional dependence. These properties provide complementary checks of the multiloop computation and establish the static interaction as a controlled input to the remaining dynamical sectors.

To present the calculation, Section~\ref{sec:ST} first fixes the ST conventions and source parameters. Section~\ref{sec:EFT} develops the static EFT rules, loop counting, reduction strategy, and factorization argument.
The potential is organized by PN order in Section~\ref{sec:results}, and its checks and scope are discussed in Section~\ref{sec:checks}. Section~\ref{sec:conclusions} contains our conclusions. Appendix~\ref{app:mass-coefficients} gives the scalar-response dictionary, Appendix~\ref{app:integrals} lists the independent topologies, propagator bases, and master bases at each loop order, and Appendix~\ref{app:feynman-rules} collects the Feynman rules.

\section{Massless scalar-tensor theories}
\label{sec:ST}

In this work, we consider massless scalar-tensor (ST) theories in which gravity is mediated by the metric and a single scalar field, with matter coupled universally to a conformally related metric. This class includes Brans--Dicke theory and the models studied by Damour and Esposito-Far\`ese, which allow for nonperturbative strong-field phenomena such as spontaneous scalarization~\cite{Brans:1961sx,Damour:1992we,Damour:1993hw}.
For the two-body calculation, it is convenient to work in the Einstein frame, where the two-derivative bulk action takes a universal form and the choice of theory enters through the scalar coupling to matter:
\begin{equation}
 S=-\frac{2}{\kappa^2}\int\dd^{d+1}x\,\sqrt{|g|}
 \left[R-2g^{\mu\nu}\partial_\mu\varphi\partial_\nu\varphi\right]
 +S_{\mathrm m}[\Psi,\mathcal A^2(\varphi)g_{\mu\nu}].
 \label{eq:ST-action}
\end{equation}
Here $\Psi$ collectively denotes the matter fields, which couple minimally to the Jordan-frame metric $\widetilde g_{\mu\nu}=\mathcal A^2(\varphi)g_{\mu\nu}$.
Thus, within the class considered here, the gravitational theory is specified by the coupling function $\mathcal A(\varphi)$.

Throughout the calculation, we use units with $c=1$ and the mostly-minus signature, $\eta_{\mu\nu}=\mathrm{diag}(1,-1,\ldots,-1)$.
Greek indices run over spacetime components, $\mu,\nu=0,\ldots,d$, while Latin indices denote spatial components, $i,j=1,\ldots,d$. Our curvature conventions are
\begin{equation}
\begin{aligned}
 R^\mu{}_{\nu\rho\sigma}
 &=
 \partial_\rho\Gamma^\mu_{\nu\sigma}
 -\partial_\sigma\Gamma^\mu_{\nu\rho}
 +\Gamma^\mu_{\alpha\rho}\Gamma^\alpha_{\nu\sigma}
 -\Gamma^\mu_{\alpha\sigma}\Gamma^\alpha_{\nu\rho},
 \\
 R_{\mu\nu}&=R^\alpha{}_{\mu\alpha\nu}.
\end{aligned}
\end{equation}
To regulate the spatial momentum integrals appearing in the two-body calculation, we continue the number of spatial dimensions to $d=3-2\varepsilon$, with $\varepsilon$ the dimensional regulator, and write the gravitational coupling as
\begin{equation}
 \kappa^2=32\pi G_d,\qquad G_d=G\ell^{-2\varepsilon},
 \label{eq:gravitational-coupling}
\end{equation}
where $G$ is the bare Einstein-frame gravitational constant and $\ell$ is an arbitrary length scale. The constant $G$ sets the normalization of the bulk action; the Newton constant measured between weakly self-gravitating bodies also includes the scalar-mediated interaction. In four spacetime dimensions, their relation is
\begin{equation}
 G_{\mathrm N}
 =G\mathcal A_0^2(1+\alpha_0^2),
 \qquad
 \mathcal A_0=\mathcal A(\varphi_0),
 \qquad
 \alpha_0=
 \left.\frac{\dd\ln\mathcal A}{\dd\varphi}\right|_{\varphi_0},
 \label{eq:measured-newton-constant}
\end{equation}
where $\varphi_0$ denotes the asymptotic scalar value.
The factor $\mathcal A_0^2$ accounts for the conversion to Jordan-frame units, while $\alpha_0^2$ gives the relative contribution of scalar exchange~\cite{Damour:1992we,Damour:1996ke}.

With these conventions established, we turn to the effective description of the compact objects. At distances large compared with their sizes, the matter sector can be replaced by a point-particle action. For nonspinning bodies, its leading contribution is
\begin{equation}
 S_{\mathrm m}\longrightarrow S_{\mathrm{pp}}
 =-\sum_{A=1}^2\int M_A(\varphi)\,\dd\tau_A,
 \qquad
 \dd\tau_A=
 \sqrt{g_{\mu\nu}\,\dd x_A^\mu\dd x_A^\nu}.
 \label{eq:point-particle-action}
\end{equation}
Here $A=1,2$ labels the two compact objects, $x_A^\mu$ their worldlines, and $\tau_A$ their proper times.
This replacement, usually called skeletonization~\cite{1975ApJEardley,Damour:1992we}, incorporates the internal structure and gravitational binding energy of each body into the scalar-dependent mass $M_A(\varphi)$.
Although $\mathcal A(\varphi)$ no longer appears explicitly, its influence is therefore retained in these mass functions and their derivatives. Determining them requires matching to compact-object solutions for the chosen theory and internal state; for neutron stars, this also requires an equation of state.

To specify the resulting source couplings, we assume an asymptotically flat background with constant scalar value $\varphi_0$ and expand the mass functions about it:
\begin{equation}
\begin{aligned}
 M_A(\varphi)
 &=m_A\sum_{q=0}^{\infty}
 \frac{a_{A,q}}{q!}(\varphi-\varphi_0)^q,
 \qquad m_A=M_A(\varphi_0),
 \\
 a_{A,q}
 &=\frac{1}{m_A}
 \left.\frac{\dd^q M_A}{\dd\varphi^q}\right|_{\varphi_0},
 \qquad a_{A,0}=1.
\end{aligned}
 \label{eq:mass-expansion}
\end{equation}
Here $m_A$ is the mass at the asymptotic background and $a_{A,q}$ is its normalized $q$th scalar derivative. The derivatives refer to a fixed internal state on a smooth branch of compact-object solutions. For a cold, nonspinning neutron star, the standard prescription holds the baryon number and equation of state fixed~\cite{Damour:1996ke}. The expansion is local on that branch; it does not assume analyticity across a scalarization transition.
These coefficients describe how each body's mass responds to the ambient scalar field and hence encode its strong-field properties in the effective description. They can equivalently be expressed in terms of the customary scalar-response parameters, also referred to as Einstein-frame sensitivities,
\begin{equation}
 \alpha_A=
 \left.\frac{\dd\ln M_A}{\dd\varphi}\right|_{\varphi_0},
 \qquad
 \beta_A=
 \left.\frac{\dd^2\ln M_A}{\dd\varphi^2}\right|_{\varphi_0},
 \qquad
 \beta'_A=
 \left.\frac{\dd^3\ln M_A}{\dd\varphi^3}\right|_{\varphi_0},\qquad\ldots .
 \label{eq:response-hierarchy}
\end{equation}
Unlike $\alpha_0$, which characterizes the coupling function $\mathcal A(\varphi)$ at infinity, these parameters depend on the compact object as well as the underlying theory. In this notation, the first three expansion coefficients are
\begin{equation}
 a_{A,1}=\alpha_A,\qquad
 a_{A,2}=\alpha_A^2+\beta_A,\qquad
 a_{A,3}=\alpha_A^3+3\alpha_A\beta_A+\beta'_A.
\end{equation}
The corresponding dictionary through $a_{A,6}$, the highest derivative required by static 5PN power counting, is collected in Appendix~\ref{app:mass-coefficients}.
Keeping the coefficients $a_{A,q}$ explicit provides a compact notation and allows the results to be applied to different theories and compact-object models once the corresponding matching has been performed.

With these source parameters left general, our calculation determines the nonspinning point-particle contribution to the static two-body dynamics through 5PN, with the effects of strong self-gravity retained through $M_A(\varphi)$. This contribution is an essential ingredient of the compact-binary dynamics at this order. Tidal effects, in turn, are described by additional worldline operators beyond Eq.~\eqref{eq:point-particle-action}; their contributions can be incorporated separately and are outside the scope of the present calculation.

This exclusion covers scalar, gravitational, and mixed tidal operators, whose contributions in ST theories have their own PN hierarchy~\cite{Bernard:2023eul}. No compact-object specialization of $M_A(\varphi)$ is needed below.

Finally, to prepare the action for the perturbative calculation described in the next section, we impose harmonic gauge by adding
\begin{equation}
 S_{\mathrm{GF}}=\frac{1}{\kappa^2}
 \int\dd^{d+1}x\,\sqrt{|g|}\,
 g_{\mu\nu}\Gamma^\mu\Gamma^\nu,
 \qquad
 \Gamma^\mu=g^{\rho\sigma}\Gamma^\mu_{\rho\sigma}.
 \label{eq:gauge-fixing}
\end{equation}
In what follows, $S_{\mathrm{bulk}}$ denotes the sum of the Einstein--Hilbert, scalar, and gauge-fixing actions.

\section{Effective Field Theory framework}
\label{sec:EFT}

\subsection{The static limit and effective action}
\label{sec:static-effective-action}
We now specialize the EFT framework introduced above to the static sector of massless ST theory~\cite{Goldberger:2004jt,Kuntz:2019zef,Almeida:2024uph}.
The hierarchy between the orbital and radiation scales allows the method of regions~\cite{beneke1998} to be applied, separating potential modes, with momenta scaling as $(k^0,|\bm k|)\sim(v/r,1/r)$, from radiation modes, for which $(k^0,|\bm k|)\sim(v/r,v/r)$.
The potential fields can then be integrated out to obtain the near-zone effective action for the two worldlines. For the present calculation, we extract its static sector by taking the worldlines and fields to be time independent, so that velocities, time derivatives, and retardation corrections drop out. The remaining terms therefore form an expansion in $GM/r$, with $M=m_1+m_2$ and the scalar-response coefficients held fixed. To assign a PN order to these terms, we use the usual bound-orbit counting $GM/r\sim v^2$, according to which a contribution proportional to $G^{n+1}$ is an $n$PN correction to the Newtonian interaction. Holding the worldlines fixed is thus a prescription for extracting this part of the potential, rather than an assumption that the bodies remain at rest under their mutual attraction.

Below we outline the elements of this EFT framework relevant to our calculation. For a more detailed account, including the extension of NRGR to ST theories and to other sectors, we refer the reader to Refs.~\cite{Kuntz:2019zef,Almeida:2024uph}.

To make the coupling expansion explicit from the outset, we normalize the perturbative fields by $\kappa$ and use the temporal Kaluza--Klein (KK) parametrization~\cite{Kol:2007bc,Kol:2010si,Almeida:2020mrg},
\begin{equation}
 g_{\mu\nu}=e^{2\kappa\Phi}
 \begin{pmatrix}
 1 & \quad & -\kappa A_j\\
 -\kappa A_i & \quad &\kappa^2 A_iA_j-e^{-c_d\kappa\Phi}\gamma_{ij}
 \end{pmatrix},
 \label{eq:KK}
\end{equation}
with $\gamma_{ij}=\delta_{ij}+\kappa\sigma_{ij}$ and $c_d=2(d-1)/(d-2)$.
Here $\Phi$ is a scalar, $A_i$ a spatial vector, and $\sigma_{ij}$ a symmetric rank-two spatial tensor, together parametrizing the $(d+1)(d+2)/2$ independent components of the metric perturbation. As we will see below, this decomposition considerably simplifies the static limit of the binary dynamics. 
The fundamental scalar is expanded with the same normalization,
\begin{equation}
 \varphi=\varphi_0+\kappa\phi.
 \label{eq:scalar-normalization}
\end{equation}
Thus $\Phi$ and $\phi$ denote distinct scalar fields, associated with the metric and the fundamental scalar, respectively.
With this normalization, the quadratic action is independent of $\kappa$.

For the PN expansion, it is natural to parametrize the worldlines by the coordinate time $t=x^0$, with $v_A^i=\dd x_A^i/\dd t$ the (spatial) velocity of body $A$. Using the KK metric, the proper-time element then takes the form
\begin{equation}
    \frac{\dd\tau_A}{\dd t}
    =
    e^{\kappa\Phi}
    \sqrt{
        (1-\kappa A_i v_A^i)^2
        -e^{-c_d\kappa\Phi}\gamma_{ij}v_A^i v_A^j
    },
\end{equation}
where all fields are evaluated on the corresponding worldline.
In the static limit, where $v_A^i=0$, this expression simplifies to $\dd\tau_A=e^{\kappa\Phi}\dd t$, so that the source action becomes
\begin{equation}
    S_{\mathrm{pp}}^{\mathrm{static}}
    =
    -\sum_{A=1}^{2}\int\dd t\,
    M_A(\varphi)e^{\kappa\Phi}.
    \label{eq:static-worldline}
\end{equation}
This form makes explicit that only the KK scalar $\Phi$ and the fundamental scalar perturbation $\phi$ couple directly to the static sources.

The absence of a vector coupling to the static sources has a further consequence. Since every worldline coupling to $A_i$ contains a velocity, it vanishes in the static limit, while every static bulk vertex involving $A_i$ has exactly two vector legs.
These legs must therefore connect exclusively to bulk vertices and, with no worldline endpoints available, necessarily form closed cycles. Such internal loops contribute only to quantum corrections and are thus excluded from the classical calculation. Consequently, no vector vertex can appear in the classical static diagrams, allowing us to set $A_i=0$ from the start.
The spatial tensor $\sigma_{ij}$, on the other hand, contributes through bulk interactions even though it does not couple directly to the static sources. Indeed, the vertices $\sigma\Phi^2$ and $\sigma\phi^2$ contain a single tensor leg, allowing a tensor propagator to connect two such vertices whose scalar legs attach to the worldlines, without forming an internal loop.

With $A_i=0$, and after dropping terms containing time derivatives as well as spatial total derivatives, the bulk action reduces to
\begin{equation}
    S_{\mathrm{bulk}}^{\mathrm{static}}
    =
    \int\dd t\,\dd^d x\,\sqrt{\gamma}\left[ \frac{2}{\kappa^2} \left(
        R[\gamma]
        -\frac{1}{2}\gamma^{ij}\mathcal C_i\mathcal C_j \right)
        -c_d\gamma^{ij}\partial_i\Phi\,\partial_j\Phi
        -4\gamma^{ij}\partial_i\phi\,\partial_j\phi
    \right],
    \label{eq:static-bulk}
\end{equation}
where $\gamma=\det\gamma_{ij}$ and $\mathcal C_i=\gamma_{ik}\gamma^{lm}\Gamma^k_{lm}[\gamma]$.
Here $R[\gamma]$ and $\Gamma^k_{lm}[\gamma]$ are the Ricci scalar and Christoffel symbols of the spatial metric $\gamma_{ij}$, which is also used to raise and lower spatial indices.

Together with the static source action, this expression determines the two-body effective action through the path integral
\begin{equation}
    \exp \left[{iS_{\mathrm{eff}}^{\mathrm{static}}}\right]
    =
    \int\mathcal D\Phi\,\mathcal D\sigma_{ij}\,\mathcal D\phi\,
    \exp\left[{i(S_{\mathrm{bulk}}^{\mathrm{static}}
          +S_{\mathrm{pp}}^{\mathrm{static}})}\right].
    \label{eq:effective-action}
\end{equation}
Integrating out the gravitational and scalar perturbations thus leaves an effective description in terms of the two worldlines alone. Its interaction part then defines the static potential according to
\begin{equation}
    S_{\mathrm{eff,int}}^{\mathrm{static}}
    =
    -\int\dd t\,\mathcal V_{\mathrm{static}}(r),
    \label{eq:potential-definition}
\end{equation}
where $\bm r = \bm x_1-\bm x_2$ is the separation between the two bodies and $r=|\bm r|$.

\subsection{Feynman rules and power counting}
\label{sec:Feynman-rules}

While the path integral provides the formal definition of $S_{\mathrm{eff}}$, its practical evaluation proceeds through the Feynman diagrams generated by its perturbative expansion.
To construct these diagrams, we derive the corresponding rules from the two parts of the action: the bulk action supplies the massless propagators and field interactions, while the expansion of $S_{\mathrm{pp}}$, including the scalar dependence of $M_A(\varphi)$, determines the couplings to the worldlines.

Since $S_{\mathrm{eff}}$ is obtained by taking the logarithm of the path integral, only connected diagrams contribute.
Restricting to the classical limit further excludes closed loops of the integrated gravitational and scalar fields, with the worldlines treated as external sources rather than propagating degrees of freedom. To isolate the interaction potential, we then retain diagrams involving both bodies and omit vacuum and one-body self-energy contributions, which are independent of their separation. These restrictions still allow nonlinear bulk interactions and multiple insertions on either worldline. In particular, excluding closed field loops does not eliminate the spatial momentum loops that arise when evaluating classical diagrams.

For the connected classical diagrams retained above, the static limit brings a further simplification: the instantaneous propagators collapse all time integrations to a single overall factor $T=\int\dd t$. Factoring it out according to $S_{\mathrm{eff,int}}^{\mathrm{static}}=-T \mathcal V_{\mathrm{static}}$ therefore leaves rules involving only spatial momenta, with no remaining time integrations or temporal Dirac delta functions.
The potential can then be written schematically as
\begin{equation}
    \mathcal V_{\mathrm{static}}(r)
    =i\lim_{d\to3}
    \int\frac{\dd^d\bm q}{(2\pi)^d}
    e^{i\bm q\cdot\bm r}
    \left(\sum_{\rm diags}\,\GenericSourceBlob\right),
    \label{eq:potential-diagrams}
\end{equation}
where ${\bm q}$ denotes the spatial momentum transfer.
The sum runs over the connected classical two-body diagrams specified above. The external legs ending above and below the gray region represent attachments to the two worldlines. The gray region stands for an arbitrary connected field interaction and its source attachments; the number and valence of the insertions are left unspecified. Each diagram includes its symmetry factor and the factors of $i$ prescribed by the Feynman rules, so that the overall factor of $i$ in Eq.~\eqref{eq:potential-diagrams} converts its contribution to $iS_{\mathrm{eff,int}}^{\mathrm{static}}/T$ into the potential defined in Eq.~\eqref{eq:potential-definition}.
The explicit momentum-space Feynman rules are collected in Appendix~\ref{app:feynman-rules}. 

The propagators are obtained by inverting the quadratic part of Eq.~\eqref{eq:static-bulk}, which is diagonal in the three fields and hence contains no kinetic mixing. Since the terms linear in the bulk fields are total derivatives, there are also no bulk one-point vertices. Having obtained the propagators from the bulk action, we next consider the couplings to the sources. These follow by expanding both the exponential and the scalar-dependent mass in Eq.~\eqref{eq:static-worldline}, which gives
\begin{equation}
    S_{\mathrm{pp}}^{\mathrm{static}}
    =
    -\sum_{A=1}^{2}m_A\int\dd t
    \sum_{p,q\geq0}
    \frac{\kappa^{p+q}a_{A,q}}{p!\,q!}\Phi^p\phi^q,
    \label{eq:worldline-expansion}
\end{equation}
where the fields are evaluated on the corresponding worldline.
Each term generates a source vertex with $p$ KK scalar legs and $q$ fundamental scalar legs, and the calculation through 5PN requires those with $1\leq p+q\leq6$. Thus $a_{A,6}$ is the highest mass derivative needed at this order; vertices involving $a_{A,q}$ with $q\geq7$ first contribute beyond 5PN.
Having specified the couplings to the worldlines, we now turn to the bulk interactions. Their structure is particularly simple in the static limit: in Eq.~\eqref{eq:static-bulk}, each scalar appears only through its own quadratic derivative term, while the spatial metric generates interactions involving arbitrary numbers of $\sigma_{ij}$ fields. Consequently, the bulk vertices fall into three families,
\begin{equation}
    \sigma^N,\qquad
    \sigma^{N-2}\Phi^2,\qquad
    \sigma^{N-2}\phi^2,
    \qquad N\geq3,
\end{equation}
where derivatives and tensor contractions are implicit.
Of these, only the following vertices are needed for the classical two-body diagrams through $\mathcal O(G^6)$:
\begin{equation}
    \sigma^3,\qquad
    \sigma\Phi^2,\qquad
    \sigma\phi^2,\qquad
    \sigma^2\Phi^2,\qquad
    \sigma^2\phi^2.
    \label{eq:required-bulk-vertices}
\end{equation}
Although the full static action contains interactions with more field legs, they do not enter the calculation at this order.

With these rules in hand, the counting in powers of the gravitational coupling follows directly from the field normalization. A worldline vertex with $p+q$ legs carries $\kappa^{p+q}\sim G_d^{(p+q)/2}$, while an $N$-leg bulk vertex carries $\kappa^{N-2}\sim G_d^{N/2-1}$, and the propagators contain no explicit power of $G_d$. Holding the masses and scalar-response coefficients fixed, we can therefore organize the static interaction as an expansion in $G$: the Newtonian term is of order $G$, and its $n$PN correction is of order $G^{n+1}$. Accordingly, the calculation through 5PN requires all connected classical two-body contributions through $\mathcal O(G^6)$. As a first check of the normalization, single exchange of $\Phi$ and $\phi$ reproduces the Newtonian potential,
\begin{equation}
    \mathcal V_{\mathrm N}^{G}(r)
    =\underbrace{\begin{gathered}
      \begin{tikzpicture}[scale=.7]
       \draw[dashed,thick] (0,-.42)--(0,.42);
       \fill (0,-.42) circle (1.5pt);\fill (0,.42) circle (1.5pt);
      \end{tikzpicture}
     \end{gathered}}_{\Phi}
    +\underbrace{\begin{gathered}
      \begin{tikzpicture}[scale=.7]
       \draw[thick] (0,-.43)--(0,.43);
       \fill (0,-.42) circle (1.5pt);\fill (0,.42) circle (1.5pt);
      \end{tikzpicture}
     \end{gathered}}_{\phi}
    =-\frac{Gm_1m_2}{r}(1+\alpha_1\alpha_2).
    \label{eq:Newtonian-potential}
\end{equation}

The exchange diagrams in Eq.~\eqref{eq:Newtonian-potential} include the Fourier transform and overall factor of $i$ in Eq.~\eqref{eq:potential-diagrams}.

More generally, the virial assignment $GM/r\sim v^2$ organizes the local $n$PN dynamics into schematic sectors
\begin{equation}
 G^p(v^2)^{n+1-p},\qquad p=0,\ldots,n+1,
 \label{eq:PN-sectors}
\end{equation}
with the necessary masses and separation factors understood. The $p=0$ term belongs to the free kinetic action. The velocity-independent interaction occupies the highest-$G$ sector, $p=n+1$, so that
\begin{equation}
 n\mathrm{PN}\big|_{\mathrm{static}}
 \quad\longleftrightarrow\quad G^{n+1}
 \quad\longleftrightarrow\quad n\text{-loop spatial integrals}.
 \label{eq:PN-G-loops}
\end{equation}
The loop count holds at fixed external momentum transfer, before the final Fourier transform to position space. Accordingly, the static 1PN, 2PN, 3PN, 4PN, and 5PN sectors require one-, two-, three-, four-, and five-loop two-point integrals, respectively. This is the highest loop order at each PN order, even though the static potential is only one part of the two-body dynamics. The independent topologies from 1PN through 5PN and the corresponding propagator bases entering the static calculation in this paper are collected in Appendix~\ref{app:integrals}.

Once the Feynman rules are inserted, the static diagrams become massless propagator-type integrals with one external spatial momentum $\bm q$. This connection with QFT two-point functions is well established in PN EFT~\cite{Foffa:2016rgu}: the attachments to each nonpropagating worldline are collected into an external vertex, leaving the field interaction as a momentum-space two-point graph. Schematically,
\begin{equation}
 \GenericSourceBlob
 \quad\longleftrightarrow\quad
 \begin{gathered}
 \begin{tikzpicture}[scale=.8,baseline=(current bounding box.center)]
  \draw[thick] (-1.8,0)--(-.7,0);
  \draw[thick] (.7,0)--(1.8,0);
  \draw[gray!20,fill=gray!20] (0,0) ellipse (.7 and .48);
 \end{tikzpicture}
 \end{gathered}
 \label{eq:diagram-two-point}
\end{equation}
On the left, the upper and lower external endpoints represent attachments to the two sources, and the gray region denotes an unspecified connected field interaction. On the right, the blob denotes the corresponding two-point integral, with its internal structure left implicit. The mapping retains the numerator, coupling, and symmetry factors of the original static diagram; the following subsection describes the reduction of these integrals to a scalar master basis.

\subsection{Diagram generation and multiloop integration}
\label{sec:multiloop}

We generate the diagrams with \texttt{QGRAF}~\cite{Nogueira:1991ex} and process its output with custom symbolic routines.
These routines remove vacuum contributions, one-body self-energies, and non-classical diagrams containing closed internal field loops, retaining only the connected classical two-body contributions specified above.
Table~\ref{tab:diagram-counts} summarizes the number of diagrams surviving this selection;
these counts include pure-GR diagrams and treat body-exchanged configurations separately.
Note that a retained graph may still vanish after tensor contraction or integration, and such graphs are included in these totals.

At $L$-loop level, the maximum number of propagators is $2L+1$ for even $L$ and $2L$ for odd $L$; diagrams saturating this bound define the \emph{top topologies}. The last column of Table~\ref{tab:diagram-counts} counts all independent topologies retained at each order, including any sub-topologies.
Diagrams with fewer propagators---typically involving quartic or higher-point bulk vertices---correspond to \emph{sub-topologies}, as they can be embedded into one of the top topologies.

\begin{table}[t]
 \centering
 \begin{tabular}{lcccc}
  \hline
  Static PN order & Order in $G$ & Spatial loops & Retained diagrams & Ind. topologies\\
  \hline
  Newtonian & $G$   & 0 & 2 & 1\\
  1PN       & $G^2$ & 1 & 6 & 1\\
  2PN       & $G^3$ & 2 & 31 & 1\\
  3PN       & $G^4$ & 3 & 134 & 3\\
  4PN       & $G^5$ & 4 & 980 & 10\\
  5PN       & $G^6$ & 5 & 6224 & 22\\
  \hline
 \end{tabular}
 \caption{Classical two-body diagrams and independent topologies after \texttt{QGRAF} generation and automated symbolic selection, before inserting Feynman rules.
 Pure-GR diagrams, separate body-exchanged configurations, and contributions that subsequently vanish are included. Spatial loops are counted at fixed external momentum transfer.}
 \label{tab:diagram-counts}
\end{table}

For each diagram, the Feynman rules and symmetry factors are inserted, and the index contractions and algebraic manipulations are carried out in \texttt{FORM}~\cite{Vermaseren:2000nd}.
For each $L$-loop topology, its physical propagators are supplemented, where necessary, with additional irreducible scalar products (ISPs) to form a complete basis of $N_L = L(L+3)/2$ independent scalar products, in terms of which any numerator can be uniquely decomposed.
At a given loop order $L$, by adopting an appropriate loop-momentum routing for each topology, we find that all relevant integrals can be mapped onto a single propagator basis, admitting the unified representation
\begin{align}\label{ints-family-L}
M^{(L)}_{a_1 \cdots a_{N_L} } = \int \Bigg( \prod_{i=1}^{L} \frac{d^d \ell_i}{\pi^{d/2}} \Bigg) \frac{1}{\prod_{j=1}^{N_L} D_j^{a_j} } \, .
\end{align}
The integer indices $a_j$ specify denominator powers when positive, omit a basis element when zero, and represent numerator powers when negative.
Complete propagator bases for one through five loops are given in Appendix~\ref{app:integrals}.
Within this integral family, each topology corresponds to a sector, comprising all integrals that share a common set of denominators.
Beyond its notational simplicity, this unified representation offers a significant advantage in the reduction process: since different top sectors frequently share common sub-sectors, a joint reduction avoids redundant computations and substantially reduces the overall workload.

\subsubsection{IBP reduction and master integrals}
\label{sec:two-point}

At fixed momentum transfer $\bm q$, the families in Eq.~\eqref{ints-family-L} are one-scale massless two-point integrals in $d=3-2\varepsilon$ spatial dimensions. These integrals satisfy in dimensional regularization the following identities \cite{Tkachov:1981wb,Chetyrkin:1981qh},
\begin{equation}
 0=\prod_{j=1}^{L}\Bigg(\int\frac{\dd^d \ell_j}{\pi^{d/2}}\Bigg)
 \frac{\partial}{\partial \ell_r^i}
 \left[{v^i \over \prod_b D_b^{a_b}}\right],
 \qquad \bm v\in\{\bm \ell_1,\ldots,\bm \ell_L,\bm q\}.
 \label{eq:IBP}
\end{equation}
Expanding these identities yields a system of homogeneous linear equations.
We solve such systems with \texttt{FIRE6.5}, expressing the full integral set at each loop order in terms of a finite collection of master integrals
\begin{equation}
 {M}^{(L)}_{a_1\cdots a_{N_L}}(\bm q;d)
 =\sum_j c_{a_1\cdots a_{N_L},j}(d,\bm q^2)\,\mathcal M_j(\bm q;d).
 \label{eq:MI-reduction}
\end{equation}
Here $\mathcal M_j$ are the master integrals and $c_{a_1\cdots a_{N_L},j}$ are rational reduction coefficients. Momentum shifts and graph symmetries are used to identify equivalent representations before this step. An integral is scaleless only when no external or internal dimensionful scale remains, in which case it vanishes in dimensional regularization. A massless propagator integral that depends nontrivially on $\bm q^2\ne0$ retains that scale and need not vanish. By contrast, propagator cancellations or factorization can produce a subintegral with no remaining dependence on $\bm q$ or any other scale; that subintegral is set to zero.
The master integrals $\mathcal{M}^{(L)}_i$ obtained from this reduction for one through five loops are displayed in Fig.~\ref{fig:master-integrals}, with their explicit propagator bases and index vectors collected in Appendix~\ref{app:integrals}.

\begin{figure}[!t]
\centering
\resizebox{\linewidth}{!}{%
\begin{tikzpicture}[x=1cm,y=1cm,line width=.48pt,line cap=round,line join=round]
\path[use as bounding box] (-7.8,-11.98) rectangle (7.8,1.52);
\node[font=\small] at (-6.4,1.31) {1PN (1 loop)};
\draw[line width=.25pt,black!45] (-7.72,1.06)--(-5.08,1.06);
\node[font=\small] at (-1.6,1.31) {2PN (2 loops)};
\draw[line width=.25pt,black!45] (-4.56,1.06)--(1.36,1.06);
\node[font=\small] at (4.8,1.31) {3PN (3 loops)};
\draw[line width=.25pt,black!45] (1.84,1.06)--(7.76,1.06);
\node[font=\small] at (0,-1.8) {4PN (4 loops)};
\draw[line width=.25pt,black!45] (-7.72,-2.05)--(7.72,-2.05);
\node[font=\small] at (0,-7.15) {5PN (5 loops)};
\draw[line width=.25pt,black!45] (-7.72,-7.4)--(7.72,-7.4);
\begin{scope}[shift={(-6.4,0)}]
\coordinate (v0) at (-1.00000,0.00000);
\coordinate (v1) at (1.00000,0.00000);
\draw (-1.28,0)--(v0);
\draw (v1)--(1.28,0);
\draw (v0) .. controls (-0.33333,-1.04000) and (0.33333,-1.04000) .. (v1); 
\draw (v0) .. controls (-0.33333,1.04000) and (0.33333,1.04000) .. (v1); 
\fill (v0) circle[radius=1.18pt];
\fill (v1) circle[radius=1.18pt];
\node[font=\small] at (0,-1.07) {$\mathcal M^{(1)}_{1}$};
\end{scope}
\begin{scope}[shift={(-3.2,0)}]
\coordinate (v0) at (-1.00000,0.00000);
\coordinate (v1) at (1.00000,0.00000);
\draw (-1.28,0)--(v0);
\draw (v1)--(1.28,0);
\draw (v0) .. controls (-0.33333,-1.04000) and (0.33333,-1.04000) .. (v1); 
\draw (v0) .. controls (-0.33333,0.00000) and (0.33333,0.00000) .. (v1); 
\draw (v0) .. controls (-0.33333,1.04000) and (0.33333,1.04000) .. (v1); 
\fill (v0) circle[radius=1.18pt];
\fill (v1) circle[radius=1.18pt];
\node[font=\small] at (0,-1.07) {$\mathcal M^{(2)}_{1}$};
\end{scope}
\begin{scope}[shift={(0,0)}]
\coordinate (v0) at (0.00000,0.00000);
\coordinate (v1) at (-1.00000,0.00000);
\coordinate (v2) at (1.00000,0.00000);
\draw (-1.28,0)--(v1);
\draw (v2)--(1.28,0);
\draw (v0) .. controls (-0.33333,0.78667) and (-0.66667,0.78667) .. (v1); 
\draw (v0) .. controls (-0.33333,-0.78667) and (-0.66667,-0.78667) .. (v1); 
\draw (v0) .. controls (0.33333,-0.78667) and (0.66667,-0.78667) .. (v2); 
\draw (v0) .. controls (0.33333,0.78667) and (0.66667,0.78667) .. (v2); 
\fill (v1) circle[radius=1.18pt];
\fill (v2) circle[radius=1.18pt];
\fill (v0) circle[radius=1.18pt];
\node[font=\small] at (0,-1.07) {$\mathcal M^{(2)}_{2}$};
\end{scope}
\begin{scope}[shift={(3.2,0)}]
\coordinate (v0) at (-1.00000,0.00000);
\coordinate (v1) at (1.00000,0.00000);
\draw (-1.28,0)--(v0);
\draw (v1)--(1.28,0);
\draw (v0) .. controls (-0.33333,-1.04000) and (0.33333,-1.04000) .. (v1); 
\draw (v0) .. controls (-0.33333,-0.34667) and (0.33333,-0.34667) .. (v1); 
\draw (v0) .. controls (-0.33333,0.34667) and (0.33333,0.34667) .. (v1); 
\draw (v0) .. controls (-0.33333,1.04000) and (0.33333,1.04000) .. (v1); 
\fill (v0) circle[radius=1.18pt];
\fill (v1) circle[radius=1.18pt];
\node[font=\small] at (0,-1.07) {$\mathcal M^{(3)}_{1}$};
\end{scope}
\begin{scope}[shift={(6.4,0)}]
\coordinate (v0) at (0.00000,0.74000);
\coordinate (v1) at (-1.00000,0.00000);
\coordinate (v2) at (1.00000,0.00000);
\draw (-1.28,0)--(v1);
\draw (v2)--(1.28,0);
\draw (v0) .. controls (-0.51575,0.73984) and (-0.84909,0.49318) .. (v1); 
\draw (v0) .. controls (-0.15091,0.24682) and (-0.48425,0.00016) .. (v1); 
\draw (v0) .. controls (0.15091,0.24682) and (0.48425,0.00016) .. (v2); 
\draw (v0) .. controls (0.51575,0.73984) and (0.84909,0.49318) .. (v2); 
\draw (v1) .. controls (-0.33333,-0.97333) and (0.33333,-0.97333) .. (v2); 
\fill (v1) circle[radius=1.18pt];
\fill (v2) circle[radius=1.18pt];
\fill (v0) circle[radius=1.18pt];
\node[font=\small] at (0,-1.07) {$\mathcal M^{(3)}_{2}$};
\end{scope}
\begin{scope}[shift={(-5.7,-3.0)}]
\coordinate (v0) at (-1.00000,0.00000);
\coordinate (v1) at (1.00000,0.00000);
\draw (-1.28,0)--(v0);
\draw (v1)--(1.28,0);
\draw (v0) .. controls (-0.33333,-1.04000) and (0.33333,-1.04000) .. (v1); 
\draw (v0) .. controls (-0.33333,-0.52000) and (0.33333,-0.52000) .. (v1); 
\draw (v0) .. controls (-0.33333,0.00000) and (0.33333,0.00000) .. (v1); 
\draw (v0) .. controls (-0.33333,0.52000) and (0.33333,0.52000) .. (v1); 
\draw (v0) .. controls (-0.33333,1.04000) and (0.33333,1.04000) .. (v1); 
\fill (v0) circle[radius=1.18pt];
\fill (v1) circle[radius=1.18pt];
\node[font=\small] at (0,-1.07) {$\mathcal M^{(4)}_{1}$};
\end{scope}
\begin{scope}[shift={(-1.9,-3.0)}]
\coordinate (v0) at (0.00000,0.00000);
\coordinate (v1) at (-1.00000,0.00000);
\coordinate (v2) at (1.00000,0.00000);
\draw (-1.28,0)--(v1);
\draw (v2)--(1.28,0);
\draw (v0) .. controls (-0.33333,0.78667) and (-0.66667,0.78667) .. (v1); 
\draw (v0) .. controls (-0.33333,-0.78667) and (-0.66667,-0.78667) .. (v1); 
\draw (v0) .. controls (0.33333,-0.78667) and (0.66667,-0.78667) .. (v2); 
\draw (v0) .. controls (0.33333,-0.26222) and (0.66667,-0.26222) .. (v2); 
\draw (v0) .. controls (0.33333,0.26222) and (0.66667,0.26222) .. (v2); 
\draw (v0) .. controls (0.33333,0.78667) and (0.66667,0.78667) .. (v2); 
\fill (v1) circle[radius=1.18pt];
\fill (v2) circle[radius=1.18pt];
\fill (v0) circle[radius=1.18pt];
\node[font=\small] at (0,-1.07) {$\mathcal M^{(4)}_{2}$};
\end{scope}
\begin{scope}[shift={(1.9,-3.0)}]
\coordinate (v0) at (0.00000,0.74000);
\coordinate (v1) at (-1.00000,0.00000);
\coordinate (v2) at (1.00000,0.00000);
\draw (-1.28,0)--(v1);
\draw (v2)--(1.28,0);
\draw (v0) .. controls (-0.51575,0.73984) and (-0.84909,0.49318) .. (v1); 
\draw (v0) .. controls (-0.33333,0.49333) and (-0.66667,0.24667) .. (v1); 
\draw (v0) .. controls (-0.15091,0.24682) and (-0.48425,0.00016) .. (v1); 
\draw (v0) .. controls (0.15091,0.24682) and (0.48425,0.00016) .. (v2); 
\draw (v0) .. controls (0.51575,0.73984) and (0.84909,0.49318) .. (v2); 
\draw (v1) .. controls (-0.33333,-0.97333) and (0.33333,-0.97333) .. (v2); 
\fill (v1) circle[radius=1.18pt];
\fill (v2) circle[radius=1.18pt];
\fill (v0) circle[radius=1.18pt];
\node[font=\small] at (0,-1.07) {$\mathcal M^{(4)}_{3}$};
\end{scope}
\begin{scope}[shift={(5.7,-3.0)}]
\coordinate (v0) at (0.00000,0.74000);
\coordinate (v1) at (-1.00000,0.00000);
\coordinate (v2) at (1.00000,0.00000);
\draw (-1.28,0)--(v1);
\draw (v2)--(1.28,0);
\draw (v0) .. controls (-0.51575,0.73984) and (-0.84909,0.49318) .. (v1); 
\draw (v0) .. controls (-0.15091,0.24682) and (-0.48425,0.00016) .. (v1); 
\draw (v0) .. controls (0.15091,0.24682) and (0.48425,0.00016) .. (v2); 
\draw (v0) .. controls (0.51575,0.73984) and (0.84909,0.49318) .. (v2); 
\draw (v1) .. controls (-0.33333,-0.50667) and (0.33333,-0.50667) .. (v2); 
\draw (v1) .. controls (-0.33333,-1.09333) and (0.33333,-1.09333) .. (v2); 
\fill (v1) circle[radius=1.18pt];
\fill (v2) circle[radius=1.18pt];
\fill (v0) circle[radius=1.18pt];
\node[font=\small] at (0,-1.07) {$\mathcal M^{(4)}_{4}$};
\end{scope}
\begin{scope}[shift={(-3.8,-5.35)}]
\coordinate (v0) at (-1.00000,0.00000);
\coordinate (v1) at (0.00000,0.00000);
\coordinate (v2) at (1.00000,0.00000);
\draw (-1.28,0)--(v0);
\draw (v2)--(1.28,0);
\draw (v0) .. controls (-0.66667,-0.78667) and (-0.33333,-0.78667) .. (v1); 
\draw (v0) .. controls (-0.66667,0.00000) and (-0.33333,0.00000) .. (v1); 
\draw (v0) .. controls (-0.66667,0.78667) and (-0.33333,0.78667) .. (v1); 
\draw (v1) .. controls (0.33333,-0.78667) and (0.66667,-0.78667) .. (v2); 
\draw (v1) .. controls (0.33333,0.00000) and (0.66667,0.00000) .. (v2); 
\draw (v1) .. controls (0.33333,0.78667) and (0.66667,0.78667) .. (v2); 
\fill (v0) circle[radius=1.18pt];
\fill (v2) circle[radius=1.18pt];
\fill (v1) circle[radius=1.18pt];
\node[font=\small] at (0,-1.07) {$\mathcal M^{(4)}_{5}$};
\end{scope}
\begin{scope}[shift={(0,-5.35)}]
\coordinate (v0) at (-1.00000,0.00000);
\coordinate (v1) at (0.00000,-0.74000);
\coordinate (v2) at (1.00000,0.00000);
\coordinate (v3) at (0.00000,0.74000);
\draw (-1.28,0)--(v0);
\draw (v2)--(1.28,0);
\draw (v0) .. controls (-0.66667,-0.24667) and (-0.33333,-0.49333) .. (v1); 
\draw (v0) .. controls (-0.54770,0.08590) and (-0.21436,0.33256) .. (v3); 
\draw (v0) .. controls (-0.78564,0.40744) and (-0.45230,0.65410) .. (v3); 
\draw (v1) .. controls (0.45230,-0.65410) and (0.78564,-0.40744) .. (v2); 
\draw (v1) .. controls (0.21436,-0.33256) and (0.54770,-0.08590) .. (v2); 
\draw (v1) .. controls (0.00000,-0.24667) and (0.00000,0.24667) .. (v3); 
\draw (v2) .. controls (0.66667,0.24667) and (0.33333,0.49333) .. (v3); 
\fill (v0) circle[radius=1.18pt];
\fill (v2) circle[radius=1.18pt];
\fill (v3) circle[radius=1.18pt];
\fill (v1) circle[radius=1.18pt];
\node[font=\small] at (0,-1.07) {$\mathcal M^{(4)}_{6}$};
\end{scope}
\begin{scope}[shift={(3.8,-5.35)}]
\coordinate (v0) at (0.00000,0.00000);
\coordinate (v1) at (-1.00000,0.00000);
\coordinate (v2) at (1.00000,0.00000);
\coordinate (v3) at (0.00000,0.78000);
\coordinate (v4) at (0.00000,-0.78000);
\draw (-1.28,0)--(v1);
\draw (v2)--(1.28,0);
\draw (v0) .. controls (-0.33333,0.00000) and (-0.66667,0.00000) .. (v1); 
\draw (v0) .. controls (0.33333,0.00000) and (0.66667,0.00000) .. (v2); 
\draw (v0) .. controls (0.00000,0.26000) and (0.00000,0.52000) .. (v3); 
\draw (v0) .. controls (0.00000,-0.26000) and (0.00000,-0.52000) .. (v4); 
\draw (v1) .. controls (-0.66667,0.26000) and (-0.33333,0.52000) .. (v3); 
\draw (v1) .. controls (-0.66667,-0.26000) and (-0.33333,-0.52000) .. (v4); 
\draw (v2) .. controls (0.66667,0.26000) and (0.33333,0.52000) .. (v3); 
\draw (v2) .. controls (0.66667,-0.26000) and (0.33333,-0.52000) .. (v4); 
\fill (v1) circle[radius=1.18pt];
\fill (v2) circle[radius=1.18pt];
\fill (v3) circle[radius=1.18pt];
\fill (v4) circle[radius=1.18pt];
\fill (v0) circle[radius=1.18pt];
\node[font=\small] at (0,-1.07) {$\mathcal M^{(4)}_{7}$};
\end{scope}
\begin{scope}[shift={(-5.7,-8.35)}]
\coordinate (v0) at (-1.00000,0.00000);
\coordinate (v1) at (1.00000,0.00000);
\draw (-1.28,0)--(v0);
\draw (v1)--(1.28,0);
\draw (v0) .. controls (-0.33333,-1.04000) and (0.33333,-1.04000) .. (v1); 
\draw (v0) .. controls (-0.33333,-0.62400) and (0.33333,-0.62400) .. (v1); 
\draw (v0) .. controls (-0.33333,-0.20800) and (0.33333,-0.20800) .. (v1); 
\draw (v0) .. controls (-0.33333,0.20800) and (0.33333,0.20800) .. (v1); 
\draw (v0) .. controls (-0.33333,0.62400) and (0.33333,0.62400) .. (v1); 
\draw (v0) .. controls (-0.33333,1.04000) and (0.33333,1.04000) .. (v1); 
\fill (v0) circle[radius=1.18pt];
\fill (v1) circle[radius=1.18pt];
\node[font=\small] at (0,-1.07) {$\mathcal M^{(5)}_{1}$};
\end{scope}
\begin{scope}[shift={(-1.9,-8.35)}]
\coordinate (v0) at (-1.00000,0.00000);
\coordinate (v1) at (0.00000,0.74000);
\coordinate (v2) at (1.00000,0.00000);
\draw (-1.28,0)--(v0);
\draw (v2)--(1.28,0);
\draw (v0) .. controls (-0.48425,0.00016) and (-0.15091,0.24682) .. (v1); 
\draw (v0) .. controls (-0.84909,0.49318) and (-0.51575,0.73984) .. (v1); 
\draw (v0) .. controls (-0.33333,-0.50667) and (0.33333,-0.50667) .. (v2); 
\draw (v0) .. controls (-0.33333,-0.80000) and (0.33333,-0.80000) .. (v2); 
\draw (v0) .. controls (-0.33333,-1.09333) and (0.33333,-1.09333) .. (v2); 
\draw (v1) .. controls (0.15091,0.24682) and (0.48425,0.00016) .. (v2); 
\draw (v1) .. controls (0.51575,0.73984) and (0.84909,0.49318) .. (v2); 
\fill (v0) circle[radius=1.18pt];
\fill (v2) circle[radius=1.18pt];
\fill (v1) circle[radius=1.18pt];
\node[font=\small] at (0,-1.07) {$\mathcal M^{(5)}_{2}$};
\end{scope}
\begin{scope}[shift={(1.9,-8.35)}]
\coordinate (v0) at (-1.00000,0.00000);
\coordinate (v1) at (0.00000,0.74000);
\coordinate (v2) at (1.00000,0.00000);
\draw (-1.28,0)--(v0);
\draw (v2)--(1.28,0);
\draw (v0) .. controls (-0.48425,0.00016) and (-0.15091,0.24682) .. (v1); 
\draw (v0) .. controls (-0.84909,0.49318) and (-0.51575,0.73984) .. (v1); 
\draw (v0) .. controls (-0.33333,-0.50667) and (0.33333,-0.50667) .. (v2); 
\draw (v0) .. controls (-0.33333,-1.09333) and (0.33333,-1.09333) .. (v2); 
\draw (v1) .. controls (0.15091,0.24682) and (0.48425,0.00016) .. (v2); 
\draw (v1) .. controls (0.33333,0.49333) and (0.66667,0.24667) .. (v2); 
\draw (v1) .. controls (0.51575,0.73984) and (0.84909,0.49318) .. (v2); 
\fill (v0) circle[radius=1.18pt];
\fill (v2) circle[radius=1.18pt];
\fill (v1) circle[radius=1.18pt];
\node[font=\small] at (0,-1.07) {$\mathcal M^{(5)}_{3}$};
\end{scope}
\begin{scope}[shift={(5.7,-8.35)}]
\coordinate (v0) at (-1.00000,0.00000);
\coordinate (v1) at (1.00000,0.00000);
\coordinate (v2) at (0.00000,0.72000);
\coordinate (v3) at (0.00000,-0.72000);
\draw (-1.28,0)--(v0);
\draw (v1)--(1.28,0);
\draw (v0) .. controls (-0.54981,0.07769) and (-0.21647,0.31769) .. (v2); 
\draw (v0) .. controls (-0.78353,0.40231) and (-0.45019,0.64231) .. (v2); 
\draw (v0) .. controls (-0.78353,-0.40231) and (-0.45019,-0.64231) .. (v3); 
\draw (v0) .. controls (-0.54981,-0.07769) and (-0.21647,-0.31769) .. (v3); 
\draw (v1) .. controls (0.78353,0.40231) and (0.45019,0.64231) .. (v2); 
\draw (v1) .. controls (0.54981,0.07769) and (0.21647,0.31769) .. (v2); 
\draw (v1) .. controls (0.54981,-0.07769) and (0.21647,-0.31769) .. (v3); 
\draw (v1) .. controls (0.78353,-0.40231) and (0.45019,-0.64231) .. (v3); 
\fill (v0) circle[radius=1.18pt];
\fill (v1) circle[radius=1.18pt];
\fill (v2) circle[radius=1.18pt];
\fill (v3) circle[radius=1.18pt];
\node[font=\small] at (0,-1.07) {$\mathcal M^{(5)}_{4}$};
\end{scope}
\begin{scope}[shift={(-5.7,-10.7)}]
\coordinate (v0) at (-1.00000,0.00000);
\coordinate (v1) at (0.00000,0.74000);
\coordinate (v2) at (1.00000,0.00000);
\draw (-1.28,0)--(v0);
\draw (v2)--(1.28,0);
\draw (v0) .. controls (-0.48425,0.00016) and (-0.15091,0.24682) .. (v1); 
\draw (v0) .. controls (-0.60586,0.16450) and (-0.27253,0.41116) .. (v1); 
\draw (v0) .. controls (-0.72747,0.32884) and (-0.39414,0.57550) .. (v1); 
\draw (v0) .. controls (-0.84909,0.49318) and (-0.51575,0.73984) .. (v1); 
\draw (v0) .. controls (-0.33333,-0.97333) and (0.33333,-0.97333) .. (v2); 
\draw (v1) .. controls (0.15091,0.24682) and (0.48425,0.00016) .. (v2); 
\draw (v1) .. controls (0.51575,0.73984) and (0.84909,0.49318) .. (v2); 
\fill (v0) circle[radius=1.18pt];
\fill (v2) circle[radius=1.18pt];
\fill (v1) circle[radius=1.18pt];
\node[font=\small] at (0,-1.07) {$\mathcal M^{(5)}_{5}$};
\end{scope}
\begin{scope}[shift={(-1.9,-10.7)}]
\coordinate (v0) at (0.00000,0.74000);
\coordinate (v1) at (-1.00000,0.00000);
\coordinate (v2) at (1.00000,0.00000);
\draw (-1.28,0)--(v1);
\draw (v2)--(1.28,0);
\draw (v0) .. controls (-0.51575,0.73984) and (-0.84909,0.49318) .. (v1); 
\draw (v0) .. controls (-0.33333,0.49333) and (-0.66667,0.24667) .. (v1); 
\draw (v0) .. controls (-0.15091,0.24682) and (-0.48425,0.00016) .. (v1); 
\draw (v0) .. controls (0.15091,0.24682) and (0.48425,0.00016) .. (v2); 
\draw (v0) .. controls (0.33333,0.49333) and (0.66667,0.24667) .. (v2); 
\draw (v0) .. controls (0.51575,0.73984) and (0.84909,0.49318) .. (v2); 
\draw (v1) .. controls (-0.33333,-0.97333) and (0.33333,-0.97333) .. (v2); 
\fill (v1) circle[radius=1.18pt];
\fill (v2) circle[radius=1.18pt];
\fill (v0) circle[radius=1.18pt];
\node[font=\small] at (0,-1.07) {$\mathcal M^{(5)}_{6}$};
\end{scope}
\begin{scope}[shift={(1.9,-10.7)}]
\coordinate (v0) at (0.00000,0.12000);
\coordinate (v1) at (0.00000,0.76000);
\coordinate (v2) at (-1.00000,0.00000);
\coordinate (v3) at (1.00000,0.00000);
\coordinate (v4) at (0.00000,-0.42000);
\draw (-1.28,0)--(v2);
\draw (v3)--(1.28,0);
\draw (v0) .. controls (0.00000,0.33333) and (0.00000,0.54667) .. (v1); 
\draw (v0) .. controls (-0.33333,0.08000) and (-0.66667,0.04000) .. (v2); 
\draw (v0) .. controls (0.33333,0.08000) and (0.66667,0.04000) .. (v3); 
\draw (v0) .. controls (0.00000,-0.06000) and (0.00000,-0.24000) .. (v4); 
\draw (v1) .. controls (-0.33333,0.50667) and (-0.66667,0.25333) .. (v2); 
\draw (v1) .. controls (0.33333,0.50667) and (0.66667,0.25333) .. (v3); 
\draw (v2) .. controls (-0.33333,-1.13333) and (0.33333,-1.13333) .. (v3); 
\draw (v2) .. controls (-0.66667,-0.14000) and (-0.33333,-0.28000) .. (v4); 
\draw (v3) .. controls (0.66667,-0.14000) and (0.33333,-0.28000) .. (v4); 
\fill (v2) circle[radius=1.18pt];
\fill (v3) circle[radius=1.18pt];
\fill (v1) circle[radius=1.18pt];
\fill (v4) circle[radius=1.18pt];
\fill (v0) circle[radius=1.18pt];
\node[font=\small] at (0,-1.07) {$\mathcal M^{(5)}_{7}$};
\end{scope}
\begin{scope}[shift={(5.7,-10.7)}]
\coordinate (v0) at (0.00000,0.76000);
\coordinate (v1) at (0.00000,-0.05000);
\coordinate (v2) at (-1.00000,0.00000);
\coordinate (v3) at (1.00000,0.00000);
\draw (-1.28,0)--(v2);
\draw (v3)--(1.28,0);
\draw (v0) .. controls (0.00000,0.49000) and (0.00000,0.22000) .. (v1); 
\draw (v0) .. controls (-0.33333,0.50667) and (-0.66667,0.25333) .. (v2); 
\draw (v0) .. controls (0.21232,0.34743) and (0.54565,0.09410) .. (v3); 
\draw (v0) .. controls (0.45435,0.66590) and (0.78768,0.41257) .. (v3); 
\draw (v1) .. controls (-0.32335,0.16642) and (-0.65668,0.18308) .. (v2); 
\draw (v1) .. controls (-0.34332,-0.23308) and (-0.67665,-0.21642) .. (v2); 
\draw (v1) .. controls (0.33333,-0.03333) and (0.66667,-0.01667) .. (v3); 
\draw (v2) .. controls (-0.33333,-1.00000) and (0.33333,-1.00000) .. (v3); 
\fill (v2) circle[radius=1.18pt];
\fill (v3) circle[radius=1.18pt];
\fill (v0) circle[radius=1.18pt];
\fill (v1) circle[radius=1.18pt];
\node[font=\small] at (0,-1.07) {$\mathcal M^{(5)}_{8}$};
\end{scope}
\end{tikzpicture}%
}
\caption{Master-integral topologies entering the static potential from 1PN through 5PN, corresponding to one- through five-loop massless two-point functions. The twenty distinct two-point topologies are labelled by the masters $\mathcal M^{(L)}_i$ defined in Appendix~\ref{app:integrals}.
External legs carry the momentum transfer $\bm q$; filled dots denote graph vertices. Products and lower-loop insertions are retained explicitly.}
\label{fig:master-integral-topologies}
\label{fig:master-integrals}
\end{figure}
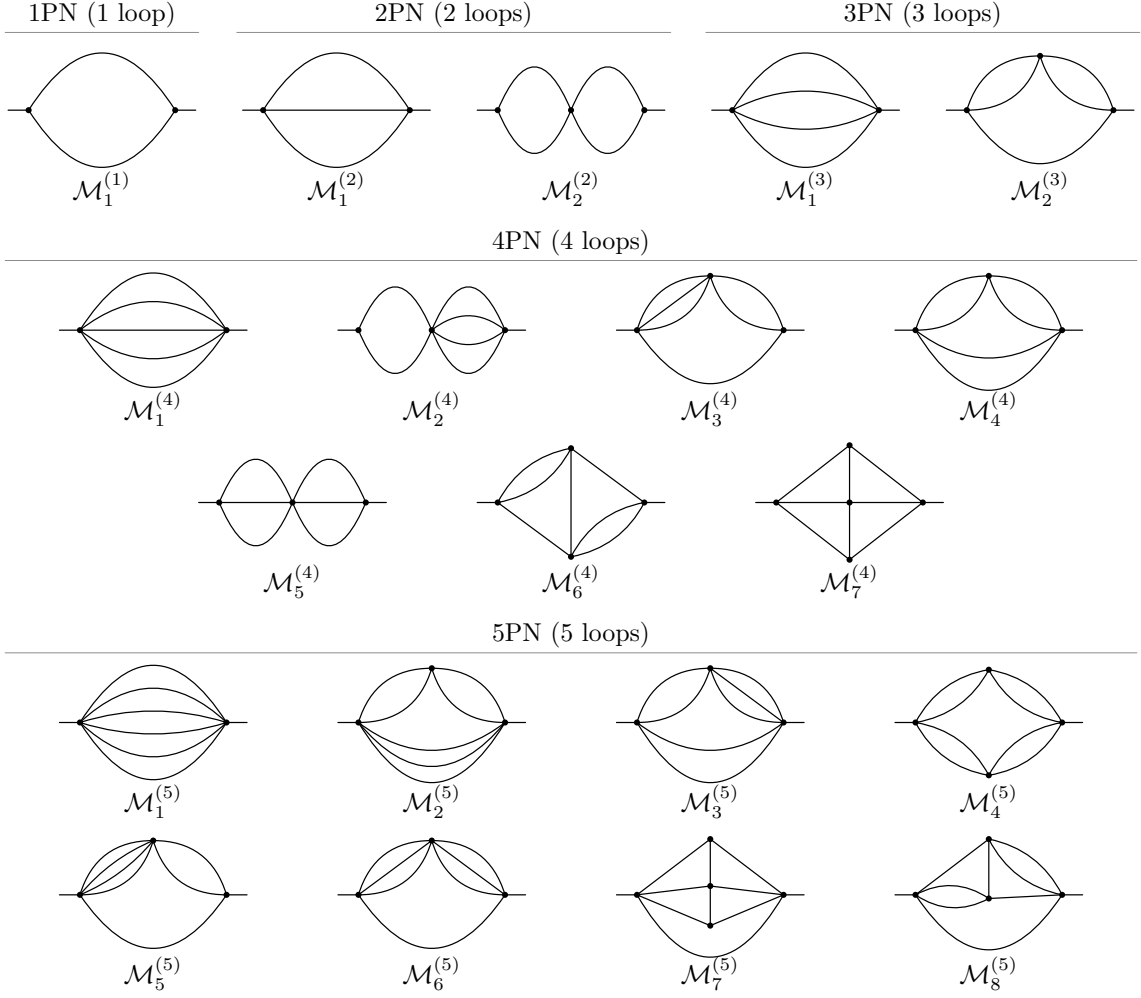

\subsubsection{Bubble integrals and the four-loop input}
\label{sec:master-input}

The basic lower-order building block is the massless bubble,
\begin{equation}
\begin{aligned}
 B_d(\alpha,\beta;\bm q^2)
 &\equiv\int\frac{\dd^d\bm k}{\pi^{d/2}}
 \frac{1}{(\bm k^2)^\alpha[(\bm k+\bm q)^2]^\beta}
 \\
 &=(\bm q^2)^{d/2-\alpha-\beta}
 \frac{\Gamma(\alpha+\beta-d/2)
       \Gamma(d/2-\alpha)\Gamma(d/2-\beta)}
      {\Gamma(\alpha)\Gamma(\beta)\Gamma(d-\alpha-\beta)}.
\end{aligned}
 \label{eq:bubble}
\end{equation}
Here $\alpha$ and $\beta$ are propagator exponents and $\Gamma$ is the Euler gamma function. The expression is understood by analytic continuation from a convergence domain. In the static calculation through 3PN, the required masters can be evaluated using this formula and its iterated application, after the necessary IBP reduction. This does not mean that every unreduced tensor integral is itself a nested bubble: the reduction first relates nontrivial numerator and propagator structures to the simpler master basis.

At 4PN, genuinely nontrivial four-loop information is required.
We use the master denoted $M_{3,6}$ in Ref.~\cite{Foffa:2016rgu}.
The subscripts on $M_{3,6}$ label that reference's master basis, not the loop order. With $s=\bm q^2$, the result reads in our conventions
\begin{align}
 {\mathcal M^{(4)}_7}(3-2\varepsilon;s)
 ={}&\frac{c(\varepsilon;s)}{s^2}
 \bigg[\frac{1}{8\varepsilon^2}+\frac{1}{4\varepsilon}
       -4+\frac{\pi^2}{24}\nonumber\\
 &\hspace{1.8cm}
 -2\varepsilon\left(-9+\pi^2\left(\frac{13}{8}-\ln2\right)
                   +\frac{77}{6}\zeta(3)\right)
 +\mathcal O(\varepsilon^2)\bigg],
 \label{eq:M36}\\
 c(\varepsilon;s)={}&
 (4\pi)^2 e^{-4\gamma_E\varepsilon}s^{-4\varepsilon}.
 \label{eq:M36-normalization}
\end{align}
Here $\gamma_E$ is the Euler--Mascheroni constant and $\zeta(s)$ the Riemann zeta function; in particular, $\zeta(3)$ is Ap\'ery's constant. The terms shown illustrate the transcendental structures, including $\pi^2\ln2$, that must be retained through the required dimensional expansion.

To extract the finite potential, the order in $\varepsilon$ to which a given master must be expanded is determined by the Laurent expansion of its complete coefficient after IBP reduction, including any lower-loop factors. In particular, the double pole in Eq.~\eqref{eq:M36} requires a regular prefactor through $\varepsilon^2$ even when only the finite part is sought; a pole in that prefactor also requires positive powers of $\varepsilon$ in the master. For comparison, the four-loop master $M_{2,2}$ defined in Ref.~\cite{Foffa:2016rgu} is finite at $d=3$. It drops out of that reference's GR amplitudes because its complete coefficient is suppressed by positive powers of $\varepsilon$. Inspection of the complete coefficient in our ST reduction shows that the same conclusion holds here.
In general,
\begin{equation}
 [C(\varepsilon)\mathcal M(\varepsilon)]_{\varepsilon^0}
 =\sum_{r+s=0}C_r\mathcal M_s,
 \label{eq:finite-product}
\end{equation}
where $C_r$ and $\mathcal M_s$ are the Laurent coefficients of the complete prefactor and master. A term proportional to a positive power of $\varepsilon$ in an integrand numerator or reduction coefficient can therefore contribute to the finite result when it multiplies a compensating $1/\varepsilon$ or $1/\varepsilon^2$ pole.
Such terms may be discarded only after the full Laurent products in Eq.~\eqref{eq:finite-product} have been inspected.

After the regulated sum is assembled, the Fourier transform gives $\mathcal V_{\mathrm{static}}(r)$. The useful scalar transform is
\begin{equation}
 \int\frac{\dd^d\bm q}{(2\pi)^d}\frac{e^{i\bm q\cdot\bm r}}{(\bm q^2)^\lambda}
 =\frac{\Gamma(d/2-\lambda)}
 {4^\lambda\pi^{d/2}\Gamma(\lambda)}
 (r^2)^{\lambda-d/2},\qquad r>0,
 \label{eq:Fourier-power}
\end{equation}
again interpreted by analytic continuation. The limit $d\to3$ is taken after multiplying the reduction coefficients, master integrals, $G_d$ factors, and Fourier kernel. This ordering retains finite terms generated by poles. Polynomial terms in $\bm q^2$ Fourier transform into $\delta^{(d)}(\bm r)$ and its derivatives, such as $\nabla^2\delta^{(d)}(\bm r)$. These contact terms are localized at coincident worldlines and do not contribute to the interaction of two distinct bodies at finite separation, $r>0$.

\subsubsection{Odd-PN factorization in scalar-tensor theory}
\label{sec:factorization}

The master-integral organization above admits a further simplification in the five-loop sector, beyond the reuse of GR denominators. In GR, odd-PN static diagrams factorize into lower-order pieces~\cite{Foffa:2019hrb}. A prime static diagram cannot be split into lower-order static pieces sewn at nonlinear worldline insertions; its source couplings are all linear. The correlation-function formulation of Ref.~\cite{Brunello:2026static} relates the absence of odd-PN prime graphs to the $\mathbb Z_2$ reflection symmetry of the static KK scalar. In the present theory, Eq.~\eqref{eq:static-bulk} has two independent $\mathbb Z_2$ symmetries,
\begin{equation}
 \Phi\longrightarrow-\Phi,\qquad
 \phi\longrightarrow-\phi,
 \label{eq:bulk-reflections}
\end{equation}
with $\sigma_{ij}$ unchanged. In terms of the unrescaled fundamental field, the second reflection is $\varphi-\varphi_0\to-(\varphi-\varphi_0)$, consistently with Eq.~\eqref{eq:scalar-normalization}. These are symmetries of the static bulk action, not of the general worldline mass functions.
Around the perturbative vacuum, a connected bulk correlator therefore obeys
\begin{equation}
 \big\langle\Phi(x_1)\cdots\Phi(x_{N_\Phi})
 \phi(y_1)\cdots\phi(y_{N_\phi})\big\rangle_{\mathrm{bulk},c}=0
 \quad\text{if $N_\Phi$ or $N_\phi$ is odd}.
 \label{eq:bulk-parity}
\end{equation}

A prime classical graph has only linear worldline insertions.
With $N=N_\Phi+N_\phi$ such insertions, tree counting gives an overall coupling $\kappa^{2(N-1)}$, hence $G^{N-1}$.
At static $n$PN, it would therefore require $N=n+2$.
For odd $n$, this number is odd, so at least one of the two selection rules in Eq.~\eqref{eq:bulk-parity} is violated.
Thus the extra massless scalar preserves the mechanism excluding a new prime odd-PN bulk contribution.

The surviving graphs contain nonlinear worldline vertices from $M_A(\varphi)e^{\kappa\Phi}$. These composite insertions sew lower-order bulk trees together, with the appropriate scalar response coefficient and combinatorial weight at each sewing point. This factorization takes place diagrammatically, before the field assignments and source coefficients are summed into a complete lower-order potential. When sewing the bulk subgraphs, one must retain the $\Phi$ or $\phi$ identity of each external scalar leg, the appropriate worldline coefficient $a_{A,q}$, the combinatorial factors, and the body labels. Consequently, the 5PN ST result cannot in general be obtained simply by multiplying the fully summed expressions for lower-PN ST potentials. At 5PN the allowed prime sectors are Newtonian, 2PN, and 4PN. No genuinely new five-loop master integral is required, although the momentum-space representation still contains five integrations.
This is the direct ST extension of the GR structural argument, without requiring a separate all-order formula for the ST potential. The distinct role of even orders, where new prime integrals can occur, is illustrated by the six-loop GR calculation of Ref.~\cite{Brunello:2025static}.

\section{Static potential through 5PN}
\label{sec:results}

We now give the complete static two-body potential through 5PN, defined by the interaction action in Eq.~\eqref{eq:potential-definition} and expressed in the variables and gauge of Sections~\ref{sec:ST} and~\ref{sec:EFT}. Its PN decomposition is
\begin{align}
 \mathcal V_{\mathrm{static}}(r)
 ={}&\mathcal V_{\mathrm N}^{G}
 +\mathcal V_{1\mathrm{PN}}^{G^2}
 +\mathcal V_{2\mathrm{PN}}^{G^3}
 +\mathcal V_{3\mathrm{PN}}^{G^4}
 +\mathcal V_{4\mathrm{PN}}^{G^5}
 +\mathcal V_{5\mathrm{PN}}^{G^6}
 +\mathcal O(G^7).
 \label{eq:Vstatic-PN-decomposition}
\end{align}
The remainder refers only to higher static orders. All expressions below are real-valued potentials: for the position-space diagram sum $\mathcal D_n$, the convention $S_{\mathrm{eff,int}}^{\mathrm{static}}=-T\mathcal V$ gives $\mathcal D_n=-i\mathcal V_{n\mathrm{PN}}^{G^{n+1}}$ and hence $\mathcal V_{n\mathrm{PN}}^{G^{n+1}}=i\mathcal D_n$, consistently with Eq.~\eqref{eq:potential-diagrams}.

Note that the static potential is a coordinate-dependent coefficient of the near-zone effective action. Its form can depend on the gauge, field parametrization, and allowed field or worldline coordinate transformations. Comparison with a potential in different variables therefore requires the corresponding transformations; a gauge-invariant observable also requires the other dynamical sectors.
In particular, the static interaction alone is not a binding-energy observable.

\subsection{Organization of the result and normalized scalar couplings}
\label{sec:result-organization}

At fixed PN order, the potential is a homogeneous polynomial in $m_1$ and $m_2$. Coefficients of mass monomials related by $1\leftrightarrow2$ are obtained by exchanging all body-labelled scalar responses. To make these relations explicit and expose the familiar scalar-tensor structure of the lower orders, we use the normalized combinations of Juli\'e et al.~\cite{Julie:2022qux}, translated into our conventions.
All responses are evaluated at the asymptotic background $\varphi_0$, so the superscript $0$ used in that reference is suppressed. For $A=1,2$ and $B\ne A$, define
\begin{equation}
\begin{aligned}
 G_{AB}&\equiv G(1+\alpha_A\alpha_B),
 &\bar\gamma_{AB}&=-\frac{2\alpha_A\alpha_B}{1+\alpha_A\alpha_B},\\[3pt]
 h&\equiv\frac{1}{1+\alpha_1\alpha_2}
 =1+\frac{\bar\gamma_{12}}{2}.
\end{aligned}
 \label{eq:normalized-G12}
\end{equation}

Here $G_{AB}$ is the dimensionful two-body coupling and already
contains the Einstein-frame gravitational constant $G$.
The symbol $h$, on the other hand, is a dimensionless shorthand, 
not an additional physical coupling.
The standard normalized scalar-tensor combinations needed for our comparison
of the lower PN sectors depend on the
dimensionless factor $1+\alpha_A\alpha_B$ and are given by
\begin{equation}
\begin{aligned}
 \bar\beta_A&=\frac{\beta_A\alpha_B^2}{2(1+\alpha_A\alpha_B)^2},
 &\delta_A&=\frac{\alpha_A^2}{(1+\alpha_A\alpha_B)^2},\\[5pt]
 \epsilon_A&=\frac{\beta'_A\alpha_B^3}{(1+\alpha_A\alpha_B)^3},
 &\zeta_{AB}&=\frac{\beta_A\beta_B\alpha_A\alpha_B}{(1+\alpha_A\alpha_B)^3},\\[5pt]
 \omega_A&=\frac{\alpha_A^2\beta_A\beta_B^2}{(1+\alpha_A\alpha_B)^4},
 &\kappa_A&=\frac{\beta''_A\alpha_B^4}{8(1+\alpha_A\alpha_B)^4},\\[5pt]
 \xi_A&=\frac{\alpha_A^2\alpha_B\beta_A\beta'_B}{(1+\alpha_A\alpha_B)^4},
 &\psi_A&=\frac{\alpha_A\alpha_B\beta_A}{(1+\alpha_A\alpha_B)^3}.
\end{aligned}
 \label{eq:normalized-couplings}
\end{equation}

The body-labelled coupling $\kappa_A$ is distinct from the field
normalization $\kappa$ in Eq.~\eqref{eq:gravitational-coupling}.
These combinations are built from the response hierarchy in
Eq.~\eqref{eq:response-hierarchy}; they are not independent
compact-object parameters.

For the PN corrections, the normalized mass-sector decomposition is
\begin{equation}
  \mathcal V_{n\mathrm{PN}}^{G^{n+1}}
 =\frac{G_{12}^{n+1}}{r^{n+1}}
 \sum_{k=1}^{n+1}m_1^k m_2^{n+2-k}\widehat C_{n,k},
 \qquad n=1,\ldots,5.
 \label{eq:Vstatic-normalized}
\end{equation}
The mass degree $n+2$ follows from the source-vertex counting of
the connected classical trees at order $G^{n+1}$. The Newtonian
term will be given separately below. We define $\mathsf E$ to
exchange every body label on scalar responses and their combinations,
so that
\begin{equation}
 \widehat C_{n,n+2-k}=\mathsf E\,\widehat C_{n,k},
 \qquad \mathsf E:\ 1\leftrightarrow2,\qquad \mathsf E^2=1.
 \label{eq:coefficient-exchange}
\end{equation}
Thus $\alpha_1\leftrightarrow\alpha_2$,
$\beta_1^{(p)}\leftrightarrow\beta_2^{(p)}$ for every $p\geq0$,
and likewise $\bar\beta_1\leftrightarrow\bar\beta_2$,
$\delta_1\leftrightarrow\delta_2$, and all other body-labelled
couplings. The symmetric quantities $G_{12}$, $h$,
$\bar\gamma_{12}$, and $\zeta_{12}$ are unchanged. The masses
are exchanged separately in the full potential. At even $n$, the
central sector $k=(n+2)/2$ is self-exchanged; this applies to
$\widehat C_{2,2}$ and $\widehat C_{4,3}$.

\subsection{Results through 3PN}
\label{sec:lower-results}

Single exchange fixes the Newtonian normalization,
\begin{equation}
  \mathcal V_{\mathrm N}^{G}=-\frac{G_{12}m_1m_2}{r},
 \label{eq:V-N}
\end{equation}
in agreement with Eq.~\eqref{eq:Newtonian-potential} since
$a_{A,1}=\alpha_A$. The first three PN corrections can then be
written using only five independent coefficient functions:
\begin{align}
 \mathcal V_{1\mathrm{PN}}^{G^2}
 &=\frac{G_{12}^2m_1m_2}{r^2}
 \bigl[m_1\widehat C_{1,2}+m_2\mathsf E\widehat C_{1,2}\bigr],
 \label{eq:V-1PN}\\
 \mathcal V_{2\mathrm{PN}}^{G^3}
 &=\frac{G_{12}^3m_1m_2}{r^3}
 \bigl[m_1^2\widehat C_{2,3}+m_1m_2\widehat C_{2,2}
       +m_2^2\mathsf E\widehat C_{2,3}\bigr],
 \label{eq:V-2PN}\\
 \mathcal V_{3\mathrm{PN}}^{G^4}
 &=\frac{G_{12}^4m_1m_2}{r^4}
 \bigl[m_1^3\widehat C_{3,4}+m_1^2m_2\widehat C_{3,3}
       +m_1m_2^2\mathsf E\widehat C_{3,3}
       +m_2^3\mathsf E\widehat C_{3,4}\bigr].
 \label{eq:V-3PN}
\end{align}
At 1PN, the sole independent coefficient is
\begin{align}
 \widehat C_{1,2} ={}& \bar\beta_{2} + \frac{1}{2}.
 \label{eq:C1-2}
\end{align}

At 2PN, the two independent sectors are
\begin{align}
 \widehat C_{2,3} ={}& - \bar\beta_{2} - \frac{\delta_{1}}{3}   - \frac{\epsilon_{2}}{6} - \frac{2 h^{2} + 1}{6},
 \label{eq:C2-3} \\
 \widehat C_{2,2} ={}& - 2 \bar\beta_{1} - 2 \bar\beta_{2}   - 2 h - \zeta_{12}   -1.
 \label{eq:C2-2}
\end{align}

The second expression is manifestly invariant under body exchange.
At 3PN, the normalized notation also accommodates the higher
responses and their mixed products without introducing additional
coupling definitions:
\begin{align}
 \widehat C_{3,4} ={}& \frac{\bar\beta_{2}^{2}}{2}
 +\frac{2\bar\beta_{2}\delta_{1}}{3}
 +\frac{\bar\beta_{2}(4h^{2}+3)}{6}
 +\frac{\delta_{1}}{3}+\frac{\epsilon_{2}}{6}
 +\frac{\kappa_{2}}{3}+\frac{8h^{2}+1}{24},
 \label{eq:C3-4} \\
%
%
 \widehat C_{3,3} ={}&
 \frac{2\bar\beta_{1}(h^{2}+3)}{3}
 +2\bar\beta_{2}^{2}+\bar\beta_{2}(8h+5)
 +\delta_{1}+\frac{\epsilon_{2}}{2}+h^{2}+4h
 \nonumber\\*
 &+\frac{\omega_{1}}{2}-\frac{\psi_{1}(h-3)}{3}
 +\frac{\xi_{1}}{2}-\frac{\zeta_{12}(h-5)}{2}+1.
 \label{eq:C3-3}
\end{align}

Together with Eq.~\eqref{eq:coefficient-exchange}, these formulas
specify all mass sectors through 3PN. Newtonian through 2PN
reproduce the previously known static ST interaction
\cite{Damour:1992we,Damour:1996field,Mirshekari:2013vb,Almeida:2024uph}, after
translating conventions. The 3PN expression supplies our
independent EFT determination of the static $G^4$ sector.
The normalized couplings facilitate comparison with the established
ST literature, but do not by themselves identify this result with
the harmonic-coordinate 3PN Lagrangian or subsequent Hamiltonians
\cite{Bernard:2018hta,Bernard:2018ivi,Jain:2022nxs,Julie:2022qux};
that comparison requires the appropriate variable transformations,
as discussed in more detail in Section~\ref{sec:comparison}.

\subsection{The 4PN and 5PN static potentials}
\label{sec:higher-results}

At 4PN and 5PN, the uniform mass-response coefficients provide an
economical explicit representation of the new static interaction.
A fully normalized extension of the lower-order couplings can also
be constructed, but requires many additional shorthand combinations
without making the expressions appreciably more compact or
providing enough additional insight to justify that extended dictionary.
We therefore present these orders directly in terms of
$\bm a_A=(a_{A,1},\ldots,a_{A,6})$, as defined in
Eq.~\eqref{eq:mass-expansion}. Writing the polynomial coefficient
functions as $F_{n,k}(\bm a_1,\bm a_2)$, the
corresponding decomposition is
\begin{equation}
  \mathcal V_{n\mathrm{PN}}^{G^{n+1}}
 =\frac{G^{n+1}}{r^{n+1}}
 \sum_{k=1}^{n+1}m_1^k m_2^{n+2-k}F_{n,k}(\bm a_1,\bm a_2).
 \label{eq:mass-sector-decomposition}
\end{equation}
For $n=1,\ldots,5$, the two representations are related exactly by
\begin{equation}
 F_{n,k}=(1+\alpha_1\alpha_2)^{n+1}\widehat C_{n,k},\qquad
 F_{n,n+2-k}=\mathsf E F_{n,k},\qquad
 \mathsf E:(\bm a_1,\bm a_2)\mapsto(\bm a_2,\bm a_1).
 \label{eq:coefficient-representations}
\end{equation}
The same potential is thus expressed either in the normalized
scalar-tensor coupling basis with powers of $G_{12}$ extracted,
or directly as polynomials in $a_{A,q}$ with explicit powers of
$G$; the two forms are algebraically equivalent. Extracting
$(1+\alpha_1\alpha_2)^{n+1}$ from $F_{n,k}$ does not imply
polynomial divisibility: $\widehat C_{n,k}$ may contain inverse
powers of $1+\alpha_1\alpha_2$, which cancel when the prefactor
is restored. The normalized coordinates require
$1+\alpha_1\alpha_2\ne0$, whereas the original polynomial
representation remains regular there as well.

At 4PN, the two exchanged pairs and the central sector give
\begin{equation}
 \mathcal V_{4\mathrm{PN}}^{G^5}
 =\frac{G^5m_1m_2}{r^5}
 \bigl[m_1^4F_{4,5}+m_1^3m_2F_{4,4}+m_1^2m_2^2F_{4,3}
 +m_1m_2^3\mathsf E F_{4,4}+m_2^4\mathsf E F_{4,5}\bigr].
 \label{eq:V-4PN}
\end{equation}

The extreme and next-to-extreme mass sectors are
\begin{align}
 F_{4,5}={}&-\frac{1}{120}\Bigl\{a_{1,1}^{5}(24a_{2,1}+20a_{2,3}+a_{2,5})+a_{1,1}^{4}(60a_{2,2}+5a_{2,4}+24)
 \nonumber\\
 &\quad +a_{1,1}^{3}(108a_{2,1}+30a_{2,3})+a_{1,1}^{2}(70a_{2,2}+68)+89a_{1,1}a_{2,1}+45\Bigr\}
 \label{eq:F4-5} 
\end{align}
and
\begin{align}
 F_{4,4}={}&-\frac{1}{6}\Bigl\{a_{1,1}^{4}\bigl[30a_{2,1}^{2}+16a_{2,1}a_{2,3}+5a_{2,2}(3a_{2,2}+2)+a_{2,4}\bigr]
 \nonumber\\
 &\quad +a_{1,1}^{3}\bigl[a_{2,1}(14a_{1,2}a_{2,2}+a_{1,2}a_{2,4}+113a_{2,2}+74)+a_{2,3}(3a_{1,2}a_{2,2}+19)\bigr]
 \nonumber\\
 &\quad +2a_{1,1}^{2}\bigl[3a_{1,2}a_{2,1}a_{2,3}+a_{2,1}^{2}(9a_{1,2}+72)+a_{2,2}(3a_{1,2}a_{2,2}+2a_{1,2}+46)+17\bigr]
 \nonumber\\
 &\quad +2a_{1,1}a_{2,1}\bigl[a_{1,2}(9a_{2,2}+4)+99\bigr]+10a_{1,2}a_{2,1}^{2}+62\Bigr\}.
 \label{eq:F4-4}
\end{align}

For the central sector, it is useful to display the exchange
invariance explicitly. Here
$(1+\mathsf E)X=X(\bm a_1,\bm a_2)+X(\bm a_2,\bm a_1)$
acts on the outer square bracket:
\begin{align}
 F_{4,3} ={}& -\frac{1}{12}\Bigl\{
 344a_{1,1}^{3}a_{2,1}^{3}
 +a_{1,1}^{2}a_{2,1}^{2}(255a_{1,2}a_{2,2}+3a_{1,3}a_{2,3}+1472)
 \nonumber\\*
 &\quad +a_{1,1}a_{2,1}(12a_{1,2}^{2}a_{2,2}^{2}
 +240a_{1,2}a_{2,2}+1499)+423
 \nonumber\\*
 &\quad +(1+\mathsf E)\Bigl[
 a_{1,1}^{3}\bigl(6a_{2,1}^{2}a_{2,3}+12a_{2,1}a_{2,2}^{2}
 +318a_{2,1}a_{2,2}+38a_{2,1}+11a_{2,3}\bigr)
 \nonumber\\*
 &\qquad +a_{1,1}^{2}\bigl(46a_{1,2}a_{2,1}^{2}
 +18a_{1,2}a_{2,1}a_{2,3}+18a_{1,2}a_{2,2}^{2}
 \nonumber\\*
 &\qquad\quad +6a_{1,3}a_{2,1}a_{2,2}^{2}
 +4a_{2,1}a_{2,3}+4a_{2,2}^{2}+273a_{2,2}+26\bigr)
 +24a_{1,1}a_{1,2}a_{2,1}\Bigr]\Bigr\}.
 \label{eq:F4-3}
\end{align}

Every term outside that square bracket is separately invariant.
Consequently $\mathsf E F_{4,3}=F_{4,3}$, while
$F_{4,1}=\mathsf E F_{4,5}$ and
$F_{4,2}=\mathsf E F_{4,4}$ complete the result without repeating
the exchanged polynomials.

At 5PN there is no self-exchanged mass sector. Three independent
polynomials suffice, with the full potential given by
\begin{equation}
\begin{aligned}
 \mathcal V_{5\mathrm{PN}}^{G^6}
 =\frac{G^6m_1m_2}{r^6}\Bigl[
 &m_1^5F_{5,6}+m_1^4m_2F_{5,5}+m_1^3m_2^2F_{5,4}\\
 &+m_1^2m_2^3\mathsf E F_{5,4}
  +m_1m_2^4\mathsf E F_{5,5}+m_2^5\mathsf E F_{5,6}\Bigr].
\end{aligned}
 \label{eq:V-5PN}
\end{equation}
The extreme mass sector is
\begin{align}
 F_{5,6}={}&\frac{1}{720}\Bigl\{a_{1,1}^{6}(184a_{2,2}+40a_{2,4}+a_{2,6})+a_{1,1}^{5}(368a_{2,1}+160a_{2,3}+6a_{2,5})
 \nonumber\\
 &\quad +a_{1,1}^{4}(608a_{2,2}+55a_{2,4}+184)+a_{1,1}^{3}(896a_{2,1}+180a_{2,3})
 \nonumber\\
 &\quad +a_{1,1}^{2}(439a_{2,2}+408)+534a_{1,1}a_{2,1}+225\Bigr\}.
 \label{eq:F5-6}
\end{align}

The coefficient multiplying $m_1^5m_2^2$ also contains products
of the responses on both bodies:
\begin{align}
 F_{5,5}={}&\frac{1}{360}\Bigl\{3a_{1,1}^{5}\bigl[a_{2,1}(1320a_{2,2}+105a_{2,4}+184)+290a_{2,2}a_{2,3}+120a_{2,3}+5a_{2,5}\bigr]
 \nonumber\\
 &\quad +a_{1,1}^{4}\Bigl(45a_{1,2}a_{2,3}^{2}+a_{2,1}^{2}(540a_{1,2}+8208)+15a_{2,1}\bigl[a_{1,2}a_{2,5}+a_{2,3}(32a_{1,2}+212)\bigr]
 \nonumber\\*
 &\quad\quad +5a_{2,2}^{2}(96a_{1,2}+561)+60a_{2,2}(a_{1,2}a_{2,4}+80)+375a_{2,4}+460\Bigr)
 \nonumber\\
 &\quad +8a_{1,1}^{3}\bigl[3a_{1,2}a_{2,1}(95a_{2,2}+5a_{2,4}+22)+2a_{2,1}(915a_{2,2}+862)
 \nonumber\\*
 &\quad\quad +a_{2,3}(45a_{1,2}a_{2,2}+15a_{1,2}+330)\bigr]
 \nonumber\\
 &\quad +8a_{1,1}^{2}\Bigl(a_{1,2}\bigl[3a_{2,1}(83a_{2,1}+25a_{2,3})
 \nonumber\\*
 &\quad\quad\quad +15a_{2,2}(5a_{2,2}+4)+10\bigr]+1901a_{2,1}^{2}+1305a_{2,2}+615\Bigr)
 \nonumber\\
 &\quad +8a_{1,1}a_{2,1}\bigl[2a_{1,2}(90a_{2,2}+53)+2385\bigr]+692a_{1,2}a_{2,1}^{2}+5460\Bigr\}.
 \label{eq:F5-5}
\end{align}

The remaining independent polynomial, multiplying $m_1^4m_2^3$,
is given by
\begin{align}
 &F_{5,4}={}\frac{1}{36}\Bigl\{a_{1,1}^{4}\Bigl(a_{2,1}^{2}(3357a_{2,2}+6a_{2,4}+1304)+2a_{2,1}a_{2,3}(18a_{2,2}+281)
 \nonumber\\*
 &\quad\quad\quad +2a_{2,2}\bigl[3a_{2,2}(2a_{2,2}+89)+86\bigr]+14a_{2,4}\Bigr)
 \nonumber\\
 &\quad\quad +a_{1,1}^{3}\Bigl[6a_{1,3}a_{2,2}^{3}+a_{2,1}^{3}(888a_{1,2}+6742)
 \nonumber\\*
 &\quad\quad\quad +a_{2,1}^{2}\bigl[3a_{1,3}(20a_{2,2}+a_{2,4})+a_{2,3}(471a_{1,2}+68)\bigr]
 \nonumber\\*
 &\quad\quad\quad +2a_{2,1}\Bigl(12a_{1,2}a_{2,4}+a_{2,2}\bigl[51a_{1,2}(9a_{2,2}+4)+9a_{1,3}a_{2,3}+66a_{2,2}+4822\bigr]
 \nonumber\\*
 &\quad\quad\quad\quad +2a_{2,4}+1200\Bigr)+a_{2,3}(72a_{1,2}a_{2,2}+12a_{2,2}+504)\Bigr]
 \nonumber\\
 &\quad\quad +a_{1,1}^{2}\Bigl[72a_{1,3}a_{2,1}^{3}+a_{2,1}^{2}\Bigl(3a_{1,2}\bigl[3a_{1,2}a_{2,4}+a_{2,2}(60a_{1,2}+1159)+584\bigr]
 \nonumber\\*
 &\quad\quad\quad\quad +27a_{1,3}a_{2,3}+336a_{2,2}+19649\Bigr)
 \nonumber\\*
 &\quad\quad\quad +6a_{2,1}\bigl[9a_{1,2}a_{2,3}(a_{1,2}a_{2,2}+9)+a_{1,3}a_{2,2}(9a_{2,2}+4)+6a_{2,3}\bigr]
 \nonumber\\*
 &\quad\quad\quad +2a_{2,2}\bigl[9a_{1,2}a_{2,2}(a_{1,2}a_{2,2}+27)+48a_{1,2}+18a_{2,2}+2528\bigr]+948\Bigr]
 \nonumber\\
 &\quad\quad +a_{1,1}a_{2,1}\Bigl(6a_{1,2}^{2}\bigl[24a_{2,1}^{2}+9a_{2,1}a_{2,3}+2a_{2,2}(9a_{2,2}+4)\bigr]+3a_{1,2}(853a_{2,1}^{2}+780a_{2,2}+200)
 \nonumber\\*
 &\quad\quad\quad +3a_{1,3}a_{2,1}(35a_{2,2}+12)+220a_{2,1}^{2}+164a_{2,2}+16176\Bigr)
 \nonumber\\
 &\quad\quad +3a_{2,1}^{2}\Bigl(a_{1,2}\bigl[a_{1,2}(35a_{2,2}+12)+629\bigr]+17a_{1,3}a_{2,1}+46\Bigr)+3918\Bigr\}.
 \label{eq:F5-4}
\end{align}

The remaining sectors are $F_{5,1}=\mathsf E F_{5,6}$,
$F_{5,2}=\mathsf E F_{5,5}$, and
$F_{5,3}=\mathsf E F_{5,4}$. Equations~\eqref{eq:V-4PN}
and~\eqref{eq:V-5PN}, with their explicit coefficient functions,
thus give the complete static $G^5$ and $G^6$ interactions.

The highest response at successive orders is $a_{A,2}$ at 1PN,
$a_{A,3}$ at 2PN, $a_{A,4}$ at 3PN, $a_{A,5}$ at 4PN, and
$a_{A,6}$ at 5PN. This hierarchy realizes the power counting of
Section~\ref{sec:Feynman-rules}, rather than an arbitrary truncation
of $M_A(\varphi)$. Finally, setting $a_{A,q}=0$ for $q\geq1$
removes the scalar responses, gives $G_{12}=G$ and $h=1$, and reduces the
displayed expressions to their GR static coefficients. The detailed
comparison and the other checks are discussed in
Section~\ref{sec:checks}.

\section{Checks and discussion}
\label{sec:checks}

\subsection{Lower-order comparison and the GR limit}
\label{sec:comparison}

With the explicit potential in hand, the first check is agreement
with the known ST static interaction
through 2PN. The most direct EFT comparison is
Ref.~\cite{Almeida:2024uph}, whose massless ST sector is obtained
by switching off the Gauss--Bonnet coupling. Its Einstein-frame
mass and scalar-response definitions coincide with ours under
$m_A^0=m_A$, $\alpha_A^0=a_{A,1}$,
$\beta_A^0=a_{A,2}-a_{A,1}^2$, and
$\beta_A^{\prime0}=a_{A,3}-3a_{A,1}a_{A,2}+2a_{A,1}^3$.
Its field normalization $\Lambda^{-1}$ corresponds to $\kappa$;
its gravitational KK scalar corresponds to our $\Phi$, while its
rescaled fundamental scalar fluctuation corresponds to our $\phi$.
The opposite metric-signature convention is accounted for at the
action level. Finally, the interaction Lagrangian is compared
with $-\mathcal V_{\mathrm{static}}$, as fixed by the definition of
$S_{\mathrm{eff,int}}^{\mathrm{static}}$ in
Eq.~\eqref{eq:potential-definition}.

These identifications make the lower-order agreement precise
without confusing the Einstein-frame constant $G$ with the
measured $G_{\mathrm N}$ in
Eq.~\eqref{eq:measured-newton-constant}. The corresponding 2PN
dynamics is also known from the Jordan-frame calculation of
Mirshekari and Will~\cite{Mirshekari:2013vb}; the Einstein-frame
EFT result uses the same point-particle mass-function description
and closely related KK conventions, so its static limit can be
compared directly using the field and response dictionary above. Agreement through 2PN has been explicitly verified.
A comparison with the existing 3PN dynamics would additionally
require consistent treatment of coordinate transformations and
acceleration-dependent terms, whose order reduction can move
contributions into the highest-$G$ sector. We therefore do not
count a 3PN comparison among the checks performed here.

A second check is obtained by removing the scalar dependence of
the source masses,
\begin{equation}
 a_{A,q}=0\quad(q\geq1),\qquad a_{A,0}=1,
 \label{eq:GR-limit}
\end{equation}
which eliminates the scalar couplings in the point-particle
sector. We have verified that the result then reduces through
5PN to the GR static potential in the corresponding EFT
variables~\cite{Gilmore:2008gq,Foffa:2011ub,Foffa:2016rgu,Foffa:2019hrb,Blumlein:2019zku}.
This comparison concerns the static potential before any
additional order reduction of the velocity-dependent dynamics.
It therefore tests the normalization and pure gravitational
subsector without conflating it with a complete ordinary
Hamiltonian at the same PN order.

\subsection{Exchange symmetry, finiteness, and rationality}
\label{sec:finiteness}

Beyond these comparisons, we have explicitly checked in the raw
output of the multiloop calculation, before using exchange symmetry
to organize the presentation, that every PN contribution is invariant
under simultaneous exchange of all body labels,
\begin{equation}
 (m_1,\bm a_1)\longleftrightarrow(m_2,\bm a_2).
 \label{eq:exchange-check}
\end{equation}
Individual diagrams need not have this symmetry, particularly
when nonlinear source vertices occur on only one body, and the
filtered sample counts exchanged configurations separately.
Recovering Eq.~\eqref{eq:exchange-check} after summing them is
therefore a useful algebraic check on the long scalar-dependent
expressions and their combinatorial weights.

At each order from 1PN through 5PN, the complete static
contribution is finite as $\varepsilon\to0$. All regulator poles
that occur in the intermediate calculation cancel in the
properly weighted sum. We have also checked separately at each
of these orders that all residual transcendental constants cancel: the final
coefficients are polynomials in the scalar-response parameters with rational coefficients in each mass sector. In the notation of
Eq.~\eqref{eq:mass-sector-decomposition}, this implies
\begin{equation}
 {F_{n,k}}\in\mathbb Q[\bm a_1,\bm a_2],
 \qquad n=1,\ldots,5,
 \label{eq:rationality}
\end{equation}
with the inverse powers of $r$ and overall powers of $G$
factored out as displayed there.

Neither property is imposed on individual diagrams or masters.
Tensor reduction produces regulator-dependent coefficients, and
master expansions can contain both poles and transcendental
constants. In particular, Eq.~\eqref{eq:M36} contains $\pi^2$,
$\zeta(3)$ and $\ln2$, including the combination $\pi^2\ln2$.
The cancellation encompasses every transcendental structure
occurring through the required expansion of $\mathcal M^{(4)}_7$ and the
other masters, after multiplication by their complete prefactors.
This provides a check on the combination of numerator
algebra, reduction, dimensional expansions, and Fourier
normalization. It is consequently essential to retain the
products in Eq.~\eqref{eq:finite-product} before taking the
three-dimensional limit. The same distinction between complicated
intermediate expressions and a finite rational static sum appears
in the four-, five-, and six-loop GR calculations
\cite{Foffa:2016rgu,Foffa:2019hrb,Blumlein:2019zku,Brunello:2025static}.
The statements here concern the strictly static point-particle
sector and do not imply rationality of complete PN Hamiltonians
or gauge-invariant observables.

\subsection{Structural implications and scope}
\label{sec:scope}

Alongside these algebraic checks, the factorization argument in
Section~\ref{sec:factorization}
explains why reaching five spatial loops does not require a new
five-loop prime master. The two independent scalar reflections
constrain the connected bulk building blocks, while the derivatives
of $M_A(\varphi)$ determine how those blocks are sewn on each
worldline. Increasing the number of scalar response parameters
therefore enlarges the algebraic result without introducing a new
odd-PN prime sector. This simplification concerns the integral
structure; the scalar-dependent sewing factors still have to be
computed, and neither exchange symmetry nor rationality can be
assumed diagram by diagram.

Together, the lower-order comparison, GR limit, body exchange, and
cancellation checks test complementary aspects of the calculation.
They support the use of the same multiloop infrastructure for
high-PN ST dynamics that has proved effective in GR. Nevertheless,
the present interaction remains one building block of that
program. A complete conservative result through 5PN also needs
the velocity-dependent near-zone sectors and the appropriate
radiation-region contributions. In GR, this combination of
velocity-dependent potential-region dynamics with hereditary
information has been developed extensively. Examples include
next-to-leading hereditary terms~\cite{Foffa:2019eeb}, conservative
tails~\cite{Almeida:2021xwn}, and recent 5PN scattering-based
analyses incorporating radiation-reaction and hereditary
effects~\cite{Porto:2026fsd}. These works address
complementary ingredients and the separation of conservative and
radiative contributions; they do not identify the static sector
with the complete high-PN dynamics. In ST theories, conservative
scalar tails already enter at 3PN~\cite{Bernard:2018hta,Julie:2022qux}.
Finite-size operators are separate inputs, and their scalar and
mixed contributions must be included when the goal is a complete
compact-object description~\cite{Bernard:2023eul}. None of those
additions is fixed by the static potential alone.
\section{Conclusions}
\label{sec:conclusions}

We have computed the nonspinning point-particle static interaction
in massless scalar-tensor theories through 5PN within the PN EFT
framework, thereby extending the high-loop static program beyond
GR. By retaining the derivatives of each body's Einstein-frame
mass function through sixth order, as required by 5PN power
counting, the result keeps the compact objects' scalar responses
explicit. The calculation covers every sector from $G$ through
$G^6$, combining tree-level Newtonian exchange with IBP reduction
of the one- through five-loop massless two-point integrals.

The resulting potential reproduces the known static ST interaction
through 2PN and supplies an independent EFT determination of the
3PN sector before extending the explicit result to 4PN and 5PN.
This extension is made tractable by a structural simplification:
the additional scalar modifies source couplings and integrand
numerators while preserving the relevant GR denominator topologies.
In addition, the independent $\mathbb Z_2$ reflections of $\Phi$
and $\phi$ in the static bulk action extend the symmetry explanation
of odd-PN factorization to the present theory. The 5PN result
therefore requires lower-order master information but no new
five-loop prime integral.

Several complementary checks test both this integral structure
and the scalar-response dependence of the result. The potential
satisfies the GR limit and body-exchange symmetry, while at every
order from 1PN through 5PN the regulator poles and residual
transcendental constants cancel in the sum. Thus, despite the
more complicated intermediate master expansions, the final static
interaction is finite and has rational coefficients.

These results isolate the highest-$G$, highest-loop point-particle
sector of the conservative dynamics. Completing that dynamics
through 5PN will also require velocity-dependent interactions,
radiation-region effects, and the relevant finite-size operators;
waveform predictions further require the corresponding radiative
information. The static potential alone therefore does not fix
observables at this accuracy, but provides a tested analytic
input for future complete predictions in scalar-tensor gravity.

\acknowledgments
The authors would like to thank Stefano Foffa for helpful discussions. 
This work was supported in part by the Start-up Research Fund of Southeast University No. RF1028624160.

\appendix
\section{Dictionary for the mass-expansion coefficients}
\label{app:mass-coefficients}

The worldline couplings are conveniently expressed in terms of
the normalized derivatives of the mass functions,
\begin{equation}
 a_{A,q}
 =\frac{1}{m_A}
 \left.\frac{\dd^q M_A(\varphi)}{\dd\varphi^q}\right|_{\varphi_0},
 \qquad
 m_A=M_A(\varphi_0).
\end{equation}
To relate these coefficients to the scalar-response parameters,
we introduce the functions
\begin{equation}
 \alpha_A(\varphi)
 =\frac{\dd\ln M_A(\varphi)}{\dd\varphi},
 \qquad
 \beta_A(\varphi)
 =\frac{\dd\alpha_A(\varphi)}{\dd\varphi}.
\end{equation}
In the expressions below, symbols without an explicit scalar
argument are evaluated at $\varphi_0$, and primes denote
derivatives with respect to $\varphi$ before this evaluation.
Thus,
\begin{equation}
 \alpha_A=\alpha_A(\varphi_0),
 \qquad
 \beta_A^{(n)}
 =\left.\frac{\dd^n\beta_A(\varphi)}{\dd\varphi^n}
 \right|_{\varphi_0}
 =\left.\frac{\dd^{n+2}\ln M_A(\varphi)}
 {\dd\varphi^{n+2}}\right|_{\varphi_0},
 \qquad n\geq0,
\end{equation}
with $\beta_A^{(0)}=\beta_A$ and
$\beta_A^{(1)}=\beta'_A$.

The derivatives use the same fixed internal state and smooth
compact-object branch as in Section~\ref{sec:ST}.

The dictionary follows directly by differentiating
$M_A=\exp(\ln M_A)$. In particular, the identity
\begin{equation}
 \frac{1}{M_A}
 \frac{\dd^{q+1}M_A}{\dd\varphi^{q+1}}
 =
 \left(\frac{\dd}{\dd\varphi}+\alpha_A(\varphi)\right)
 \left[
 \frac{1}{M_A}\frac{\dd^q M_A}{\dd\varphi^q}
 \right]
\end{equation}
generates the coefficients recursively. Evaluating at
$\varphi_0$ gives the expressions required through 5PN:
\begin{align}
 a_{A,0} &= 1,
 \\
 a_{A,1} &= \alpha_A,
 \\
 a_{A,2} &= \alpha_A^2+\beta_A,
 \\
 a_{A,3} &= \alpha_A^3
             +3\alpha_A\beta_A+\beta'_A,
 \\
 a_{A,4} &= \alpha_A^4
             +6\alpha_A^2\beta_A
             +3\beta_A^2
             +4\alpha_A\beta'_A
             +\beta''_A,
 \\
 a_{A,5} &= \alpha_A^5
             +10\alpha_A^3\beta_A
             +15\alpha_A\beta_A^2
             +10\alpha_A^2\beta'_A
 \nonumber\\
          &\quad
             +10\beta_A\beta'_A
             +5\alpha_A\beta''_A
             +\beta'''_A,
 \\
 a_{A,6} &= \alpha_A^6
             +15\alpha_A^4\beta_A
             +45\alpha_A^2\beta_A^2
             +15\beta_A^3
 \nonumber\\
          &\quad
             +20\alpha_A^3\beta'_A
             +60\alpha_A\beta_A\beta'_A
             +10(\beta'_A)^2
 \nonumber\\
          &\quad
             +15\alpha_A^2\beta''_A
             +15\beta_A\beta''_A
             +6\alpha_A\beta'''_A
             +\beta''''_A.
 \label{eq:mass-coefficient-dictionary}
\end{align}
These relations hold separately for each body and require no
specific choice of the coupling function $\mathcal A(\varphi)$.
The dependence on the theory and compact-object structure
enters through the matched mass functions and their derivatives.
\section{Topology families and master bases}
\label{app:integrals}
\suppressfloats[t]

In this appendix, we collect all diagram topologies entering our
static loop calculations through 5PN order, together with the integral
families and master bases used to evaluate them. The results are
organized by post-Newtonian order, from one to five loops.

In the figures, large dots in the upper and lower rows denote source
insertions on the two bodies, while smaller dots denote bulk vertices.
The lines indicate connectivity, with field-species assignments
suppressed. Identifying all source insertions on each body with a
single external vertex gives the corresponding two-point topology
and its loop count.

Below we list the independent topologies required at each order and the propagator bases used for their two-point integrals.

\subsection{One loop: 1PN}
\label{app:family-1PN}

\begin{figure}[h]
\centering
\setlength{\tabcolsep}{3pt}
\begin{tabular}{c}
\begin{tikzpicture}[x=1.05cm,y=.85cm,baseline=(current bounding box.center)]
 \path[use as bounding box] (-1.05,-1.12) rectangle (1.05,1.12);
 \coordinate (U0) at (0,1);
 \coordinate (L0) at (-0.7,-1);
 \coordinate (L1) at (0.7,-1);
 \draw[line width=.6pt] (U0)--(L0);
 \draw[line width=.6pt] (U0)--(L1);
 \fill (U0) circle (1.9pt);
 \fill (L0) circle (1.9pt);
 \fill (L1) circle (1.9pt);
\end{tikzpicture}
\end{tabular}
\caption{Diagram topology contributing to the static 1PN calculation (one loop).}
\label{fig:topologies-1PN}
\end{figure}
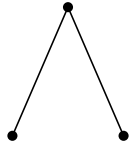

\leavevmode{At 1PN, the single topology in Fig.~\ref{fig:topologies-1PN} becomes a one-loop bubble when the insertions on each body are identified. Its two propagators span the two independent loop-dependent scalar products:}
\begin{equation}
\begin{aligned}
 D_{1}&=\bm \ell_1^2, &\quad D_{2}&=(\bm \ell_1+\bm q)^2.
\end{aligned}
 \label{eq:family-1PN}
\end{equation}
The single master is the standard bubble,
\begin{equation}
 \mathcal M^{(1)}_1={M}^{(1)}_{1,1}=B_d(1,1;\bm q^2),
 \label{eq:masters-1PN}
\end{equation}
with $B_d$ given in Eq.~\eqref{eq:bubble}.

\subsection{Two loops: 2PN}
\label{app:family-2PN}

\begin{figure}[h]
\centering
\setlength{\tabcolsep}{3pt}
\begin{tabular}{c}
\begin{tikzpicture}[x=1.05cm,y=.85cm,baseline=(current bounding box.center)]
 \path[use as bounding box] (-1.05,-1.12) rectangle (1.05,1.12);
 \coordinate (U0) at (-0.65,1);
 \coordinate (U1) at (0.65,1);
 \coordinate (L0) at (-0.65,-1);
 \coordinate (L1) at (0.65,-1);
 \coordinate (B0) at (-0.65,0);
 \coordinate (B1) at (0.65,0);
 \draw[line width=.6pt] (U0)--(B0);
 \draw[line width=.6pt] (B0)--(L0);
 \draw[line width=.6pt] (U1)--(B1);
 \draw[line width=.6pt] (B1)--(L1);
 \draw[line width=.6pt] (B0)--(B1);
 \fill (U0) circle (1.9pt);
 \fill (U1) circle (1.9pt);
 \fill (L0) circle (1.9pt);
 \fill (L1) circle (1.9pt);
 \fill (B0) circle (1.1pt);
 \fill (B1) circle (1.1pt);
\end{tikzpicture}
\end{tabular}
\caption{{Independent topology contributing to the static 2PN calculation (two loops).}}
\label{fig:topologies-2PN}
\end{figure}
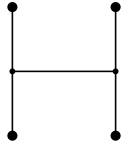

\leavevmode{At 2PN, the H-shaped topology in Fig.~\ref{fig:topologies-2PN} becomes the two-loop kite when the insertions on each body are identified. Its five propagators span the full $N_2=5$ scalar-product basis}
\begin{equation}
\begin{aligned}
 D_{1}&=\bm \ell_1^2, &\quad D_{2}&=\bm \ell_2^2,\\
 D_{3}&=(\bm \ell_1+\bm \ell_2)^2, &\quad D_{4}&=(\bm \ell_1+\bm q)^2,\\
 D_{5}&=(\bm \ell_2-\bm q)^2.
\end{aligned}
 \label{eq:family-2PN}
\end{equation}
All integrals in this family can be reduced to the two master integrals below
\begin{equation}
\begin{aligned}
 \mathcal M^{(2)}_1&={M}^{(2)}_{0,1,1,1,0},\\
 \mathcal M^{(2)}_2&={M}^{(2)}_{1,1,0,1,1}.
\end{aligned}
 \label{eq:masters-2PN}
\end{equation}
\leavevmode{The former, known as the two-loop sunrise integral, can be evaluated by successive bubble integrations, while the latter factorizes into two one-loop bubbles.}

\subsection{Three loops: 3PN}
\label{app:family-3PN}

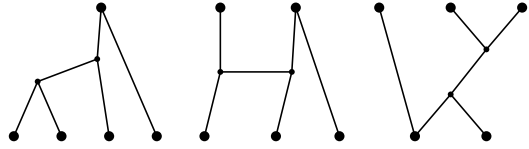
\begin{figure}[h]
\centering
\setlength{\tabcolsep}{3pt}
\begin{tabular}{ccc}
\begin{tikzpicture}[x=1.05cm,y=.85cm,baseline=(current bounding box.center)]
 \path[use as bounding box] (-1.05,-1.12) rectangle (1.05,1.12);
 \coordinate (U0) at (0.2,1);
 \coordinate (L0) at (-0.9,-1);
 \coordinate (L1) at (-0.3,-1);
 \coordinate (L2) at (0.3,-1);
 \coordinate (L3) at (0.9,-1);
 \coordinate (B0) at (-0.6,-0.15);
 \coordinate (B1) at (0.15,0.2);
 \draw[line width=.6pt] (U0)--(B1);
 \draw[line width=.6pt] (B1)--(L2);
 \draw[line width=.6pt] (U0)--(L3);
 \draw[line width=.6pt] (B0)--(L0);
 \draw[line width=.6pt] (B0)--(L1);
 \draw[line width=.6pt] (B0)--(B1);
 \fill (U0) circle (1.9pt);
 \fill (L0) circle (1.9pt);
 \fill (L1) circle (1.9pt);
 \fill (L2) circle (1.9pt);
 \fill (L3) circle (1.9pt);
 \fill (B0) circle (1.1pt);
 \fill (B1) circle (1.1pt);
\end{tikzpicture} &
\begin{tikzpicture}[x=1.05cm,y=.85cm,baseline=(current bounding box.center)]
 \path[use as bounding box] (-1.05,-1.12) rectangle (1.05,1.12);
 \coordinate (U0) at (-0.6,1);
 \coordinate (U1) at (0.35,1);
 \coordinate (L0) at (-0.8,-1);
 \coordinate (L1) at (0.1,-1);
 \coordinate (L2) at (0.9,-1);
 \coordinate (B0) at (-0.6,0);
 \coordinate (B1) at (0.3,0);
 \draw[line width=.6pt] (U0)--(B0);
 \draw[line width=.6pt] (B0)--(L0);
 \draw[line width=.6pt] (U1)--(B1);
 \draw[line width=.6pt] (B1)--(L1);
 \draw[line width=.6pt] (U1)--(L2);
 \draw[line width=.6pt] (B0)--(B1);
 \fill (U0) circle (1.9pt);
 \fill (U1) circle (1.9pt);
 \fill (L0) circle (1.9pt);
 \fill (L1) circle (1.9pt);
 \fill (L2) circle (1.9pt);
 \fill (B0) circle (1.1pt);
 \fill (B1) circle (1.1pt);
\end{tikzpicture} &
\begin{tikzpicture}[x=1.05cm,y=.85cm,baseline=(current bounding box.center)]
 \path[use as bounding box] (-1.05,-1.12) rectangle (1.05,1.12);
 \coordinate (U0) at (-0.9,1);
 \coordinate (U1) at (0,1);
 \coordinate (U2) at (0.9,1);
 \coordinate (L0) at (-0.45,-1);
 \coordinate (L1) at (0.45,-1);
 \coordinate (B0) at (0.45,0.35);
 \coordinate (B1) at (0,-0.35);
 \draw[line width=.6pt] (U0)--(L0);
 \draw[line width=.6pt] (U1)--(B0);
 \draw[line width=.6pt] (U2)--(B0);
 \draw[line width=.6pt] (B0)--(B1);
 \draw[line width=.6pt] (B1)--(L0);
 \draw[line width=.6pt] (B1)--(L1);
 \fill (U0) circle (1.9pt);
 \fill (U1) circle (1.9pt);
 \fill (U2) circle (1.9pt);
 \fill (L0) circle (1.9pt);
 \fill (L1) circle (1.9pt);
 \fill (B0) circle (1.1pt);
 \fill (B1) circle (1.1pt);
\end{tikzpicture}
\end{tabular}
\caption{Independent topologies contributing to the static 3PN calculation (three loops).}
\label{fig:topologies-3PN}
\end{figure}

\leavevmode{The three independent 3PN top topologies in Fig.~\ref{fig:topologies-3PN} each have six propagators. Their integrals are represented as distinct sectors of a common three-loop family}
\begin{equation}
\begin{aligned}
 D_{1}&=\bm \ell_1^2, &\quad D_{2}&=\bm \ell_2^2,\\
 D_{3}&=\bm \ell_3^2, &\quad D_{4}&=(\bm \ell_1-\bm \ell_2)^2,\\
 D_{5}&=(\bm \ell_1+\bm \ell_3)^2, &\quad D_{6}&=(\bm \ell_2+\bm \ell_3)^2,\\
 D_{7}&=(\bm \ell_1-\bm q)^2, &\quad D_{8}&=\bm \ell_2\cdot \bm q,\\
 D_{9}&=\bm \ell_3\cdot \bm q.
\end{aligned}
 \label{eq:family-3PN}
\end{equation}
\leavevmode{For each topology, six of the seven quadratic forms $D_1,\ldots,D_7$ serve as propagator denominators. The unused quadratic form and the explicit scalar products $D_8$ and $D_9$ are needed to complete the $N_3=9$ scalar-product basis.}
The master basis used in the reduction is
\begin{equation}
\begin{aligned}
 \mathcal M^{(3)}_1&={M}^{(3)}_{0,0,1,1,0,1,1,0,0},\\
 \mathcal M^{(3)}_2&={M}^{(3)}_{0,1,1,1,1,0,1,0,0}.
\end{aligned}
 \label{eq:masters-3PN}
\end{equation}
\leavevmode{The two masters in Eq.~\eqref{eq:masters-3PN} occupy sectors with four and five active denominators, respectively. Their indices refer to the $D_j$ basis in Eq.~\eqref{eq:family-3PN}; a zero index omits the corresponding $D_j$.}

\subsection{Four loops: 4PN}
\label{app:family-4PN}

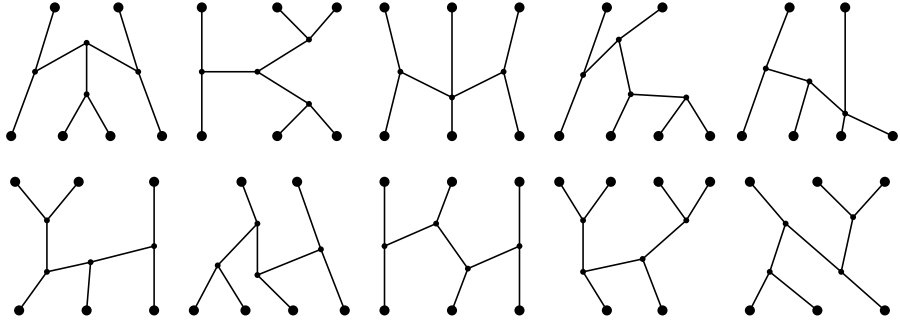
\begin{figure}[h]
\centering
\setlength{\tabcolsep}{3pt}
\begin{tabular}{ccccc}
\begin{tikzpicture}[x=1.05cm,y=.85cm,baseline=(current bounding box.center)]
 \path[use as bounding box] (-1.05,-1.12) rectangle (1.05,1.12);
 \coordinate (U0) at (-0.4,1);
 \coordinate (U1) at (0.4,1);
 \coordinate (L0) at (-0.95,-1);
 \coordinate (L1) at (-0.3,-1);
 \coordinate (L2) at (0.3,-1);
 \coordinate (L3) at (0.95,-1);
 \coordinate (B0) at (-0.65,0);
 \coordinate (B1) at (0.65,0);
 \coordinate (B2) at (0,0.45);
 \coordinate (B3) at (0,-0.35);
 \draw[line width=.6pt] (U0)--(B0);
 \draw[line width=.6pt] (B0)--(L0);
 \draw[line width=.6pt] (U1)--(B1);
 \draw[line width=.6pt] (B1)--(L3);
 \draw[line width=.6pt] (B0)--(B2);
 \draw[line width=.6pt] (B1)--(B2);
 \draw[line width=.6pt] (B2)--(B3);
 \draw[line width=.6pt] (B3)--(L1);
 \draw[line width=.6pt] (B3)--(L2);
 \fill (U0) circle (1.9pt);
 \fill (U1) circle (1.9pt);
 \fill (L0) circle (1.9pt);
 \fill (L1) circle (1.9pt);
 \fill (L2) circle (1.9pt);
 \fill (L3) circle (1.9pt);
 \fill (B0) circle (1.1pt);
 \fill (B1) circle (1.1pt);
 \fill (B2) circle (1.1pt);
 \fill (B3) circle (1.1pt);
\end{tikzpicture} &
\begin{tikzpicture}[x=1.05cm,y=.85cm,baseline=(current bounding box.center)]
 \path[use as bounding box] (-1.05,-1.12) rectangle (1.05,1.12);
 \coordinate (U0) at (-0.85,1);
 \coordinate (U1) at (0.1,1);
 \coordinate (U2) at (0.85,1);
 \coordinate (L0) at (-0.85,-1);
 \coordinate (L1) at (0.1,-1);
 \coordinate (L2) at (0.85,-1);
 \coordinate (B0) at (-0.85,0);
 \coordinate (B1) at (-0.15,0);
 \coordinate (B2) at (0.5,0.5);
 \coordinate (B3) at (0.5,-0.5);
 \draw[line width=.6pt] (U0)--(B0);
 \draw[line width=.6pt] (B0)--(L0);
 \draw[line width=.6pt] (B0)--(B1);
 \draw[line width=.6pt] (B1)--(B2);
 \draw[line width=.6pt] (B1)--(B3);
 \draw[line width=.6pt] (B2)--(U1);
 \draw[line width=.6pt] (B2)--(U2);
 \draw[line width=.6pt] (B3)--(L1);
 \draw[line width=.6pt] (B3)--(L2);
 \fill (U0) circle (1.9pt);
 \fill (U1) circle (1.9pt);
 \fill (U2) circle (1.9pt);
 \fill (L0) circle (1.9pt);
 \fill (L1) circle (1.9pt);
 \fill (L2) circle (1.9pt);
 \fill (B0) circle (1.1pt);
 \fill (B1) circle (1.1pt);
 \fill (B2) circle (1.1pt);
 \fill (B3) circle (1.1pt);
\end{tikzpicture} &
\begin{tikzpicture}[x=1.05cm,y=.85cm,baseline=(current bounding box.center)]
 \path[use as bounding box] (-1.05,-1.12) rectangle (1.05,1.12);
 \coordinate (U0) at (-0.85,1);
 \coordinate (U1) at (0,1);
 \coordinate (U2) at (0.85,1);
 \coordinate (L0) at (-0.85,-1);
 \coordinate (L1) at (0,-1);
 \coordinate (L2) at (0.85,-1);
 \coordinate (B0) at (-0.65,0);
 \coordinate (B1) at (0,-0.4);
 \coordinate (B2) at (0.65,0);
 \draw[line width=.6pt] (U0)--(B0);
 \draw[line width=.6pt] (B0)--(L0);
 \draw[line width=.6pt] (U2)--(B2);
 \draw[line width=.6pt] (B2)--(L2);
 \draw[line width=.6pt] (B0)--(B1);
 \draw[line width=.6pt] (B2)--(B1);
 \draw[line width=.6pt] (U1)--(B1);
 \draw[line width=.6pt] (B1)--(L1);
 \fill (U0) circle (1.9pt);
 \fill (U1) circle (1.9pt);
 \fill (U2) circle (1.9pt);
 \fill (L0) circle (1.9pt);
 \fill (L1) circle (1.9pt);
 \fill (L2) circle (1.9pt);
 \fill (B0) circle (1.1pt);
 \fill (B1) circle (1.1pt);
 \fill (B2) circle (1.1pt);
\end{tikzpicture} &
\begin{tikzpicture}[x=1.05cm,y=.85cm,baseline=(current bounding box.center)]
 \path[use as bounding box] (-1.05,-1.12) rectangle (1.05,1.12);
 \coordinate (U0) at (-0.35,1);
 \coordinate (U1) at (0.35,1);
 \coordinate (L0) at (-0.95,-1);
 \coordinate (L1) at (-0.3,-1);
 \coordinate (L2) at (0.3,-1);
 \coordinate (L3) at (0.95,-1);
 \coordinate (B0) at (-0.65,-0.05);
 \coordinate (B1) at (-0.2,0.5);
 \coordinate (B2) at (-0.05,-0.35);
 \coordinate (B3) at (0.65,-0.4);
 \draw[line width=.6pt] (U0)--(B0);
 \draw[line width=.6pt] (B0)--(L0);
 \draw[line width=.6pt] (B0)--(B1);
 \draw[line width=.6pt] (B1)--(U1);
 \draw[line width=.6pt] (B1)--(B2);
 \draw[line width=.6pt] (B2)--(L1);
 \draw[line width=.6pt] (B2)--(B3);
 \draw[line width=.6pt] (B3)--(L2);
 \draw[line width=.6pt] (B3)--(L3);
 \fill (U0) circle (1.9pt);
 \fill (U1) circle (1.9pt);
 \fill (L0) circle (1.9pt);
 \fill (L1) circle (1.9pt);
 \fill (L2) circle (1.9pt);
 \fill (L3) circle (1.9pt);
 \fill (B0) circle (1.1pt);
 \fill (B1) circle (1.1pt);
 \fill (B2) circle (1.1pt);
 \fill (B3) circle (1.1pt);
\end{tikzpicture} &
\begin{tikzpicture}[x=1.05cm,y=.85cm,baseline=(current bounding box.center)]
 \path[use as bounding box] (-1.05,-1.12) rectangle (1.05,1.12);
 \coordinate (U0) at (-0.35,1);
 \coordinate (U1) at (0.35,1);
 \coordinate (L0) at (-0.95,-1);
 \coordinate (L1) at (-0.3,-1);
 \coordinate (L2) at (0.3,-1);
 \coordinate (L3) at (0.95,-1);
 \coordinate (B0) at (-0.65,0.05);
 \coordinate (B1) at (-0.1,-0.15);
 \coordinate (B2) at (0.35,-0.65);
 \draw[line width=.6pt] (U0)--(B0);
 \draw[line width=.6pt] (B0)--(L0);
 \draw[line width=.6pt] (B0)--(B1);
 \draw[line width=.6pt] (B1)--(L1);
 \draw[line width=.6pt] (B1)--(B2);
 \draw[line width=.6pt] (U1)--(B2);
 \draw[line width=.6pt] (B2)--(L2);
 \draw[line width=.6pt] (B2)--(L3);
 \fill (U0) circle (1.9pt);
 \fill (U1) circle (1.9pt);
 \fill (L0) circle (1.9pt);
 \fill (L1) circle (1.9pt);
 \fill (L2) circle (1.9pt);
 \fill (L3) circle (1.9pt);
 \fill (B0) circle (1.1pt);
 \fill (B1) circle (1.1pt);
 \fill (B2) circle (1.1pt);
\end{tikzpicture}\\[34pt]
\begin{tikzpicture}[x=1.05cm,y=.85cm,baseline=(current bounding box.center)]
 \path[use as bounding box] (-1.05,-1.12) rectangle (1.05,1.12);
 \coordinate (U0) at (-0.9,1);
 \coordinate (U1) at (-0.1,1);
 \coordinate (U2) at (0.85,1);
 \coordinate (L0) at (-0.85,-1);
 \coordinate (L1) at (0,-1);
 \coordinate (L2) at (0.85,-1);
 \coordinate (B0) at (-0.5,0.4);
 \coordinate (B1) at (-0.5,-0.4);
 \coordinate (B2) at (0.05,-0.25);
 \coordinate (B3) at (0.85,0);
 \draw[line width=.6pt] (U0)--(B0);
 \draw[line width=.6pt] (U1)--(B0);
 \draw[line width=.6pt] (B0)--(B1);
 \draw[line width=.6pt] (B1)--(L0);
 \draw[line width=.6pt] (B1)--(B2);
 \draw[line width=.6pt] (B2)--(L1);
 \draw[line width=.6pt] (B2)--(B3);
 \draw[line width=.6pt] (B3)--(U2);
 \draw[line width=.6pt] (B3)--(L2);
 \fill (U0) circle (1.9pt);
 \fill (U1) circle (1.9pt);
 \fill (U2) circle (1.9pt);
 \fill (L0) circle (1.9pt);
 \fill (L1) circle (1.9pt);
 \fill (L2) circle (1.9pt);
 \fill (B0) circle (1.1pt);
 \fill (B1) circle (1.1pt);
 \fill (B2) circle (1.1pt);
 \fill (B3) circle (1.1pt);
\end{tikzpicture} &
\begin{tikzpicture}[x=1.05cm,y=.85cm,baseline=(current bounding box.center)]
 \path[use as bounding box] (-1.05,-1.12) rectangle (1.05,1.12);
 \coordinate (U0) at (-0.35,1);
 \coordinate (U1) at (0.35,1);
 \coordinate (L0) at (-0.95,-1);
 \coordinate (L1) at (-0.3,-1);
 \coordinate (L2) at (0.3,-1);
 \coordinate (L3) at (0.95,-1);
 \coordinate (B0) at (-0.65,-0.3);
 \coordinate (B1) at (-0.15,0.35);
 \coordinate (B2) at (-0.15,-0.45);
 \coordinate (B3) at (0.65,-0.05);
 \draw[line width=.6pt] (U0)--(B1);
 \draw[line width=.6pt] (B1)--(B0);
 \draw[line width=.6pt] (B0)--(L0);
 \draw[line width=.6pt] (B0)--(L1);
 \draw[line width=.6pt] (B1)--(B2);
 \draw[line width=.6pt] (B2)--(L2);
 \draw[line width=.6pt] (B2)--(B3);
 \draw[line width=.6pt] (B3)--(U1);
 \draw[line width=.6pt] (B3)--(L3);
 \fill (U0) circle (1.9pt);
 \fill (U1) circle (1.9pt);
 \fill (L0) circle (1.9pt);
 \fill (L1) circle (1.9pt);
 \fill (L2) circle (1.9pt);
 \fill (L3) circle (1.9pt);
 \fill (B0) circle (1.1pt);
 \fill (B1) circle (1.1pt);
 \fill (B2) circle (1.1pt);
 \fill (B3) circle (1.1pt);
\end{tikzpicture} &
\begin{tikzpicture}[x=1.05cm,y=.85cm,baseline=(current bounding box.center)]
 \path[use as bounding box] (-1.05,-1.12) rectangle (1.05,1.12);
 \coordinate (U0) at (-0.85,1);
 \coordinate (U1) at (0,1);
 \coordinate (U2) at (0.85,1);
 \coordinate (L0) at (-0.85,-1);
 \coordinate (L1) at (0,-1);
 \coordinate (L2) at (0.85,-1);
 \coordinate (B0) at (-0.85,0);
 \coordinate (B1) at (-0.2,0.35);
 \coordinate (B2) at (0.2,-0.35);
 \coordinate (B3) at (0.85,0);
 \draw[line width=.6pt] (U0)--(B0);
 \draw[line width=.6pt] (B0)--(L0);
 \draw[line width=.6pt] (B0)--(B1);
 \draw[line width=.6pt] (B1)--(U1);
 \draw[line width=.6pt] (B1)--(B2);
 \draw[line width=.6pt] (B2)--(L1);
 \draw[line width=.6pt] (B2)--(B3);
 \draw[line width=.6pt] (B3)--(U2);
 \draw[line width=.6pt] (B3)--(L2);
 \fill (U0) circle (1.9pt);
 \fill (U1) circle (1.9pt);
 \fill (U2) circle (1.9pt);
 \fill (L0) circle (1.9pt);
 \fill (L1) circle (1.9pt);
 \fill (L2) circle (1.9pt);
 \fill (B0) circle (1.1pt);
 \fill (B1) circle (1.1pt);
 \fill (B2) circle (1.1pt);
 \fill (B3) circle (1.1pt);
\end{tikzpicture} &
\begin{tikzpicture}[x=1.05cm,y=.85cm,baseline=(current bounding box.center)]
 \path[use as bounding box] (-1.05,-1.12) rectangle (1.05,1.12);
 \coordinate (U0) at (-0.95,1);
 \coordinate (U1) at (-0.3,1);
 \coordinate (U2) at (0.3,1);
 \coordinate (U3) at (0.95,1);
 \coordinate (L0) at (-0.35,-1);
 \coordinate (L1) at (0.35,-1);
 \coordinate (B0) at (-0.65,0.4);
 \coordinate (B1) at (-0.65,-0.4);
 \coordinate (B2) at (0.1,-0.2);
 \coordinate (B3) at (0.65,0.4);
 \draw[line width=.6pt] (U0)--(B0);
 \draw[line width=.6pt] (U1)--(B0);
 \draw[line width=.6pt] (B0)--(B1);
 \draw[line width=.6pt] (B1)--(L0);
 \draw[line width=.6pt] (B1)--(B2);
 \draw[line width=.6pt] (B2)--(L1);
 \draw[line width=.6pt] (B2)--(B3);
 \draw[line width=.6pt] (B3)--(U2);
 \draw[line width=.6pt] (B3)--(U3);
 \fill (U0) circle (1.9pt);
 \fill (U1) circle (1.9pt);
 \fill (U2) circle (1.9pt);
 \fill (U3) circle (1.9pt);
 \fill (L0) circle (1.9pt);
 \fill (L1) circle (1.9pt);
 \fill (B0) circle (1.1pt);
 \fill (B1) circle (1.1pt);
 \fill (B2) circle (1.1pt);
 \fill (B3) circle (1.1pt);
\end{tikzpicture} &
\begin{tikzpicture}[x=1.05cm,y=.85cm,baseline=(current bounding box.center)]
 \path[use as bounding box] (-1.05,-1.12) rectangle (1.05,1.12);
 \coordinate (U0) at (-0.85,1);
 \coordinate (U1) at (0,1);
 \coordinate (U2) at (0.85,1);
 \coordinate (L0) at (-0.85,-1);
 \coordinate (L1) at (0,-1);
 \coordinate (L2) at (0.85,-1);
 \coordinate (B0) at (-0.4,0.35);
 \coordinate (B1) at (-0.6,-0.4);
 \coordinate (B2) at (0.3,-0.4);
 \coordinate (B3) at (0.45,0.45);
 \draw[line width=.6pt] (U0)--(B0);
 \draw[line width=.6pt] (B0)--(B1);
 \draw[line width=.6pt] (B1)--(L0);
 \draw[line width=.6pt] (B1)--(L1);
 \draw[line width=.6pt] (B0)--(B2);
 \draw[line width=.6pt] (B2)--(L2);
 \draw[line width=.6pt] (B2)--(B3);
 \draw[line width=.6pt] (B3)--(U1);
 \draw[line width=.6pt] (B3)--(U2);
 \fill (U0) circle (1.9pt);
 \fill (U1) circle (1.9pt);
 \fill (U2) circle (1.9pt);
 \fill (L0) circle (1.9pt);
 \fill (L1) circle (1.9pt);
 \fill (L2) circle (1.9pt);
 \fill (B0) circle (1.1pt);
 \fill (B1) circle (1.1pt);
 \fill (B2) circle (1.1pt);
 \fill (B3) circle (1.1pt);
\end{tikzpicture}
\end{tabular}
\caption{Independent topologies contributing to the static 4PN calculation (four loops).}
\label{fig:topologies-4PN}
\end{figure}

\leavevmode{The ten independent 4PN topologies in Fig.~\ref{fig:topologies-4PN} comprise eight nine-propagator top topologies and two eight-propagator sub-topologies. Their integrals are represented as sectors of a common four-loop family.}

\leavevmode{The common propagator set and auxiliary scalar products are}
\begin{equation}
\begin{aligned}
 D_{1}&=\bm \ell_1^2, &\quad D_{2}&=\bm \ell_2^2,\\
 D_{3}&=\bm \ell_3^2, &\quad D_{4}&=\bm \ell_4^2,\\
 D_{5}&=(\bm \ell_1-\bm \ell_2)^2, &\quad D_{6}&=(\bm \ell_1+\bm \ell_3)^2,\\
 D_{7}&=(\bm \ell_1-\bm \ell_4)^2, &\quad D_{8}&=(\bm \ell_2+\bm \ell_3)^2,\\
 D_{9}&=(\bm \ell_2+\bm q)^2, &\quad D_{10}&=(\bm \ell_3-\bm q)^2,\\
 D_{11}&=(-\bm \ell_1+\bm \ell_2+\bm \ell_4)^2, &\quad D_{12}&=(\bm \ell_3+\bm \ell_4-\bm q)^2,\\
 D_{13}&=\bm \ell_3\cdot \bm \ell_4, &\quad D_{14}&=\bm \ell_1\cdot \bm q.
\end{aligned}
 \label{eq:family-4PN}
\end{equation}
\leavevmode{For each topology, eight or nine of the twelve quadratic forms $D_1,\ldots,D_{12}$ serve as propagator denominators. The unused quadratic forms, together with the explicit scalar products $D_{13}$ and $D_{14}$, provide the ISPs needed to complete the $N_4=14$ scalar-product basis.}
The master basis used in the reduction is
\begin{equation}
\begin{aligned}
 \mathcal M^{(4)}_1&={M}^{(4)}_{0,0,0,1,0,1,1,1,1,0,0,0,0,0},\\
 \mathcal M^{(4)}_2&={M}^{(4)}_{0,0,1,1,1,0,1,0,1,1,0,0,0,0},\\
 \mathcal M^{(4)}_3&={M}^{(4)}_{0,0,1,1,1,0,1,1,1,0,0,0,0,0},\\
 \mathcal M^{(4)}_4&={M}^{(4)}_{0,0,1,1,1,1,1,0,1,0,0,0,0,0},\\
 \mathcal M^{(4)}_5&={M}^{(4)}_{1,0,0,0,0,1,1,0,1,1,1,0,0,0},\\
 \mathcal M^{(4)}_6&={M}^{(4)}_{1,0,0,0,1,1,1,0,1,1,1,0,0,0},\\
 \mathcal M^{(4)}_7&={M}^{(4)}_{1,0,0,1,1,1,1,1,1,0,0,1,0,0}.
\end{aligned}
 \label{eq:masters-4PN}
\end{equation}
\leavevmode{The seven master index vectors in Eq.~\eqref{eq:masters-4PN} refer to the common $D_j$ basis of Eq.~\eqref{eq:family-4PN}; they do not enumerate the ten diagram topologies. The required nontrivial four-loop dimensional input is discussed in Sec.~\ref{sec:master-input}.}

\subsection{Five loops: 5PN}
\label{app:family-5PN}

\begin{figure}[h]
\centering
\setlength{\tabcolsep}{6pt}
\begin{tabular}{cccc}
\begin{tikzpicture}[x=1.1cm,y=.75cm,baseline=(current bounding box.center)]
 \path[use as bounding box] (-1.1,-1.15) rectangle (1.1,1.15);
 \coordinate (U0) at (-0.95,1.0);
 \coordinate (U1) at (0.95,1.0);
 \coordinate (L0) at (-0.95,-1.0);
 \coordinate (L4) at (-0.475,-1.0);
 \coordinate (L1) at (0.0,-1.0);
 \coordinate (L2) at (0.475,-1.0);
 \coordinate (L3) at (0.95,-1.0);
 \coordinate (B0) at (0.3023,-0.7712);
 \coordinate (B1) at (-0.3023,-0.7806);
 \coordinate (B2) at (0.2591,-0.9748);
 \coordinate (B3) at (-0.7341,-0.3543);
 \draw[line width=.6pt] (L0)--(B3); 
 \draw[line width=.6pt] (B3)--(U0); 
 \draw[line width=.6pt] (B2)--(L1); 
 \draw[line width=.6pt] (B3)--(B1); 
 \draw[line width=.6pt] (B0)--(B2); 
 \draw[line width=.6pt] (B2)--(L2); 
 \draw[line width=.6pt] (L3)--(B0); 
 \draw[line width=.6pt] (L4)--(B1); 
 \draw[line width=.6pt] (U1)--(L3); 
 \draw[line width=.6pt] (B1)--(B0); 
 \fill (U0) circle (1.9pt);
 \fill (U1) circle (1.9pt);
 \fill (L0) circle (1.9pt);
 \fill (L4) circle (1.9pt);
 \fill (L1) circle (1.9pt);
 \fill (L2) circle (1.9pt);
 \fill (L3) circle (1.9pt);
 \fill (B0) circle (1.1pt);
 \fill (B1) circle (1.1pt);
 \fill (B2) circle (1.1pt);
 \fill (B3) circle (1.1pt);
\end{tikzpicture} &
\begin{tikzpicture}[x=1.1cm,y=.75cm,baseline=(current bounding box.center)]
 \path[use as bounding box] (-1.1,-1.15) rectangle (1.1,1.15);
 \coordinate (U1) at (-0.95,1.0);
 \coordinate (U0) at (0.95,1.0);
 \coordinate (L0) at (-0.95,-1.0);
 \coordinate (L4) at (-0.475,-1.0);
 \coordinate (L2) at (0.0,-1.0);
 \coordinate (L1) at (0.475,-1.0);
 \coordinate (L3) at (0.95,-1.0);
 \coordinate (B0) at (0.1583,-0.7249);
 \coordinate (B1) at (-0.1056,-0.8042);
 \coordinate (B2) at (0.5278,-0.7881);
 \coordinate (B3) at (0.0528,-0.217);
 \draw[line width=.6pt] (B3)--(U0); 
 \draw[line width=.6pt] (L0)--(B3); 
 \draw[line width=.6pt] (L1)--(B2); 
 \draw[line width=.6pt] (B1)--(L2); 
 \draw[line width=.6pt] (B2)--(B0); 
 \draw[line width=.6pt] (L3)--(B2); 
 \draw[line width=.6pt] (B0)--(B3); 
 \draw[line width=.6pt] (B1)--(L4); 
 \draw[line width=.6pt] (U1)--(L0); 
 \draw[line width=.6pt] (B0)--(B1); 
 \fill (U1) circle (1.9pt);
 \fill (U0) circle (1.9pt);
 \fill (L0) circle (1.9pt);
 \fill (L4) circle (1.9pt);
 \fill (L2) circle (1.9pt);
 \fill (L1) circle (1.9pt);
 \fill (L3) circle (1.9pt);
 \fill (B0) circle (1.1pt);
 \fill (B1) circle (1.1pt);
 \fill (B2) circle (1.1pt);
 \fill (B3) circle (1.1pt);
\end{tikzpicture} &
\begin{tikzpicture}[x=1.1cm,y=.75cm,baseline=(current bounding box.center)]
 \path[use as bounding box] (-1.1,-1.15) rectangle (1.1,1.15);
 \coordinate (U0) at (-0.95,1.0);
 \coordinate (U1) at (0.95,1.0);
 \coordinate (L0) at (-0.95,-1.0);
 \coordinate (L1) at (-0.475,-1.0);
 \coordinate (L2) at (0.0,-1.0);
 \coordinate (L3) at (0.475,-1.0);
 \coordinate (L4) at (0.95,-1.0);
 \coordinate (B0) at (0.2418,-0.8702);
 \coordinate (B1) at (0.2505,-0.5643);
 \coordinate (B2) at (-0.6218,-0.7096);
 \coordinate (B3) at (-0.4405,-0.0207);
 \draw[line width=.6pt] (B2)--(B3); 
 \draw[line width=.6pt] (L0)--(B2); 
 \draw[line width=.6pt] (B3)--(U0); 
 \draw[line width=.6pt] (L1)--(B2); 
 \draw[line width=.6pt] (B3)--(B1); 
 \draw[line width=.6pt] (B0)--(L2); 
 \draw[line width=.6pt] (L3)--(B0); 
 \draw[line width=.6pt] (L4)--(B1); 
 \draw[line width=.6pt] (U1)--(L4); 
 \draw[line width=.6pt] (B1)--(B0); 
 \fill (U0) circle (1.9pt);
 \fill (U1) circle (1.9pt);
 \fill (L0) circle (1.9pt);
 \fill (L1) circle (1.9pt);
 \fill (L2) circle (1.9pt);
 \fill (L3) circle (1.9pt);
 \fill (L4) circle (1.9pt);
 \fill (B0) circle (1.1pt);
 \fill (B1) circle (1.1pt);
 \fill (B2) circle (1.1pt);
 \fill (B3) circle (1.1pt);
\end{tikzpicture} &
\begin{tikzpicture}[x=1.1cm,y=.75cm,baseline=(current bounding box.center)]
 \path[use as bounding box] (-1.1,-1.15) rectangle (1.1,1.15);
 \coordinate (U0) at (-0.95,1.0);
 \coordinate (U1) at (0.95,1.0);
 \coordinate (L3) at (-0.95,-1.0);
 \coordinate (L4) at (-0.475,-1.0);
 \coordinate (L0) at (0.0,-1.0);
 \coordinate (L1) at (0.475,-1.0);
 \coordinate (L2) at (0.95,-1.0);
 \coordinate (B0) at (0.5344,-0.1486);
 \coordinate (B1) at (-0.6531,-0.6845);
 \coordinate (B2) at (0.6531,-0.6759);
 \coordinate (B3) at (-0.5344,-0.2675);
 \draw[line width=.6pt] (L0)--(B3); 
 \draw[line width=.6pt] (B3)--(U0); 
 \draw[line width=.6pt] (B2)--(L1); 
 \draw[line width=.6pt] (B3)--(B1); 
 \draw[line width=.6pt] (L2)--(B2); 
 \draw[line width=.6pt] (B2)--(B0); 
 \draw[line width=.6pt] (L3)--(B1); 
 \draw[line width=.6pt] (U1)--(B0); 
 \draw[line width=.6pt] (B1)--(L4); 
 \draw[line width=.6pt] (B0)--(L0); 
 \fill (U0) circle (1.9pt);
 \fill (U1) circle (1.9pt);
 \fill (L3) circle (1.9pt);
 \fill (L4) circle (1.9pt);
 \fill (L0) circle (1.9pt);
 \fill (L1) circle (1.9pt);
 \fill (L2) circle (1.9pt);
 \fill (B0) circle (1.1pt);
 \fill (B1) circle (1.1pt);
 \fill (B2) circle (1.1pt);
 \fill (B3) circle (1.1pt);
\end{tikzpicture}\\[34pt]
\begin{tikzpicture}[x=1.1cm,y=.75cm,baseline=(current bounding box.center)]
 \path[use as bounding box] (-1.1,-1.15) rectangle (1.1,1.15);
 \coordinate (U0) at (-0.95,1.0);
 \coordinate (U1) at (0.95,1.0);
 \coordinate (L4) at (-0.95,-1.0);
 \coordinate (L0) at (-0.475,-1.0);
 \coordinate (L1) at (0.0,-1.0);
 \coordinate (L2) at (0.475,-1.0);
 \coordinate (L3) at (0.95,-1.0);
 \coordinate (B0) at (0.3023,-0.3335);
 \coordinate (B1) at (0.7341,-0.1698);
 \coordinate (B2) at (-0.2591,-0.5884);
 \coordinate (B3) at (-0.3023,0.0024);
 \draw[line width=.6pt] (B3)--(U0); 
 \draw[line width=.6pt] (L0)--(B2); 
 \draw[line width=.6pt] (B3)--(B2); 
 \draw[line width=.6pt] (U1)--(B1); 
 \draw[line width=.6pt] (B0)--(B3); 
 \draw[line width=.6pt] (B2)--(L1); 
 \draw[line width=.6pt] (L2)--(B0); 
 \draw[line width=.6pt] (L3)--(B1); 
 \draw[line width=.6pt] (U0)--(L4); 
 \draw[line width=.6pt] (B1)--(B0); 
 \fill (U0) circle (1.9pt);
 \fill (U1) circle (1.9pt);
 \fill (L4) circle (1.9pt);
 \fill (L0) circle (1.9pt);
 \fill (L1) circle (1.9pt);
 \fill (L2) circle (1.9pt);
 \fill (L3) circle (1.9pt);
 \fill (B0) circle (1.1pt);
 \fill (B1) circle (1.1pt);
 \fill (B2) circle (1.1pt);
 \fill (B3) circle (1.1pt);
\end{tikzpicture} &
\begin{tikzpicture}[x=1.1cm,y=.75cm,baseline=(current bounding box.center)]
 \path[use as bounding box] (-1.1,-1.15) rectangle (1.1,1.15);
 \coordinate (U1) at (-0.95,1.0);
 \coordinate (U0) at (0.95,1.0);
 \coordinate (L4) at (-0.95,-1.0);
 \coordinate (L3) at (-0.475,-1.0);
 \coordinate (L2) at (0.0,-1.0);
 \coordinate (L1) at (0.475,-1.0);
 \coordinate (L0) at (0.95,-1.0);
 \coordinate (B0) at (-0.0259,-0.4905);
 \coordinate (B1) at (-0.4836,-0.1448);
 \coordinate (B2) at (0.4059,-0.5804);
 \coordinate (B3) at (0.7686,-0.1697);
 \draw[line width=.6pt] (B3)--(U0); 
 \draw[line width=.6pt] (L0)--(B3); 
 \draw[line width=.6pt] (B2)--(B3); 
 \draw[line width=.6pt] (U1)--(B1); 
 \draw[line width=.6pt] (B0)--(B2); 
 \draw[line width=.6pt] (B2)--(L1); 
 \draw[line width=.6pt] (L2)--(B0); 
 \draw[line width=.6pt] (L3)--(B1); 
 \draw[line width=.6pt] (U1)--(L4); 
 \draw[line width=.6pt] (B1)--(B0); 
 \fill (U1) circle (1.9pt);
 \fill (U0) circle (1.9pt);
 \fill (L4) circle (1.9pt);
 \fill (L3) circle (1.9pt);
 \fill (L2) circle (1.9pt);
 \fill (L1) circle (1.9pt);
 \fill (L0) circle (1.9pt);
 \fill (B0) circle (1.1pt);
 \fill (B1) circle (1.1pt);
 \fill (B2) circle (1.1pt);
 \fill (B3) circle (1.1pt);
\end{tikzpicture} &
\begin{tikzpicture}[x=1.1cm,y=.75cm,baseline=(current bounding box.center)]
 \path[use as bounding box] (-1.1,-1.15) rectangle (1.1,1.15);
 \coordinate (U1) at (-0.95,1.0);
 \coordinate (U0) at (0.95,1.0);
 \coordinate (L2) at (-0.95,-1.0);
 \coordinate (L1) at (-0.475,-1.0);
 \coordinate (L0) at (0.0,-1.0);
 \coordinate (L4) at (0.475,-1.0);
 \coordinate (L3) at (0.95,-1.0);
 \coordinate (B0) at (-0.1583,-0.2502);
 \coordinate (B1) at (-0.6861,-0.0286);
 \coordinate (B2) at (-0.2111,-0.6778);
 \coordinate (B3) at (0.4222,-0.1941);
 \draw[line width=.6pt] (B3)--(U0); 
 \draw[line width=.6pt] (B2)--(L0); 
 \draw[line width=.6pt] (U1)--(B1); 
 \draw[line width=.6pt] (B0)--(B2); 
 \draw[line width=.6pt] (B2)--(L1); 
 \draw[line width=.6pt] (B3)--(B0); 
 \draw[line width=.6pt] (L2)--(B1); 
 \draw[line width=.6pt] (U0)--(L3); 
 \draw[line width=.6pt] (L4)--(B3); 
 \draw[line width=.6pt] (B1)--(B0); 
 \fill (U1) circle (1.9pt);
 \fill (U0) circle (1.9pt);
 \fill (L2) circle (1.9pt);
 \fill (L1) circle (1.9pt);
 \fill (L0) circle (1.9pt);
 \fill (L4) circle (1.9pt);
 \fill (L3) circle (1.9pt);
 \fill (B0) circle (1.1pt);
 \fill (B1) circle (1.1pt);
 \fill (B2) circle (1.1pt);
 \fill (B3) circle (1.1pt);
\end{tikzpicture} &
\begin{tikzpicture}[x=1.1cm,y=.75cm,baseline=(current bounding box.center)]
 \path[use as bounding box] (-1.1,-1.15) rectangle (1.1,1.15);
 \coordinate (U0) at (-0.95,1.0);
 \coordinate (U1) at (0.0,1.0);
 \coordinate (U2) at (0.95,1.0);
 \coordinate (L0) at (-0.95,-1.0);
 \coordinate (L1) at (-0.3167,-1.0);
 \coordinate (L2) at (0.3167,-1.0);
 \coordinate (L3) at (0.95,-1.0);
 \coordinate (B0) at (0.2073,-0.6667);
 \coordinate (B1) at (0.0691,-0.9241);
 \coordinate (B2) at (-0.3973,-0.4772);
 \coordinate (B3) at (-0.4491,0.6148);
 \draw[line width=.6pt] (B3)--(U0); 
 \draw[line width=.6pt] (B2)--(B3); 
 \draw[line width=.6pt] (L0)--(B2); 
 \draw[line width=.6pt] (L1)--(B1); 
 \draw[line width=.6pt] (U1)--(B3); 
 \draw[line width=.6pt] (B1)--(B0); 
 \draw[line width=.6pt] (L2)--(B1); 
 \draw[line width=.6pt] (B0)--(B2); 
 \draw[line width=.6pt] (U2)--(L3); 
 \draw[line width=.6pt] (B0)--(L3); 
 \fill (U0) circle (1.9pt);
 \fill (U1) circle (1.9pt);
 \fill (U2) circle (1.9pt);
 \fill (L0) circle (1.9pt);
 \fill (L1) circle (1.9pt);
 \fill (L2) circle (1.9pt);
 \fill (L3) circle (1.9pt);
 \fill (B0) circle (1.1pt);
 \fill (B1) circle (1.1pt);
 \fill (B2) circle (1.1pt);
 \fill (B3) circle (1.1pt);
\end{tikzpicture}\\[34pt]
\begin{tikzpicture}[x=1.1cm,y=.75cm,baseline=(current bounding box.center)]
 \path[use as bounding box] (-1.1,-1.15) rectangle (1.1,1.15);
 \coordinate (U0) at (-0.95,1.0);
 \coordinate (U1) at (0.95,1.0);
 \coordinate (L0) at (-0.95,-1.0);
 \coordinate (L1) at (-0.475,-1.0);
 \coordinate (L2) at (0.0,-1.0);
 \coordinate (L4) at (0.475,-1.0);
 \coordinate (L3) at (0.95,-1.0);
 \coordinate (B0) at (0.1056,-0.7024);
 \coordinate (B1) at (-0.0528,0.5254);
 \coordinate (B2) at (-0.5278,-0.6659);
 \coordinate (B3) at (-0.1583,-0.3707);
 \draw[line width=.6pt] (B2)--(B3); 
 \draw[line width=.6pt] (L0)--(B2); 
 \draw[line width=.6pt] (B3)--(B1); 
 \draw[line width=.6pt] (B1)--(U0); 
 \draw[line width=.6pt] (L1)--(B2); 
 \draw[line width=.6pt] (B3)--(B0); 
 \draw[line width=.6pt] (B1)--(U1); 
 \draw[line width=.6pt] (L2)--(B0); 
 \draw[line width=.6pt] (U1)--(L3); 
 \draw[line width=.6pt] (B0)--(L4); 
 \fill (U0) circle (1.9pt);
 \fill (U1) circle (1.9pt);
 \fill (L0) circle (1.9pt);
 \fill (L1) circle (1.9pt);
 \fill (L2) circle (1.9pt);
 \fill (L4) circle (1.9pt);
 \fill (L3) circle (1.9pt);
 \fill (B0) circle (1.1pt);
 \fill (B1) circle (1.1pt);
 \fill (B2) circle (1.1pt);
 \fill (B3) circle (1.1pt);
\end{tikzpicture} &
\begin{tikzpicture}[x=1.1cm,y=.75cm,baseline=(current bounding box.center)]
 \path[use as bounding box] (-1.1,-1.15) rectangle (1.1,1.15);
 \coordinate (U1) at (-0.95,1.0);
 \coordinate (U0) at (0.0,1.0);
 \coordinate (U2) at (0.95,1.0);
 \coordinate (L2) at (-0.95,-1.0);
 \coordinate (L3) at (-0.3167,-1.0);
 \coordinate (L0) at (0.3167,-1.0);
 \coordinate (L1) at (0.95,-1.0);
 \coordinate (B0) at (-0.5988,-0.5533);
 \coordinate (B1) at (-0.5297,0.0975);
 \coordinate (B2) at (0.4088,-0.4944);
 \coordinate (B3) at (-0.0403,0.301);
 \draw[line width=.6pt] (B2)--(B3); 
 \draw[line width=.6pt] (L0)--(B2); 
 \draw[line width=.6pt] (B3)--(U0); 
 \draw[line width=.6pt] (L1)--(B2); 
 \draw[line width=.6pt] (B3)--(B1); 
 \draw[line width=.6pt] (B0)--(L2); 
 \draw[line width=.6pt] (L3)--(B0); 
 \draw[line width=.6pt] (U1)--(B1); 
 \draw[line width=.6pt] (B1)--(B0); 
 \draw[line width=.6pt] (L1)--(U2); 
 \fill (U1) circle (1.9pt);
 \fill (U0) circle (1.9pt);
 \fill (U2) circle (1.9pt);
 \fill (L2) circle (1.9pt);
 \fill (L3) circle (1.9pt);
 \fill (L0) circle (1.9pt);
 \fill (L1) circle (1.9pt);
 \fill (B0) circle (1.1pt);
 \fill (B1) circle (1.1pt);
 \fill (B2) circle (1.1pt);
 \fill (B3) circle (1.1pt);
\end{tikzpicture} &
\begin{tikzpicture}[x=1.1cm,y=.75cm,baseline=(current bounding box.center)]
 \path[use as bounding box] (-1.1,-1.15) rectangle (1.1,1.15);
 \coordinate (U2) at (-0.95,1.0);
 \coordinate (U1) at (0.0,1.0);
 \coordinate (U0) at (0.95,1.0);
 \coordinate (L3) at (-0.95,-1.0);
 \coordinate (L1) at (-0.3167,-1.0);
 \coordinate (L2) at (0.3167,-1.0);
 \coordinate (L0) at (0.95,-1.0);
 \coordinate (B0) at (-0.3109,-0.5141);
 \coordinate (B1) at (0.1209,0.0512);
 \coordinate (B2) at (-0.1036,-0.7515);
 \coordinate (B3) at (0.6736,0.1249);
 \draw[line width=.6pt] (L0)--(B3); 
 \draw[line width=.6pt] (B3)--(U0); 
 \draw[line width=.6pt] (B2)--(L1); 
 \draw[line width=.6pt] (B3)--(B1); 
 \draw[line width=.6pt] (B0)--(B2); 
 \draw[line width=.6pt] (B2)--(L2); 
 \draw[line width=.6pt] (L3)--(B0); 
 \draw[line width=.6pt] (U1)--(B1); 
 \draw[line width=.6pt] (B1)--(B0); 
 \draw[line width=.6pt] (L3)--(U2); 
 \fill (U2) circle (1.9pt);
 \fill (U1) circle (1.9pt);
 \fill (U0) circle (1.9pt);
 \fill (L3) circle (1.9pt);
 \fill (L1) circle (1.9pt);
 \fill (L2) circle (1.9pt);
 \fill (L0) circle (1.9pt);
 \fill (B0) circle (1.1pt);
 \fill (B1) circle (1.1pt);
 \fill (B2) circle (1.1pt);
 \fill (B3) circle (1.1pt);
\end{tikzpicture} &
\begin{tikzpicture}[x=1.1cm,y=.75cm,baseline=(current bounding box.center)]
 \path[use as bounding box] (-1.1,-1.15) rectangle (1.1,1.15);
 \coordinate (U0) at (-0.95,1.0);
 \coordinate (U1) at (0.95,1.0);
 \coordinate (L3) at (-0.95,-1.0);
 \coordinate (L0) at (-0.475,-1.0);
 \coordinate (L2) at (0.0,-1.0);
 \coordinate (L4) at (0.475,-1.0);
 \coordinate (L1) at (0.95,-1.0);
 \coordinate (B0) at (0.0594,-0.069);
 \coordinate (B1) at (0.6531,0.0801);
 \coordinate (B2) at (-0.6531,-0.6324);
 \coordinate (B3) at (-0.5344,-0.2332);
 \draw[line width=.6pt] (B3)--(U0); 
 \draw[line width=.6pt] (L0)--(B2); 
 \draw[line width=.6pt] (B3)--(B2); 
 \draw[line width=.6pt] (B1)--(L1); 
 \draw[line width=.6pt] (L2)--(B3); 
 \draw[line width=.6pt] (B2)--(L3); 
 \draw[line width=.6pt] (B1)--(B0); 
 \draw[line width=.6pt] (B0)--(L4); 
 \draw[line width=.6pt] (U1)--(B1); 
 \draw[line width=.6pt] (B0)--(U0); 
 \fill (U0) circle (1.9pt);
 \fill (U1) circle (1.9pt);
 \fill (L3) circle (1.9pt);
 \fill (L0) circle (1.9pt);
 \fill (L2) circle (1.9pt);
 \fill (L4) circle (1.9pt);
 \fill (L1) circle (1.9pt);
 \fill (B0) circle (1.1pt);
 \fill (B1) circle (1.1pt);
 \fill (B2) circle (1.1pt);
 \fill (B3) circle (1.1pt);
\end{tikzpicture}\\[34pt]
\begin{tikzpicture}[x=1.1cm,y=.75cm,baseline=(current bounding box.center)]
 \path[use as bounding box] (-1.1,-1.15) rectangle (1.1,1.15);
 \coordinate (U1) at (-0.95,1.0);
 \coordinate (U2) at (-0.3167,1.0);
 \coordinate (U0) at (0.3167,1.0);
 \coordinate (U3) at (0.95,1.0);
 \coordinate (L1) at (-0.95,-1.0);
 \coordinate (L0) at (0.0,-1.0);
 \coordinate (L2) at (0.95,-1.0);
 \coordinate (B0) at (-0.0403,-0.4272);
 \coordinate (B1) at (-0.5297,-0.243);
 \coordinate (B2) at (-0.5988,0.5507);
 \coordinate (B3) at (0.4088,-0.19);
 \draw[line width=.6pt] (B3)--(U0); 
 \draw[line width=.6pt] (U1)--(B2); 
 \draw[line width=.6pt] (B2)--(B1); 
 \draw[line width=.6pt] (B0)--(L0); 
 \draw[line width=.6pt] (U2)--(B2); 
 \draw[line width=.6pt] (B3)--(B0); 
 \draw[line width=.6pt] (L1)--(B1); 
 \draw[line width=.6pt] (U3)--(L2); 
 \draw[line width=.6pt] (L2)--(B3); 
 \draw[line width=.6pt] (B1)--(B0); 
 \fill (U1) circle (1.9pt);
 \fill (U2) circle (1.9pt);
 \fill (U0) circle (1.9pt);
 \fill (U3) circle (1.9pt);
 \fill (L1) circle (1.9pt);
 \fill (L0) circle (1.9pt);
 \fill (L2) circle (1.9pt);
 \fill (B0) circle (1.1pt);
 \fill (B1) circle (1.1pt);
 \fill (B2) circle (1.1pt);
 \fill (B3) circle (1.1pt);
\end{tikzpicture} &
\begin{tikzpicture}[x=1.1cm,y=.75cm,baseline=(current bounding box.center)]
 \path[use as bounding box] (-1.1,-1.15) rectangle (1.1,1.15);
 \coordinate (U0) at (-0.95,1.0);
 \coordinate (U2) at (0.0,1.0);
 \coordinate (U1) at (0.95,1.0);
 \coordinate (L1) at (-0.95,-1.0);
 \coordinate (L2) at (-0.3167,-1.0);
 \coordinate (L0) at (0.3167,-1.0);
 \coordinate (L3) at (0.95,-1.0);
 \coordinate (B0) at (0.7485,0.0032);
 \coordinate (B1) at (-0.0288,0.38);
 \coordinate (B2) at (0.3455,-0.3431);
 \coordinate (B3) at (-0.4318,0.0398);
 \draw[line width=.6pt] (B3)--(U0); 
 \draw[line width=.6pt] (B2)--(L0); 
 \draw[line width=.6pt] (B1)--(B2); 
 \draw[line width=.6pt] (B2)--(B0); 
 \draw[line width=.6pt] (B3)--(B1); 
 \draw[line width=.6pt] (L1)--(U0); 
 \draw[line width=.6pt] (U1)--(B0); 
 \draw[line width=.6pt] (L2)--(B3); 
 \draw[line width=.6pt] (U2)--(B1); 
 \draw[line width=.6pt] (B0)--(L3); 
 \fill (U0) circle (1.9pt);
 \fill (U2) circle (1.9pt);
 \fill (U1) circle (1.9pt);
 \fill (L1) circle (1.9pt);
 \fill (L2) circle (1.9pt);
 \fill (L0) circle (1.9pt);
 \fill (L3) circle (1.9pt);
 \fill (B0) circle (1.1pt);
 \fill (B1) circle (1.1pt);
 \fill (B2) circle (1.1pt);
 \fill (B3) circle (1.1pt);
\end{tikzpicture} &
\begin{tikzpicture}[x=1.1cm,y=.75cm,baseline=(current bounding box.center)]
 \path[use as bounding box] (-1.1,-1.15) rectangle (1.1,1.15);
 \coordinate (U3) at (-0.95,1.0);
 \coordinate (U2) at (-0.3167,1.0);
 \coordinate (U1) at (0.3167,1.0);
 \coordinate (U0) at (0.95,1.0);
 \coordinate (L0) at (-0.95,-1.0);
 \coordinate (L1) at (0.0,-1.0);
 \coordinate (L2) at (0.95,-1.0);
 \coordinate (B0) at (-0.05,0.35);
 \coordinate (B1) at (0.35,-0.3);
 \coordinate (B2) at (0.72,0.25);
 \coordinate (B3) at (-0.5,-0.1);
 \draw[line width=.6pt] (L0)--(B3); 
 \draw[line width=.6pt] (U0)--(B2); 
 \draw[line width=.6pt] (B1)--(L1); 
 \draw[line width=.6pt] (B2)--(B0); 
 \draw[line width=.6pt] (L2)--(B2); 
 \draw[line width=.6pt] (B0)--(B3); 
 \draw[line width=.6pt] (B3)--(U2); 
 \draw[line width=.6pt] (B0)--(B1); 
 \draw[line width=.6pt] (U3)--(L0); 
 \draw[white,line width=2.8pt] (B1)--(U1); 
 \draw[line width=.6pt] (B1)--(U1); 
 \fill (U3) circle (1.9pt);
 \fill (U2) circle (1.9pt);
 \fill (U1) circle (1.9pt);
 \fill (U0) circle (1.9pt);
 \fill (L0) circle (1.9pt);
 \fill (L1) circle (1.9pt);
 \fill (L2) circle (1.9pt);
 \fill (B0) circle (1.1pt);
 \fill (B1) circle (1.1pt);
 \fill (B2) circle (1.1pt);
 \fill (B3) circle (1.1pt);
\end{tikzpicture} &
\begin{tikzpicture}[x=1.1cm,y=.75cm,baseline=(current bounding box.center)]
 \path[use as bounding box] (-1.1,-1.15) rectangle (1.1,1.15);
 \coordinate (U0) at (-0.95,1.0);
 \coordinate (U1) at (0.0,1.0);
 \coordinate (U2) at (0.95,1.0);
 \coordinate (L2) at (-0.95,-1.0);
 \coordinate (L1) at (-0.3167,-1.0);
 \coordinate (L3) at (0.3167,-1.0);
 \coordinate (L0) at (0.95,-1.0);
 \coordinate (B0) at (0.1036,-0.7137);
 \coordinate (B1) at (0.2764,0.7681);
 \coordinate (B2) at (0.3109,-0.3705);
 \coordinate (B3) at (-0.1209,0.5302);
 \draw[line width=.6pt] (B3)--(U0); 
 \draw[line width=.6pt] (B2)--(B3); 
 \draw[line width=.6pt] (L0)--(B2); 
 \draw[line width=.6pt] (U1)--(B1); 
 \draw[line width=.6pt] (B1)--(B3); 
 \draw[line width=.6pt] (L1)--(B0); 
 \draw[line width=.6pt] (U2)--(B1); 
 \draw[line width=.6pt] (B0)--(B2); 
 \draw[line width=.6pt] (U0)--(L2); 
 \draw[line width=.6pt] (B0)--(L3); 
 \fill (U0) circle (1.9pt);
 \fill (U1) circle (1.9pt);
 \fill (U2) circle (1.9pt);
 \fill (L2) circle (1.9pt);
 \fill (L1) circle (1.9pt);
 \fill (L3) circle (1.9pt);
 \fill (L0) circle (1.9pt);
 \fill (B0) circle (1.1pt);
 \fill (B1) circle (1.1pt);
 \fill (B2) circle (1.1pt);
 \fill (B3) circle (1.1pt);
\end{tikzpicture}\\[34pt]
\begin{tikzpicture}[x=1.1cm,y=.75cm,baseline=(current bounding box.center)]
 \path[use as bounding box] (-1.1,-1.15) rectangle (1.1,1.15);
 \coordinate (U0) at (-0.95,1.0);
 \coordinate (U1) at (0.0,1.0);
 \coordinate (U2) at (0.95,1.0);
 \coordinate (L1) at (-0.95,-1.0);
 \coordinate (L0) at (-0.3167,-1.0);
 \coordinate (L3) at (0.3167,-1.0);
 \coordinate (L2) at (0.95,-1.0);
 \coordinate (B0) at (0.5542,-0.6362);
 \coordinate (B1) at (0.3958,-0.159);
 \coordinate (B2) at (-0.5146,0.4199);
 \coordinate (B3) at (-0.5938,-0.4358);
 \draw[line width=.6pt] (L0)--(B3); 
 \draw[line width=.6pt] (B3)--(B2); 
 \draw[line width=.6pt] (B2)--(U0); 
 \draw[line width=.6pt] (B1)--(B0); 
 \draw[line width=.6pt] (B3)--(L1); 
 \draw[line width=.6pt] (B0)--(L2); 
 \draw[line width=.6pt] (L3)--(B0); 
 \draw[line width=.6pt] (B2)--(U1); 
 \draw[line width=.6pt] (U2)--(B1); 
 \draw[line width=.6pt] (L0)--(B1); 
 \fill (U0) circle (1.9pt);
 \fill (U1) circle (1.9pt);
 \fill (U2) circle (1.9pt);
 \fill (L1) circle (1.9pt);
 \fill (L0) circle (1.9pt);
 \fill (L3) circle (1.9pt);
 \fill (L2) circle (1.9pt);
 \fill (B0) circle (1.1pt);
 \fill (B1) circle (1.1pt);
 \fill (B2) circle (1.1pt);
 \fill (B3) circle (1.1pt);
\end{tikzpicture} &
\begin{tikzpicture}[x=1.1cm,y=.75cm,baseline=(current bounding box.center)]
 \path[use as bounding box] (-1.1,-1.15) rectangle (1.1,1.15);
 \coordinate (U2) at (-0.95,1.0);
 \coordinate (U1) at (0.0,1.0);
 \coordinate (U0) at (0.95,1.0);
 \coordinate (L3) at (-0.95,-1.0);
 \coordinate (L2) at (-0.3167,-1.0);
 \coordinate (L1) at (0.3167,-1.0);
 \coordinate (L0) at (0.95,-1.0);
 \coordinate (B0) at (-0.6861,0.0491);
 \coordinate (B1) at (0.2639,0.6061);
 \coordinate (B2) at (-0.1583,0.0369);
 \coordinate (B3) at (-0.0528,-0.5635);
 \draw[line width=.6pt] (L0)--(U0); 
 \draw[line width=.6pt] (B3)--(L1); 
 \draw[line width=.6pt] (B2)--(B3); 
 \draw[line width=.6pt] (U0)--(B1); 
 \draw[line width=.6pt] (B3)--(L2); 
 \draw[line width=.6pt] (B1)--(U1); 
 \draw[line width=.6pt] (B2)--(B0); 
 \draw[line width=.6pt] (B0)--(L3); 
 \draw[line width=.6pt] (B1)--(B2); 
 \draw[line width=.6pt] (B0)--(U2); 
 \fill (U2) circle (1.9pt);
 \fill (U1) circle (1.9pt);
 \fill (U0) circle (1.9pt);
 \fill (L3) circle (1.9pt);
 \fill (L2) circle (1.9pt);
 \fill (L1) circle (1.9pt);
 \fill (L0) circle (1.9pt);
 \fill (B0) circle (1.1pt);
 \fill (B1) circle (1.1pt);
 \fill (B2) circle (1.1pt);
 \fill (B3) circle (1.1pt);
\end{tikzpicture} &
\begin{tikzpicture}[x=1.1cm,y=.75cm,baseline=(current bounding box.center)]
 \path[use as bounding box] (-1.1,-1.15) rectangle (1.1,1.15);
 \coordinate (U2) at (-0.95,1.0);
 \coordinate (U1) at (0.0,1.0);
 \coordinate (U0) at (0.95,1.0);
 \coordinate (L3) at (-0.95,-1.0);
 \coordinate (L1) at (-0.3167,-1.0);
 \coordinate (L2) at (0.3167,-1.0);
 \coordinate (L0) at (0.95,-1.0);
 \coordinate (B0) at (-0.4952,0.1495);
 \coordinate (B1) at (0.3339,0.5529);
 \coordinate (B2) at (0.0518,-0.1792);
 \coordinate (B3) at (-0.5873,-0.5181);
 \draw[line width=.6pt] (L0)--(U0); 
 \draw[line width=.6pt] (B3)--(L1); 
 \draw[line width=.6pt] (B2)--(L2); 
 \draw[line width=.6pt] (U0)--(B1); 
 \draw[line width=.6pt] (B1)--(U1); 
 \draw[line width=.6pt] (B2)--(B0); 
 \draw[line width=.6pt] (B0)--(B3); 
 \draw[line width=.6pt] (B1)--(B2); 
 \draw[line width=.6pt] (L3)--(B3); 
 \draw[line width=.6pt] (B0)--(U2); 
 \fill (U2) circle (1.9pt);
 \fill (U1) circle (1.9pt);
 \fill (U0) circle (1.9pt);
 \fill (L3) circle (1.9pt);
 \fill (L1) circle (1.9pt);
 \fill (L2) circle (1.9pt);
 \fill (L0) circle (1.9pt);
 \fill (B0) circle (1.1pt);
 \fill (B1) circle (1.1pt);
 \fill (B2) circle (1.1pt);
 \fill (B3) circle (1.1pt);
\end{tikzpicture} &
\begin{tikzpicture}[x=1.1cm,y=.75cm,baseline=(current bounding box.center)]
 \path[use as bounding box] (-1.1,-1.15) rectangle (1.1,1.15);
 \coordinate (U0) at (-0.95,1.0);
 \coordinate (U3) at (-0.3167,1.0);
 \coordinate (U1) at (0.3167,1.0);
 \coordinate (U2) at (0.95,1.0);
 \coordinate (L0) at (-0.95,-1.0);
 \coordinate (L2) at (0.0,-1.0);
 \coordinate (L1) at (0.95,-1.0);
 \coordinate (B0) at (-0.3562,-0.095);
 \coordinate (B1) at (0.3562,-0.1152);
 \coordinate (B2) at (0.7521,0.0819);
 \coordinate (B3) at (-0.7521,0.0757);
 \draw[line width=.6pt] (L0)--(B3); 
 \draw[line width=.6pt] (U0)--(B3); 
 \draw[line width=.6pt] (B2)--(L1); 
 \draw[line width=.6pt] (L2)--(B1); 
 \draw[line width=.6pt] (B3)--(B0); 
 \draw[line width=.6pt] (B1)--(U1); 
 \draw[line width=.6pt] (B2)--(U2); 
 \draw[line width=.6pt] (U3)--(B0); 
 \draw[line width=.6pt] (B1)--(B2); 
 \draw[line width=.6pt] (B0)--(L2); 
 \fill (U0) circle (1.9pt);
 \fill (U3) circle (1.9pt);
 \fill (U1) circle (1.9pt);
 \fill (U2) circle (1.9pt);
 \fill (L0) circle (1.9pt);
 \fill (L2) circle (1.9pt);
 \fill (L1) circle (1.9pt);
 \fill (B0) circle (1.1pt);
 \fill (B1) circle (1.1pt);
 \fill (B2) circle (1.1pt);
 \fill (B3) circle (1.1pt);
\end{tikzpicture}\\[34pt]
\begin{tikzpicture}[x=1.1cm,y=.75cm,baseline=(current bounding box.center)]
 \path[use as bounding box] (-1.1,-1.15) rectangle (1.1,1.15);
 \coordinate (U2) at (-0.95,1.0);
 \coordinate (U0) at (0.0,1.0);
 \coordinate (U1) at (0.95,1.0);
 \coordinate (L2) at (-0.95,-1.0);
 \coordinate (L0) at (-0.3167,-1.0);
 \coordinate (L3) at (0.3167,-1.0);
 \coordinate (L1) at (0.95,-1.0);
 \coordinate (B0) at (-0.7521,-0.1189);
 \coordinate (B1) at (-0.3562,0.1118);
 \coordinate (B2) at (0.5146,0.4157);
 \coordinate (B3) at (0.5938,-0.5716);
 \draw[line width=.6pt] (B3)--(B2); 
 \draw[line width=.6pt] (B2)--(U0); 
 \draw[line width=.6pt] (B2)--(U1); 
 \draw[line width=.6pt] (B1)--(L0); 
 \draw[line width=.6pt] (B3)--(L1); 
 \draw[line width=.6pt] (B1)--(B0); 
 \draw[line width=.6pt] (B0)--(L2); 
 \draw[line width=.6pt] (L3)--(B3); 
 \draw[line width=.6pt] (U0)--(B1); 
 \draw[line width=.6pt] (B0)--(U2); 
 \fill (U2) circle (1.9pt);
 \fill (U0) circle (1.9pt);
 \fill (U1) circle (1.9pt);
 \fill (L2) circle (1.9pt);
 \fill (L0) circle (1.9pt);
 \fill (L3) circle (1.9pt);
 \fill (L1) circle (1.9pt);
 \fill (B0) circle (1.1pt);
 \fill (B1) circle (1.1pt);
 \fill (B2) circle (1.1pt);
 \fill (B3) circle (1.1pt);
\end{tikzpicture} &
\begin{tikzpicture}[x=1.1cm,y=.75cm,baseline=(current bounding box.center)]
 \path[use as bounding box] (-1.1,-1.15) rectangle (1.1,1.15);
 \coordinate (U2) at (-0.95,1.0);
 \coordinate (U1) at (0.0,1.0);
 \coordinate (U0) at (0.95,1.0);
 \coordinate (L2) at (-0.95,-1.0);
 \coordinate (L3) at (-0.3167,-1.0);
 \coordinate (L0) at (0.3167,-1.0);
 \coordinate (L1) at (0.95,-1.0);
 \coordinate (B0) at (-0.5938,-0.4161);
 \coordinate (B1) at (-0.5146,0.5213);
 \coordinate (B2) at (0.5146,0.5032);
 \coordinate (B3) at (0.5938,-0.3895);
 \draw[line width=.6pt] (L0)--(B3); 
 \draw[line width=.6pt] (B3)--(B2); 
 \draw[line width=.6pt] (B2)--(U0); 
 \draw[line width=.6pt] (U1)--(B1); 
 \draw[line width=.6pt] (B3)--(L1); 
 \draw[line width=.6pt] (L2)--(B0); 
 \draw[line width=.6pt] (B2)--(U1); 
 \draw[line width=.6pt] (B1)--(B0); 
 \draw[line width=.6pt] (B1)--(U2); 
 \draw[line width=.6pt] (B0)--(L3); 
 \fill (U2) circle (1.9pt);
 \fill (U1) circle (1.9pt);
 \fill (U0) circle (1.9pt);
 \fill (L2) circle (1.9pt);
 \fill (L3) circle (1.9pt);
 \fill (L0) circle (1.9pt);
 \fill (L1) circle (1.9pt);
 \fill (B0) circle (1.1pt);
 \fill (B1) circle (1.1pt);
 \fill (B2) circle (1.1pt);
 \fill (B3) circle (1.1pt);
\end{tikzpicture}
\end{tabular}
\caption{Independent topologies contributing to the static 5PN calculation (five loops).}
\label{fig:topologies-5PN}
\end{figure}
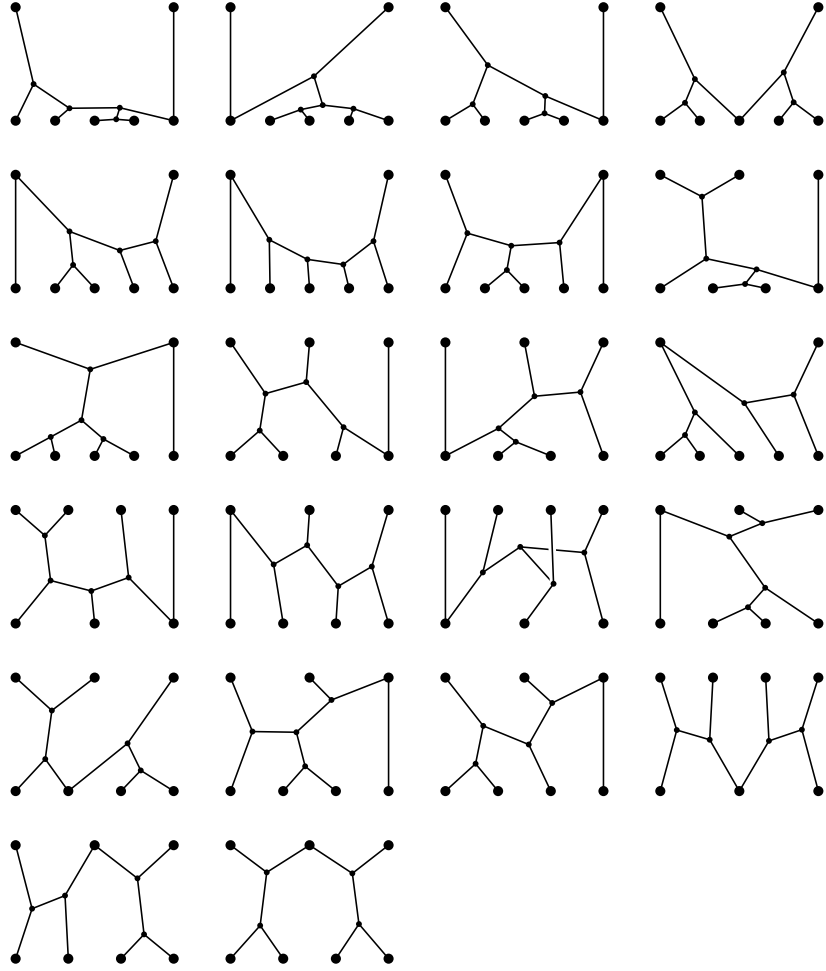

\leavevmode{The twenty-two independent 5PN top topologies in Fig.~\ref{fig:topologies-5PN} each have ten propagators. Their integrals are represented as sectors of a common five-loop family}
\begin{equation}
\begin{aligned}
 D_{1}&=\bm \ell_1^2, &\quad D_{2}&=\bm \ell_2^2,\\
 D_{3}&=\bm \ell_3^2, &\quad D_{4}&=\bm \ell_4^2,\\
 D_{5}&=\bm \ell_5^2, &\quad D_{6}&=(\bm \ell_1-\bm \ell_2)^2,\\
 D_{7}&=(\bm \ell_1-\bm \ell_3)^2, &\quad D_{8}&=(\bm \ell_1+\bm \ell_4)^2,\\
 D_{9}&=(\bm \ell_2-\bm \ell_3)^2, &\quad D_{10}&=(\bm \ell_2+\bm \ell_4)^2,\\
 D_{11}&=(\bm \ell_3-\bm \ell_5)^2, &\quad D_{12}&=(\bm \ell_4+\bm \ell_5)^2,\\
 D_{13}&=(\bm \ell_3-\bm q)^2, &\quad D_{14}&=(\bm \ell_1+\bm \ell_3-\bm \ell_5)^2,\\
 D_{15}&=(\bm \ell_1-\bm \ell_3+\bm \ell_4+\bm \ell_5)^2, &\quad D_{16}&=(\bm \ell_1-\bm \ell_2-\bm \ell_3+\bm q)^2,\\
 D_{17}&=(\bm \ell_2+\bm \ell_3+\bm \ell_4-\bm q)^2, &\quad D_{18}&=(-\bm \ell_3+\bm \ell_4+\bm \ell_5+\bm q)^2,\\
 D_{19}&=\bm \ell_2\cdot \bm \ell_5, &\quad D_{20}&=\bm \ell_1\cdot \bm q.
\end{aligned}
 \label{eq:family-5PN}
\end{equation}
\begingroup\clubpenalty=10000\widowpenalty=10000
\leavevmode{For each topology, ten of the eighteen quadratic forms $D_1,\ldots,D_{18}$ serve as propagator denominators. The master basis is}\par\endgroup\nopagebreak
\begin{equation}
\begin{aligned}
 \mathcal M^{(5)}_1&={M}^{(5)}_{0,0,0,0,0,0,0,0,0,1,1,1,0,1,1,0,1,0,0,0},\\
 \mathcal M^{(5)}_2&={M}^{(5)}_{0,0,0,0,0,0,1,1,0,1,1,1,0,1,0,0,1,0,0,0},\\
 \mathcal M^{(5)}_3&={M}^{(5)}_{0,0,0,0,0,1,0,0,0,0,1,1,1,1,1,0,1,0,0,0},\\
 \mathcal M^{(5)}_4&={M}^{(5)}_{0,1,0,0,0,1,1,0,0,0,1,0,1,1,1,0,0,1,0,0},\\
 \mathcal M^{(5)}_5&={M}^{(5)}_{0,1,0,0,1,1,0,0,0,0,1,1,1,0,1,0,0,0,0,0},\\
 \mathcal M^{(5)}_6&={M}^{(5)}_{1,0,0,0,0,1,0,0,0,0,0,1,1,1,1,0,1,0,0,0},\\
 \mathcal M^{(5)}_7&={M}^{(5)}_{1,0,0,0,0,1,0,0,0,1,1,1,1,1,1,0,1,0,0,0},\\
 \mathcal M^{(5)}_8&={M}^{(5)}_{1,0,0,0,0,1,0,1,0,0,0,1,1,1,1,0,1,0,0,0}.
\end{aligned}
 \label{eq:masters-5PN}
\end{equation}
\begingroup\interlinepenalty=10000
\leavevmode{As explained in Sec.~\ref{sec:factorization}, odd-PN factorization reduces the required five-loop input to lower-order building blocks, even though the momentum-space representation still carries five integrations. No new prime five-loop master integral is therefore needed.}\par\endgroup

\section{Static Feynman rules}
\label{app:feynman-rules}

This appendix collects the Feynman rules used in the diagram calculation through 5PN.
The rules below follow from Eqs.~\eqref{eq:static-bulk}
and~\eqref{eq:static-worldline}, with the conventions of
Sec.~\ref{sec:Feynman-rules}. The momentum-conservation delta
functions at bulk vertices are understood. Dashed, solid, and
double-wavy lines denote $\Phi$, $\phi$, and $\sigma_{ij}$,
respectively.

With the field normalization introduced
above, the propagators therefore read
	\begin{align}	
		\braket{\sigma_{ij} (\K) \sigma_{kl} (-\K)} = 
		\begin{gathered}
		\begin{tikzpicture}
			\begin{feynman}
			\vertex (a) {};
			\vertex[left=0.4 of a] (a1) ; 
			\vertex[right=1.8cm of a] (b) {};
			\vertex[right=0.4 of b] (b1) ; 
			\diagram*{
			(a) -- [line width=0.7mm, photon] (b),
			(a) -- [line width=0.25mm,photon,white] (b),						
			};
			\end{feynman}
		\end{tikzpicture}
		\end{gathered}	
		&= -\frac{i}{2\K^2}  \left( \delta_{ik} \delta_{jl} + \delta_{il} \delta_{jk} -\frac{2}{d-2} \delta_{ij} \delta_{kl} \right),
	\\
		\braket{\Phi (\K) \Phi (-\K)} = 	
		\begin{gathered}
		\begin{tikzpicture}
			\begin{feynman}
			\vertex (a) {};
			\vertex[right=1.8cm of a] (b) {};
			\diagram*{
			(a) -- [thick,dashed] (b),
			};
			\end{feynman}
		\end{tikzpicture}
		\end{gathered}	
		&= -\frac{i}{2c_d \K^2} ,	
	\\
		\braket{\phi (\K) \phi (-\K)} = 
		\begin{gathered}
		\begin{tikzpicture}
			\begin{feynman}
			\vertex (a) {};
			\vertex[right=1.8cm of a] (b) {};
			\diagram*{
			(a) -- [thick] (b),
			};
			\end{feynman}
		\end{tikzpicture}
		\end{gathered}	
		&= -\frac{i}{8\K^2}. 											
	\end{align}	

To obtain the vertex factor, we differentiate the interaction
term with respect to its $p$ fields $\Phi$ and $q$ fields $\phi$.
This produces a factor $p!\,q!$, canceling the denominator in
Eq.~\eqref{eq:worldline-expansion}, and multiplication by $i$
from the expansion of $e^{iS}$ then yields the rule below.
Any symmetry factor associated with a complete diagram is
accounted for separately when assembling its contribution.
		\begin{align}
		\begin{gathered}
		\begin{tikzpicture}
			\node[draw,circle,white] (A) at (0,-1) {};
			\begin{feynman}
			\vertex (a) {};
			\vertex[above=0.07 of a] (a0) {};
			\vertex[dot,below=1cm of a] (b) {};
			\vertex[right=1.31cm of b] (c) {};
			\vertex[left=1.31cm of b] (d) {};						
			\vertex[below=0.35cm of b] (e) {$m_A$};
			\vertex[right=0.6cm of a] (a2) {};
			\vertex[right=1cm of a] (a3) {};			
			\diagram*{
			(c) -- [thick] (d) ,
			};
			\end{feynman}
			\draw[thick, rotate around={70:(A.center)}] (A.center) -- ++(1.2,0);	
			\draw[thick, rotate around={55:(A.center)}] (A.center) -- ++(1.2,0);	
			\draw[thick, rotate around={20:(A.center)}] (A.center) -- ++(1.2,0);
			\draw[thick, dashed, rotate around={160:(A.center)}] (A.center) -- ++(1.2,0);	
			\draw[thick, dashed, rotate around={145:(A.center)}] (A.center) -- ++(1.2,0);	
			\draw[thick, dashed, rotate around={110:(A.center)}] (A.center) -- ++(1.2,0);
			\node[thick,rotate=-45] (m) at (0.57,-0.56) {...};											
			\node[thick,rotate=45] (m) at (-0.43,-0.42) {...};														
		\end{tikzpicture}
		\end{gathered}		
			&= - i \kappa^{p+q} m_A a_{A,q}  , \qquad \qquad 1 \le p+q \le 6.
		\end{align}

Starting with the cubic interactions, expanding the scalar
kinetic terms to first order in $\sigma_{ij}$ gives the
$\sigma\Phi^2$ and $\sigma\phi^2$ vertices, while the cubic
expansion of the Einstein--Hilbert and gauge-fixing terms
supplies the $\sigma^3$ vertex. The corresponding
momentum-space rules are
	\begin{align}
		\begin{gathered}
		\begin{tikzpicture}
			\begin{feynman}
			\vertex[dot] (a) {};
			\vertex[above=1.15cm of a] (b) {};
			\vertex[below left=1.4cm of a] (c) {};
			\vertex[below right=1.4cm of a] (d) {};	
			\vertex[above=1.25cm of a] (b1) {$i,j$};								
			\diagram*{
			(a) -- [momentum=\(\K_3\)] (b),
			(a) -- [line width=0.7mm,photon,] (b),
			(a) -- [line width=0.25mm,photon,white] (b),			
			(a) -- [thick,dashed,momentum'=\(\K_1\)] (c),
			(a) -- [thick,dashed,momentum=\(\K_2\)] (d),						
			};
			\end{feynman}
		\end{tikzpicture}
		\end{gathered}	
		&= - i \kappa c_d \left( k_1^i k_2^j + k_1^j k_2^i - \delta^{ij} \K_1\cdot \K_2 \right),
		\\
		\begin{gathered}
		\begin{tikzpicture}
			\begin{feynman}
			\vertex[dot] (a) {};
			\vertex[above=1.15cm of a] (b) {};
			\vertex[below left=1.4cm of a] (c) {};
			\vertex[below right=1.4cm of a] (d) {};	
			\vertex[above=1.25cm of a] (b1) {$i,j$};								
			\diagram*{
			(a) -- [momentum=\(\K_3\)] (b),
			(a) -- [line width=0.7mm,photon,] (b),
			(a) -- [line width=0.25mm,photon,white] (b),			
			(a) -- [thick,momentum'=\(\K_1\)] (c),
			(a) -- [thick,momentum=\(\K_2\)] (d),						
			};
			\end{feynman}
		\end{tikzpicture}
		\end{gathered}	
		&= - 4 i \kappa \left( k_1^i k_2^j + k_1^j k_2^i - \delta^{ij} \K_1\cdot \K_2 \right),
		\\
		\begin{gathered}
		\begin{tikzpicture}
			\begin{feynman}
			\vertex[dot] (a) {};
			\vertex[above=1.15cm of a] (b) {};
			\vertex[below left=1.15cm of a] (c) {};
			\vertex[below right=1.15cm of a] (d) {};	
			\vertex[above=1.25cm of a] (b1) {$m,n$};
			\vertex[below left=1.35cm of a] (c1) {$i,j$};
			\vertex[below right=1.35cm of a] (d1) {$k,l$};														
			\diagram*{
			(a) -- [momentum=\(\K_3\)] (b),
			(a) -- [line width=0.7mm,photon] (b),
			(a) -- [line width=0.25mm,photon,white] (b),
			(a) -- [momentum'=\(\K_1\)] (c),
			(a) -- [line width=0.7mm,photon] (c),
			(a) -- [line width=0.25mm,photon,white] (c),			
			(a) -- [momentum=\(\K_2\)] (d),						
			(a) -- [line width=0.7mm,photon] (d),						
			(a) -- [line width=0.25mm,photon,white] (d),									
			};
			\end{feynman}
		\end{tikzpicture}
		\end{gathered}	
		&=  
		\begin{gathered}
		- i \kappa \Big[
        k_{2a}k_{3b}\mathcal T^{(3)}_{(ij)(kl)(mn)ab}
        +k_{3a}k_{2b}\mathcal T^{(3)}_{(ij)(mn)(kl)ab}\\
        +k_{1a}k_{3b}\mathcal T^{(3)}_{(kl)(ij)(mn)ab}
        +k_{3a}k_{1b}\mathcal T^{(3)}_{(kl)(mn)(ij)ab}\\
        +k_{1a}k_{2b}\mathcal T^{(3)}_{(mn)(ij)(kl)ab}
        +k_{2a}k_{1b}\mathcal T^{(3)}_{(mn)(kl)(ij)ab}
        \Big]. 
		\end{gathered}
\end{align}	
Here the tensor $\mathcal T^{(3)}_{ijklmnab}$ collects the
Einstein--Hilbert and harmonic gauge-fixing contributions
to the three-$\sigma$ vertex and is given by
		\begin{align}
			\mathcal T^{(3)}_{ijklmnab} &= \frac18 \delta^{ij} \delta^{kl} \delta^{mn} \delta^{ab}  -  \frac14 \delta^{ia} \delta^{jb} \delta^{kl} \delta^{mn}   
			+ \frac12 \delta^{ia} \delta^{jb} \delta^{km} \delta^{ln} -\frac14 \delta^{ij} \delta^{ab} \delta^{km} \delta^{ln} \nonumber\\
			&   + \frac12 \delta^{ij} \delta^{km} \delta^{an} \delta^{bl}  + \delta^{ab} \delta^{im} \delta^{jk} \delta^{ln}   
			- \delta^{an} \delta^{bl} \delta^{im} \delta^{jk} + \delta^{ik} \delta^{jm} \delta^{al} \delta^{bn}  \nonumber \\
			&  - \frac12 \delta^{ab} \delta^{im} \delta^{jn} \delta^{kl}  -2  \delta^{ja} \delta^{bl} \delta^{im} \delta^{kn}  - \frac12  \delta^{ak} \delta^{bn} \delta^{ij} \delta^{lm} +2 \delta^{ja} \delta^{bn} \delta^{ik} \delta^{lm} . 
		\label{eq:tensor-cubic}
\end{align}

Parentheses on an index pair in the cubic vertex denote
symmetrization with weight $1/2$, for example
\[
 A^{(ab)}\equiv\frac12\left(A^{ab}+A^{ba}\right).
\]
Distinct parenthesized pairs are symmetrized independently.
All momenta are incoming, with $\bm k_1+\bm k_2+\bm k_3=0$.
The six terms in the cubic rule assign the three legs to the undifferentiated field and the two differentiated fields in all possible ways; the corresponding index pairs must be permuted together with their momenta.

Proceeding to the quartic interactions, the two remaining vertices follow by expanding $\sqrt{\gamma}\gamma^{ij}$ to second order in $\sigma_{ij}$ in the scalar kinetic terms.
Since these terms differ only in their overall coefficients, the $\sigma^2\Phi^2$ and $\sigma^2\phi^2$ vertices share a common momentum and tensor structure, which we denote by $\mathcal T^{(4)}_{ijkl}$. Their rules therefore take the form
	\begin{align}
		\begin{gathered}
		\begin{tikzpicture}
			\begin{feynman}
			\vertex[dot] (a) {};
			\vertex[above left=1.4cm of a] (b) {};
			\vertex[above right=1.4cm of a] (b2) {};			
			\vertex[below left=1.4cm of a] (c) {};
			\vertex[below right=1.4cm of a] (d) {};	
			\vertex[above=0.15cm of b] (b1) {$i,j$};	
			\vertex[above=0.15cm of b2] (b1) {$k,l$};														
			\diagram*{
			(a) -- [momentum'={[arrow shorten=0.1mm,arrow distance=-0.15cm]}] (b),
			(a) -- [momentum={[arrow shorten=0.1mm,arrow distance=-0.15cm]}] (b2),										
			(a) -- [edge label={\(\K_3\)},near end] (b) ,			
			(a) -- [line width=0.7mm,photon] (b),			
			(a) -- [line width=0.25mm,photon,white] (b),			
			(a) -- [edge label'={\(\K_4\)},near end] (b2),			
			(a) -- [line width=0.7mm,photon] (b2),			
			(a) -- [line width=0.25mm,photon,white] (b2),			
			(a) -- [thick,dashed,edge label'={\(\K_1\)},near end,momentum={[arrow shorten=0.1mm,arrow distance=-0.15cm]}] (c),
			(a) -- [thick,dashed,edge label={\(\K_2\)},near end,momentum'={[arrow shorten=0.1mm,arrow distance=-0.15cm]}] (d),				
			};
			\end{feynman}
		\end{tikzpicture}
		\end{gathered}	
		& = 
		\begin{gathered}
		i \kappa^2 c_d \mathcal T^{(4)}_{ijkl}(\K_1,\K_2),
		\end{gathered}
	\\
%
		\begin{gathered}
		\begin{tikzpicture}
			\begin{feynman}
			\vertex[dot] (a) {};
			\vertex[above left=1.4cm of a] (b) {};
			\vertex[above right=1.4cm of a] (b2) {};			
			\vertex[below left=1.4cm of a] (c) {};
			\vertex[below right=1.4cm of a] (d) {};	
			\vertex[above=0.15cm of b] (b1) {$i,j$};	
			\vertex[above=0.15cm of b2] (b1) {$k,l$};														
			\diagram*{
			(a) -- [momentum'={[arrow shorten=0.1mm,arrow distance=-0.15cm]}] (b),
			(a) -- [momentum={[arrow shorten=0.1mm,arrow distance=-0.15cm]}] (b2),										
			(a) -- [edge label={\(\K_3\)},near end] (b) ,			
			(a) -- [line width=0.7mm,photon] (b),			
			(a) -- [line width=0.25mm,photon,white] (b),			
			(a) -- [edge label'={\(\K_4\)},near end] (b2),			
			(a) -- [line width=0.7mm,photon] (b2),			
			(a) -- [line width=0.25mm,photon,white] (b2),			
			(a) -- [thick,edge label'={\(\K_1\)},near end,momentum={[arrow shorten=0.1mm,arrow distance=-0.15cm]}] (c),
			(a) -- [thick,edge label={\(\K_2\)},near end,momentum'={[arrow shorten=0.1mm,arrow distance=-0.15cm]}] (d),				
			};
			\end{feynman}
		\end{tikzpicture}
		\end{gathered}	
		& = 
		\begin{gathered}
		4i \kappa^2 \mathcal T^{(4)}_{ijkl}(\K_1,\K_2),
		\end{gathered}
		\end{align}
where the common tensor, analogous to the cubic structure in Eq.~\eqref{eq:tensor-cubic}, is defined as
\begin{equation}
\begin{aligned}
 \mathcal T^{(4)}_{ijkl}(\K_1,\K_2)
 ={}&k_1^{(i}k_2^{k)}\delta^{jl}
 +k_1^{(i}k_2^{l)}\delta^{jk}
 +k_1^{(j}k_2^{k)}\delta^{il}
 +k_1^{(j}k_2^{l)}\delta^{ik}
 \\
 &-k_1^{(k}k_2^{l)}\delta^{ij}
 -k_1^{(i}k_2^{j)}\delta^{kl}
 -\frac12\bigl(\delta^{ik}\delta^{jl}
 +\delta^{jk}\delta^{il}-\delta^{ij}\delta^{kl}\bigr)
 \K_1\cdot\K_2.
\end{aligned}
\label{eq:tensor-quartic}
\end{equation}

\providecommand{\href}[2]{#2}\begingroup\raggedright\endgroup

\begin{thebibliography}{99}

\bibitem{LIGOScientific:2016aoc}
{\scshape LIGO Scientific, Virgo} collaboration, \emph{{Observation of
  Gravitational Waves from a Binary Black Hole Merger}},
  \href{https://doi.org/10.1103/PhysRevLett.116.061102}{\emph{Phys. Rev. Lett.}
  {\bfseries 116} (2016) 061102}
  [\href{https://arxiv.org/abs/1602.03837}{{\ttfamily 1602.03837}}].

\bibitem{LIGOScientific:2021djp}
{\scshape KAGRA, VIRGO, LIGO Scientific} collaboration, \emph{{GWTC-3: Compact
  Binary Coalescences Observed by LIGO and Virgo during the Second Part of the
  Third Observing Run}},
  \href{https://doi.org/10.1103/PhysRevX.13.041039}{\emph{Phys. Rev. X}
  {\bfseries 13} (2023) 041039}
  [\href{https://arxiv.org/abs/2111.03606}{{\ttfamily 2111.03606}}].

\bibitem{Maggiore:2019uih}
M.~Maggiore et~al., \emph{{Science Case for the Einstein Telescope}},
  \href{https://doi.org/10.1088/1475-7516/2020/03/050}{\emph{JCAP} {\bfseries
  03} (2020) 050} [\href{https://arxiv.org/abs/1912.02622}{{\ttfamily
  1912.02622}}].

\bibitem{LISA:2022kgy}
{\scshape LISA} collaboration, \emph{{New horizons for fundamental physics with
  LISA}}, \href{https://doi.org/10.1007/s41114-022-00036-9}{\emph{Living Rev.
  Rel.} {\bfseries 25} (2022) 4}
  [\href{https://arxiv.org/abs/2205.01597}{{\ttfamily 2205.01597}}].

\bibitem{Goldberger:2004jt}
W.D.~Goldberger and I.Z.~Rothstein, \emph{{An Effective field theory of gravity
  for extended objects}},
  \href{https://doi.org/10.1103/PhysRevD.73.104029}{\emph{Phys. Rev. D}
  {\bfseries 73} (2006) 104029}
  [\href{https://arxiv.org/abs/hep-th/0409156}{{\ttfamily hep-th/0409156}}].

\bibitem{Foffa:2016rgu}
S.~Foffa, P.~Mastrolia, R.~Sturani and C.~Sturm, \emph{{Effective field theory
  approach to the gravitational two-body dynamics, at fourth post-Newtonian
  order and quintic in the Newton constant}},
  \href{https://doi.org/10.1103/PhysRevD.95.104009}{\emph{Phys. Rev. D}
  {\bfseries 95} (2017) 104009}
  [\href{https://arxiv.org/abs/1612.00482}{{\ttfamily 1612.00482}}].

\bibitem{Foffa:2019hrb}
S.~Foffa, P.~Mastrolia, R.~Sturani, C.~Sturm and W.J.~Torres~Bobadilla,
  \emph{{Static two-body potential at fifth post-Newtonian order}},
  \href{https://doi.org/10.1103/PhysRevLett.122.241605}{\emph{Phys. Rev. Lett.}
  {\bfseries 122} (2019) 241605}
  [\href{https://arxiv.org/abs/1902.10571}{{\ttfamily 1902.10571}}].

\bibitem{Blumlein:2019zku}
J.~Bl{\"u}mlein, A.~Maier and P.~Marquard, \emph{{Five-Loop Static Contribution
  to the Gravitational Interaction Potential of Two Point Masses}},
  \href{https://doi.org/10.1016/j.physletb.2019.135100}{\emph{Phys. Lett. B}
  {\bfseries 800} (2020) 135100}
  [\href{https://arxiv.org/abs/1902.11180}{{\ttfamily 1902.11180}}].

\bibitem{Brunello:2025static}
G.~Brunello, M.K.~Mandal, P.~Mastrolia, R.~Patil, M.~Pegorin, J.~Ronca et~al.,
  \emph{{Six-Loop Gravitational Interactions at the Sixth Post-Newtonian
  Order}}, \href{https://doi.org/10.1103/c2r3-cqtf}{\emph{Phys. Rev. Lett.}
  {\bfseries 137} (2026) 111401}
  [\href{https://arxiv.org/abs/2512.19498}{{\ttfamily 2512.19498}}].

\bibitem{Brunello:2026static}
\leavevmode{G.~Brunello, M.K.~Mandal, P.~Mastrolia, R.~Patil, M.~Pegorin, S.~Smith et~al.,
  \emph{{All-order structure of static gravitational interactions and the
  seventh post-Newtonian potential}},
  \href{https://doi.org/10.1103/kcxq-7zsk}{\emph{Phys. Rev. Lett.}
  (accepted 25 September 2026)}
  [\href{https://arxiv.org/abs/2604.14134}{{\ttfamily 2604.14134}}].}

\bibitem{Brans:1961sx}
C.~Brans and R.H.~Dicke, \emph{{Mach's principle and a relativistic theory of
  gravitation}}, \href{https://doi.org/10.1103/PhysRev.124.925}{\emph{Phys.
  Rev.} {\bfseries 124} (1961) 925}.

\bibitem{Damour:1992we}
T.~Damour and G.~Esposito-Farese, \emph{{Tensor multiscalar theories of
  gravitation}}, \href{https://doi.org/10.1088/0264-9381/9/9/015}{\emph{Class.
  Quant. Grav.} {\bfseries 9} (1992) 2093}.

\bibitem{Damour:1993hw}
T.~Damour and G.~Esposito-Farese, \emph{{Nonperturbative strong field effects
  in tensor - scalar theories of gravitation}},
  \href{https://doi.org/10.1103/PhysRevLett.70.2220}{\emph{Phys. Rev. Lett.}
  {\bfseries 70} (1993) 2220}.

\bibitem{Damour:1996ke}
T.~Damour and G.~Esposito-Farese, \emph{{Tensor - scalar gravity and binary
  pulsar experiments}},
  \href{https://doi.org/10.1103/PhysRevD.54.1474}{\emph{Phys. Rev. D}
  {\bfseries 54} (1996) 1474}
  [\href{https://arxiv.org/abs/gr-qc/9602056}{{\ttfamily gr-qc/9602056}}].

\bibitem{1975ApJEardley}
D.M.~Eardley, \emph{Observable effects of a scalar gravitational field in a
  binary pulsar.}, \href{https://doi.org/10.1086/181744}{\emph{Astrophys. J.
  Lett.} {\bfseries 196} (1975) L59}.

\bibitem{Damour:1996field}
\leavevmode{T.~Damour and G.~Esposito-Far\`ese, \emph{{Testing gravity to second
  post-Newtonian order: a field-theory approach}},
  \href{https://doi.org/10.1103/PhysRevD.53.5541}{\emph{Phys. Rev. D}
  {\bfseries 53} (1996) 5541--5578}
  [\href{https://arxiv.org/abs/gr-qc/9506063}{{\ttfamily gr-qc/9506063}}].}

\bibitem{Mirshekari:2013vb}
S.~Mirshekari and C.M.~Will, \emph{{Compact binary systems in scalar-tensor
  gravity: Equations of motion to 2.5 post-Newtonian order}},
  \href{https://doi.org/10.1103/PhysRevD.87.084070}{\emph{Phys. Rev. D}
  {\bfseries 87} (2013) 084070}
  [\href{https://arxiv.org/abs/1301.4680}{{\ttfamily 1301.4680}}].

\bibitem{Bernard:2018hta}
L.~Bernard, \emph{{Dynamics of compact binary systems in scalar-tensor
  theories: Equations of motion to the third post-Newtonian order}},
  \href{https://doi.org/10.1103/PhysRevD.98.044004}{\emph{Phys. Rev. D}
  {\bfseries 98} (2018) 044004}
  [\href{https://arxiv.org/abs/1802.10201}{{\ttfamily 1802.10201}}].

\bibitem{Bernard:2018ivi}
L.~Bernard, \emph{{Dynamics of compact binary systems in scalar-tensor
  theories: II. Center-of-mass and conserved quantities to 3PN order}},
  \href{https://doi.org/10.1103/PhysRevD.99.044047}{\emph{Phys. Rev. D}
  {\bfseries 99} (2019) 044047}
  [\href{https://arxiv.org/abs/1812.04169}{{\ttfamily 1812.04169}}].

\bibitem{Jain:2022nxs}
T.~Jain, P.~Rettegno, M.~Agathos, A.~Nagar and L.~Turco,
  \emph{{Effective-one-body Hamiltonian in scalar-tensor gravity at third
  post-Newtonian order}},
  \href{https://doi.org/10.1103/PhysRevD.107.084017}{\emph{Phys. Rev. D}
  {\bfseries 107} (2023) 084017}
  [\href{https://arxiv.org/abs/2211.15580}{{\ttfamily 2211.15580}}].

\bibitem{Julie:2022qux}
F.-L.~Juli\'e, V.~Baibhav, E.~Berti and A.~Buonanno, \emph{{Third
  post-Newtonian effective-one-body Hamiltonian in scalar-tensor and
  Einstein-scalar-Gauss-Bonnet gravity}},
  \href{https://doi.org/10.1103/PhysRevD.107.104044}{\emph{Phys. Rev. D}
  {\bfseries 107} (2023) 104044}
  [\href{https://arxiv.org/abs/2212.13802}{{\ttfamily 2212.13802}}].

\bibitem{Kuntz:2019zef}
A.~Kuntz, F.~Piazza and F.~Vernizzi, \emph{{Effective field theory for
  gravitational radiation in scalar-tensor gravity}},
  \href{https://doi.org/10.1088/1475-7516/2019/05/052}{\emph{JCAP} {\bfseries
  05} (2019) 052} [\href{https://arxiv.org/abs/1902.04941}{{\ttfamily
  1902.04941}}].

\bibitem{Almeida:2024uph}
G.L.~Almeida, \emph{{Binary dynamics to second post-Newtonian order in
  scalar-tensor and Einstein-scalar-Gauss-Bonnet gravity from effective field
  theory}}, \href{https://doi.org/10.1103/PhysRevD.109.084060}{\emph{Phys. Rev.
  D} {\bfseries 109} (2024) 084060}
  [\href{https://arxiv.org/abs/2402.13996}{{\ttfamily 2402.13996}}].

\bibitem{BertiJulie2019}
F.-L.~Juli\'e and E.~Berti, \emph{Post-newtonian dynamics and black hole
  thermodynamics in einstein-scalar-gauss-bonnet gravity},
  \href{https://doi.org/10.1103/PhysRevD.100.104061}{\emph{Phys. Rev. D}
  {\bfseries 100} (2019) 104061}
  [\href{https://arxiv.org/abs/1909.05258}{{\ttfamily 1909.05258}}].

\bibitem{Almeida:2024cqz}
G.L.~Almeida and S.-Y.~Zhou, \emph{{Post-Newtonian dynamics of spinning black
  hole binaries in Einstein-scalar-Gauss-Bonnet gravity}},
  \href{https://doi.org/10.1103/PhysRevD.110.124016}{\emph{Phys. Rev. D}
  {\bfseries 110} (2024) 124016}
  [\href{https://arxiv.org/abs/2408.14196}{{\ttfamily 2408.14196}}].

\bibitem{WilsonGerow:2025conservative}
\leavevmode{J.~Wilson-Gerow, \emph{{Conservative Dynamics of Relativistic Binaries
  Beyond Einstein Gravity}},
  \href{https://doi.org/10.1103/99k7-whlt}{\emph{Phys. Rev. D}
  {\bfseries 112} (2025) 044004}
  [\href{https://arxiv.org/abs/2503.02867}{{\ttfamily 2503.02867}}].}

\bibitem{Bernard:2026PMEFT}
\leavevmode{L.~Bernard, T.~Jain and S.~Mougiakakos, \emph{{Scattering angle at 3PM in
  scalar-tensor theories using the PM-EFT formalism}},
  \href{https://doi.org/10.1007/JHEP06(2026)252}{\emph{JHEP} {\bfseries
  06} (2026) 252}
  [\href{https://arxiv.org/abs/2601.17503}{{\ttfamily 2601.17503}}].}

\bibitem{Almeida:2026abr}
G.L.~Almeida, Y.~Du, Z.~Liu and H.~Wang, \emph{{Conservative binary dynamics to
  third post-Minkowskian order beyond General Relativity}},
  \href{https://arxiv.org/abs/2602.08876}{{\ttfamily 2602.08876}}.

\bibitem{Barausse:2012da}
E.~Barausse, C.~Palenzuela, M.~Ponce and L.~Lehner, \emph{{Neutron-star mergers
  in scalar-tensor theories of gravity}},
  \href{https://doi.org/10.1103/PhysRevD.87.081506}{\emph{Phys. Rev. D}
  {\bfseries 87} (2013) 081506}
  [\href{https://arxiv.org/abs/1212.5053}{{\ttfamily 1212.5053}}].

\bibitem{Shibata:2013pra}
M.~Shibata, K.~Taniguchi, H.~Okawa and A.~Buonanno, \emph{{Coalescence of
  binary neutron stars in a scalar-tensor theory of gravity}},
  \href{https://doi.org/10.1103/PhysRevD.89.084005}{\emph{Phys. Rev. D}
  {\bfseries 89} (2014) 084005}
  [\href{https://arxiv.org/abs/1310.0627}{{\ttfamily 1310.0627}}].

\bibitem{Ma:2023sok}
S.~Ma, V.~Varma, L.C.~Stein, F.~Foucart, M.D.~Duez, L.E.~Kidder et~al.,
  \emph{{Numerical simulations of black hole-neutron star mergers in
  scalar-tensor gravity}},
  \href{https://doi.org/10.1103/PhysRevD.107.124051}{\emph{Phys. Rev. D}
  {\bfseries 107} (2023) 124051}
  [\href{https://arxiv.org/abs/2304.11836}{{\ttfamily 2304.11836}}].

\bibitem{Kuan:2023hrh}
H.-J.~Kuan, K.~Van~Aelst, A.T.-L.~Lam and M.~Shibata, \emph{{Binary neutron
  star mergers in massive scalar-tensor theory: Quasiequilibrium states and
  dynamical enhancement of the scalarization}},
  \href{https://doi.org/10.1103/PhysRevD.108.064057}{\emph{Phys. Rev. D}
  {\bfseries 108} (2023) 064057}
  [\href{https://arxiv.org/abs/2309.01709}{{\ttfamily 2309.01709}}].

\bibitem{Lam:2024wpq}
A.T.-L.~Lam, H.-J.~Kuan, M.~Shibata, K.~Van~Aelst and K.~Kiuchi, \emph{{Binary
  neutron star mergers in massive scalar-tensor theory: Properties of
  postmerger remnants}},
  \href{https://doi.org/10.1103/PhysRevD.110.104018}{\emph{Phys. Rev. D}
  {\bfseries 110} (2024) 104018}
  [\href{https://arxiv.org/abs/2406.05211}{{\ttfamily 2406.05211}}].

\bibitem{Franciolini:2018uyq}
G.~Franciolini, L.~Hui, R.~Penco, L.~Santoni and E.~Trincherini,
  \emph{{Effective Field Theory of Black Hole Quasinormal Modes in
  Scalar-Tensor Theories}},
  \href{https://doi.org/10.1007/JHEP02(2019)127}{\emph{JHEP} {\bfseries 02}
  (2019) 127} [\href{https://arxiv.org/abs/1810.07706}{{\ttfamily
  1810.07706}}].

\bibitem{Nogueira:1991ex}
P.~Nogueira, \emph{{Automatic Feynman Graph Generation}},
  \href{https://doi.org/10.1006/jcph.1993.1074}{\emph{J. Comput. Phys.}
  {\bfseries 105} (1993) 279}.

\bibitem{Vermaseren:2000nd}
J.A.M.~Vermaseren, \emph{{New features of FORM}},
  \href{https://arxiv.org/abs/math-ph/0010025}{{\ttfamily math-ph/0010025}}.

\bibitem{Smirnov:2023FIRE65}
A.V.~Smirnov and M.~Zeng, \emph{{FIRE 6.5: Feynman integral reduction with new
  simplification library}},
  \href{https://doi.org/10.1016/j.cpc.2024.109261}{\emph{Comput. Phys. Commun.}
  {\bfseries 302} (2024) 109261}
  [\href{https://arxiv.org/abs/2311.02370}{{\ttfamily 2311.02370}}].

\bibitem{Bernard:2023eul}
L.~Bernard, E.~Dones and S.~Mougiakakos, \emph{{Tidal effects up to
  next-to-next-to-leading post-Newtonian order in massless scalar-tensor
  theories}}, \href{https://doi.org/10.1103/PhysRevD.109.044006}{\emph{Phys.
  Rev. D} {\bfseries 109} (2024) 044006}
  [\href{https://arxiv.org/abs/2310.19679}{{\ttfamily 2310.19679}}].

\bibitem{beneke1998}
M.~Beneke and V.A.~Smirnov, \emph{{Asymptotic expansion of Feynman integrals
  near threshold}},
  \href{https://doi.org/10.1016/S0550-3213(98)00138-2}{\emph{Nucl. Phys. B}
  {\bfseries 522} (1998) 321}
  [\href{https://arxiv.org/abs/hep-ph/9711391}{{\ttfamily hep-ph/9711391}}].

\bibitem{Kol:2007bc}
B.~Kol and M.~Smolkin, \emph{{Non-Relativistic Gravitation: From Newton to
  Einstein and Back}},
  \href{https://doi.org/10.1088/0264-9381/25/14/145011}{\emph{Class. Quant.
  Grav.} {\bfseries 25} (2008) 145011}
  [\href{https://arxiv.org/abs/0712.4116}{{\ttfamily 0712.4116}}].

\bibitem{Kol:2010si}
B.~Kol and M.~Smolkin, \emph{{Einstein's action and the harmonic gauge in terms
  of Newtonian fields}},
  \href{https://doi.org/10.1103/PhysRevD.85.044029}{\emph{Phys. Rev. D}
  {\bfseries 85} (2012) 044029}
  [\href{https://arxiv.org/abs/1009.1876}{{\ttfamily 1009.1876}}].

\bibitem{Almeida:2020mrg}
G.L.~Almeida, S.~Foffa and R.~Sturani, \emph{{Classical Gravitational
  Self-Energy from Double Copy}},
  \href{https://doi.org/10.1007/JHEP11(2020)165}{\emph{JHEP} {\bfseries 11}
  (2020) 165} [\href{https://arxiv.org/abs/2008.06195}{{\ttfamily
  2008.06195}}].

\bibitem{Tkachov:1981wb}
F.V.~Tkachov, \emph{{A theorem on analytical calculability of 4-loop
  renormalization group functions}},
  \href{https://doi.org/10.1016/0370-2693(81)90288-4}{\emph{Phys. Lett. B}
  {\bfseries 100} (1981) 65}.

\bibitem{Chetyrkin:1981qh}
K.G.~Chetyrkin and F.V.~Tkachov, \emph{{Integration by parts: The algorithm to
  calculate $\beta$-functions in 4 loops}},
  \href{https://doi.org/10.1016/0550-3213(81)90199-1}{\emph{Nucl. Phys. B}
  {\bfseries 192} (1981) 159}.

\bibitem{Gilmore:2008gq}
J.B.~Gilmore and A.~Ross, \emph{{Effective field theory calculation of second
  post-Newtonian binary dynamics}},
  \href{https://doi.org/10.1103/PhysRevD.78.124021}{\emph{Phys. Rev. D}
  {\bfseries 78} (2008) 124021}
  [\href{https://arxiv.org/abs/0810.1328}{{\ttfamily 0810.1328}}].

\bibitem{Foffa:2011ub}
S.~Foffa and R.~Sturani, \emph{{Effective field theory calculation of
  conservative binary dynamics at third post-Newtonian order}},
  \href{https://doi.org/10.1103/PhysRevD.84.044031}{\emph{Phys. Rev. D}
  {\bfseries 84} (2011) 044031}
  [\href{https://arxiv.org/abs/1104.1122}{{\ttfamily 1104.1122}}].

\bibitem{Foffa:2019eeb}
S.~Foffa and R.~Sturani, \emph{{Hereditary terms at next-to-leading order in
  two-body gravitational dynamics}},
  \href{https://doi.org/10.1103/PhysRevD.101.064033}{\emph{Phys. Rev. D}
  {\bfseries 101} (2020) 064033}
  [\href{https://arxiv.org/abs/1907.02869}{{\ttfamily 1907.02869}}].

\bibitem{Almeida:2021xwn}
G.L.~Almeida, S.~Foffa and R.~Sturani, \emph{{Tail contributions to
  gravitational conservative dynamics}},
  \href{https://doi.org/10.1103/PhysRevD.104.124075}{\emph{Phys. Rev. D}
  {\bfseries 104} (2021) 124075}
  [\href{https://arxiv.org/abs/2110.14146}{{\ttfamily 2110.14146}}].

\bibitem{Porto:2026fsd}
R.A.~Porto and M.M.~Riva, \emph{{Black hole dynamics at fifth post-Newtonian
  order}}, \href{https://doi.org/10.1007/JHEP09(2026)075}{\emph{JHEP}
  {\bfseries 09} (2026) 075}
  [\href{https://arxiv.org/abs/2604.09545}{{\ttfamily 2604.09545}}].

\end{thebibliography}
\end{document}